%% file: arxiv.tex
\pdfoutput=1
\documentclass[11pt, twoside]{procdoc2}

\RequirePackage{booktabs}
\RequirePackage{amsmath,amsfonts,bm,mathrsfs,dsfont}
\RequirePackage{microtype}
\RequirePackage{IEEEtrantools}
\RequirePackage[shortlabels, inline]{enumitem}

\DeclareMathOperator*{\argmin}{arg\,min}

\newcommand*{\kld}[2]{D_{\mathrm{KL}}\left\{#1;\,#2\right\}}

\DeclareMathOperator{\correlation}{corr}
\DeclareMathOperator{\elpd}{elpd}
\newcommand*{\elpdHatPlain}{{\ensuremath{\widehat{\mathrm{elpd}}}}}
\DeclareMathOperator{\loo}{loo}

\DeclareMathOperator{\normal}{normal}
\DeclareMathOperator{\betadist}{beta}
\DeclareMathOperator{\dirichlet}{Dirichlet}
\DeclareMathOperator{\student}{Student-}
\newcommand{\studentt}{\student t}
\newcommand{\refmodel}{\mathcal{M}_\ast}
\newcommand{\modelp}{\mathcal{M}_\perp}
\newcommand{\thetap}{\theta_\perp}
\newcommand{\data}{\mathbf{D}}
\newcommand{\E}[2]{\mathbb{E}_{#1}\left[#2\right]}
\newcommand{\dist}[2]{\text{dist}\left(#1, #2\right)}

\newcommand{\pkg}[1]{{\fontseries{b}\selectfont #1}}
\let\proglang=\textsf

\RequirePackage[svgnames]{xcolor}
\definecolor{paleaqua}{RGB}{204, 238, 255}
\definecolor{palechestnut}{RGB}{255, 204, 204}
\definecolor{palegoldenrod}{rgb}{1.0, 1.0, 0.88}
\definecolor{palegreen}{RGB}{204, 221, 170}
\definecolor{pastelgray}{rgb}{0.81, 0.81, 0.77}
\definecolor{darkaqua}{RGB}{34, 85, 85}
\definecolor{darkchestnut}{RGB}{102, 51, 51}
\definecolor{darkgoldenrod}{RGB}{238, 238, 187}
\definecolor{darkgreen}{RGB}{204, 221, 170}
\definecolor{darkgray}{rgb}{0.81, 0.81, 0.77}

\RequirePackage{tikz}
\RequirePackage{forest}
\tikzset{
    decision/.style={circle, minimum height=10pt, minimum width=10pt, draw=black, fill=none, thick, inner sep=0pt},
    chance/.style={circle, minimum width=10pt, draw=black, fill=none, thick, inner sep=0pt},
    dots/.style={minimum width=10pt, draw=none, fill=none, thick, inner sep=0pt},
  }
\usetikzlibrary{shapes.geometric,fit,arrows,arrows.meta,calc,positioning}

\tikzstyle{ref} = [rectangle, draw, fill=paleaqua, node distance=1cm, text width=8em, text centered, rounded corners, minimum height=6em, thick, font=\Large]
\tikzstyle{search} = [rectangle, draw, fill=palegoldenrod, node distance=1cm, text width=8em, text centered, rounded corners, minimum height=6em, thick, font=\Large]
\tikzstyle{cv} = [rectangle, draw, fill=palechestnut, node distance=1cm, text width=8em, text centered, rounded corners, minimum height=6em, thick, font=\Large]
\tikzstyle{select} = [rectangle, draw, fill=palegreen, node distance=1cm, text width=8em, text centered, rounded corners, minimum height=6em, thick, font=\Large]
\tikzstyle{posterior} = [rectangle, draw, fill=pastelgray, node distance=1cm, text width=8em, text centered, rounded corners, minimum height=6em, thick, font=\Large]
\tikzstyle{c} = [rectangle, draw, inner sep=2mm, label distance=3mm, dashed]

\newcommand\mybox[2][]{\tikz[overlay]\node[fill=blue!20,inner sep=2pt, anchor=text, rectangle, rounded corners=1mm,#1] {#2};\phantom{#2}}

\title{Advances in projection predictive inference}
\author[1]{Yann McLatchie}
\author[1,2]{S\"olvi R\"ognvaldsson}
\author[3]{Frank Weber}
\author[1]{Aki Vehtari}
\affil[1]{Aalto University, Finland}
\affil[2]{University of Helsinki, Finland}
\affil[3]{Rostock University Medical Centre, Germany}

\makeatletter
\def\runauthor{McLatchie, R\"ognvaldsson, Weber, and Vehtari}
\let\inserttitle\@title
\def\runtitle{\inserttitle}
\makeatother

\begin{document}
\thispagestyle{empty}
\twocolumn[
    \maketitle
    \begin{onecolabstract}
        \input{abstract}
    \end{onecolabstract}
]
\input{body}
\subsubsection*{Acknowledgments}
We acknowledge the computational resources provided by the Aalto Science-IT project. This paper was partially supported by the Research Council of Finland Flagship programme: Finnish Center for Artificial Intelligence, and Research Council of Finland project ``Safe iterative model building'' (340721).
\bibliography{practical-projpred}
%
\onecolumn
\appendix
\input{appendix}
\end{document}

%% file: abstract.tex
The concepts of Bayesian prediction, model comparison, and model selection have developed significantly over the last decade. As a result, the Bayesian community has witnessed a rapid growth in theoretical and applied contributions to building and selecting predictive models. Projection predictive inference in particular has shown promise to this end, finding application across a broad range of fields. It is less prone to over-fitting than na\"ive selection based purely on cross-validation or information criteria performance metrics, and has been known to out-perform other methods in terms of predictive performance. We survey the core concept and contemporary contributions to projection predictive inference, and present a safe, efficient, and modular workflow for prediction-oriented model selection therein. We also provide an interpretation of the projected posteriors achieved by projection predictive inference in terms of their limitations in causal settings.

%% file: body.tex
\section{Introduction}
Suppose we have response observations $y = (y_1, \dotsc, y_n)^\top$ to which we fit a model $\refmodel$ under some prior $p(\theta)$ to achieve the posterior $p(\theta \mid y, \refmodel)$.
This posterior induces a posterior predictive distribution
\begin{equation*}
    p(\tilde{y}\mid y, \refmodel) = \int p(\tilde{y} \mid \theta, \refmodel)p(\theta\mid y, \refmodel)\,\mathrm{d}\theta,
\end{equation*}
where by $\tilde{y} = (\tilde{y}_1, \dotsc, \tilde{y}_n)^\top$, we denote \emph{unobserved} response values. 
The parameter space $\Theta$ can be large, and thus we want to replace it with some restricted parameter (potentially sub-) set $\thetap \in \Theta_\perp$, where $\Theta_\perp$ has lower dimension than $\Theta$.
Following the notation of \citet{vehtari2012}, our aim then is to produce a smaller model $\modelp$ (which we will refer to as the ``restricted model'' herein) in terms of $\thetap$, capable of approximately replicating the predictive performance of the so-called ``reference model'', $\refmodel$. 

In particular we want to identify a distribution $q(\thetap\mid\modelp)$ which induces a posterior predictive distribution, $q(\tilde{y}\mid \modelp) = \int p(\tilde{y}\mid\thetap,\modelp)q(\thetap\mid\modelp)\,\mathrm{d}\thetap$, as similar in Kullback-Leibler (KL) divergence to $p(\tilde{y} \mid y, \refmodel)$ as possible.
Going forward, we use $q(\thetap\mid\modelp)$ to denote the ``projected posterior'' and $p(\tilde{y}\mid\thetap,\modelp)$ the ``restricted data model'', in contrast to the full data model $p(\tilde{y} \mid \theta, \refmodel)$.
We look to achieve $q(\thetap\mid\modelp)$ according to
\begin{multline}\label{eq:projpred-theory}
    q(\thetap\mid\modelp) = \argmin_{\nu} D_\mathrm{KL}\bigg\{p(\tilde{y} \mid y, \refmodel);\\
    \int p(\tilde{y}\mid \thetap, \modelp)\nu(\thetap)\,\mathrm{d}\thetap \bigg\}
\end{multline}
where $\nu$ denotes a distribution over $\thetap$ and $\kld{p}{q} \coloneqq \E{\tilde{y}\sim p}{\log \frac{p(\tilde{y})}{q(\tilde{y})}}$ denotes the KL divergence of $q$ from $p$.

The process of replacing $p(\theta \mid y, \refmodel)$ with $q(\thetap\mid\modelp)$ we refer to as \textit{projection predictive inference}.\footnote{In previous literature, the parameters of the reference model have occasionally been denoted $\theta_\ast$ or similar. We omit this indexing in favour of conditioning on the model for convenience.}

\citet{Lindley_1968} derived an analytic solution to the projection for simple linear regression models, and \citet{goutis1998} and \citet{dupuis2003} proposed a generic projection approach for exponential family models and when posterior inference was made using MCMC. 
Their approach rests on a computationally efficient approximation of the KL divergence by swapping the order of integration and minimisation.
\subsection{Contributions and structure of this paper}
\begin{figure}[t!]
    \centering
    \input{tikz/flowchart.tex}
    \caption{An efficient projection predictive workflow. The nodes shown in \mybox[fill=paleaqua]{blue} relate to fitting and diagnosing a reference model, and are covered in Section~\ref{sec:ref-model}. Once we have this reference model, we perform an initial full-data search (the node shown in \mybox[fill=palegoldenrod]{yellow}), which we deal with in Section~\ref{sec:search}. We then diagnose this initial search, and either cross-validate over multiple search paths if it does not pass our checks, or move directly to selecting a submodel size. These steps are shown in the \mybox[fill=palechestnut]{red} nodes, and will be discussed in Section~\ref{sec:cv}. Finally we discuss in Section~\ref{sec:posteriors} what can be done with the projected posterior (the final node shown in \mybox[fill=palegreen]{green}). Steps that are not always required are indicated by dashed arrows.}
    \label{fig:flow}
\end{figure}
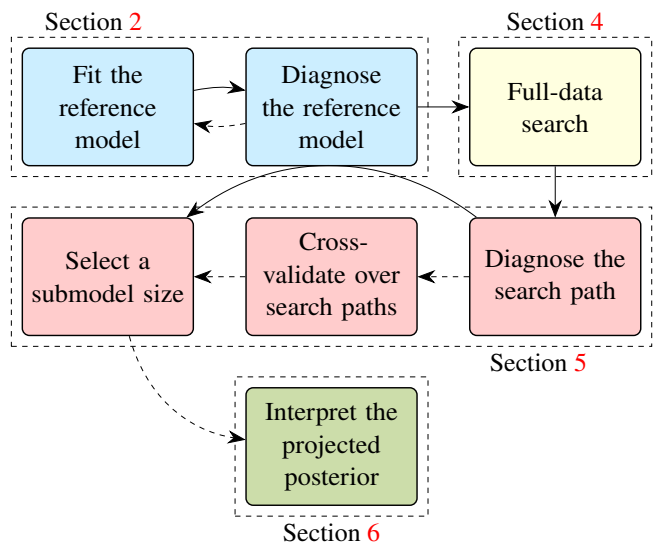
Unlike previous papers that have primarily sought to establish the theoretical foundations of projection predictive inference, we survey the contemporary landscape and detail an efficient workflow for practical projection predictive inference. We aim to bring structure and clarity to the model selection workflow of statisticians, especially those using \proglang{Stan}, and introduce the notion of projection predictive inference to the uninitiated. 

Concretely, we:
\begin{enumerate*}[label=(\arabic*), ref=\arabic*]
    \item decompose the projective model selection workflow into clear and modular components (see Figure~\ref{fig:flow}); \label{contrib:decomp}
    \item propose heuristics to achieve the most robust (in a loose sense, meaning that the selected model size and predictive performance are not too sensitive to procedural choices) results possible with the least computational effort;
    \item provide an interpretation of the ``projected posterior'' (the distribution resulting from the projection of the reference model's posterior onto the parameter space of a given submodel).\label{contrib:causal}
\end{enumerate*}

Section~\ref{sec:ref-model} recapitulates the role of a reference model in model selection, before we present the underlying theory of projection predictive inference specifically in Section~\ref{sec:projpred}; 
Section~\ref{sec:search} deals with the search component of our efficient workflow which is then followed by model validation and selection in Section~\ref{sec:cv}; 
Section~\ref{sec:case-studies} illustrates our workflow in various simulated and real-data expositions; 
finally Section~\ref{sec:discussion} completes this paper with further discussion on the role of projection predictive inference in a Bayesian workflow.
\subsection{Relation to previous literature}
\label{sec:literature}
Bayesian inference has gained traction in recent years as computational availability has made it more broadly accessible. Alongside this surge in popularity and accessibility, the application of Bayesian models to predictive tasks has also become more widespread. In practice, we are often interested in identifying a model whose out-of-sample predictive performance is best while minimising measurement and inferential cost. This view of Bayesian inference was already discussed by \citet{Lindley_1968} who developed a predictive model selection paradigm in which one constructs small models capable of approximately replicating the predictive performance of a so-called ``reference model''. This is the best-performing model (in terms of its predictive performance) available which represents all uncertainty related to the task. Essentially, we fit smaller models to the fit of the reference model (we call this procedure the ``projection'') and use these for their improved interpretability, or to decrease data collection cost \citep{piironen2018}. The computational approximation proposed by \citet{goutis1998} and \citet{dupuis2003} has made this technique feasible both computationally and in terms of a practical workflow. This procedure has become known as projection predictive model selection \citep{vehtari2012}, or, more generally, projection predictive inference \citep{piironen2018}.

Due to its excellent performance, computational efficiency, and easy-to-use software, the projection predictive approach has gained popularity in many real-world problems (see Appendix~\ref{sec:references} for an enumeration of some of these). The \pkg{projpred} package  \citep{piironen_projpred_2022}, based on projection predictive inference, is the most popular Bayesian variable selection package for \proglang{R} when comparing its download numbers to its alternatives (e.g. those recorded in Appendix~\ref{sec:references}) with the help of \pkg{cranlogs} \citep{csardi_cranlogs_2019} (last check: April 26, 2023).

The use of a reference model in model selection (i.e. substituting the observed data with the in-sample predictions made by the reference model) is not in itself a novel idea. In addition to \citet{Lindley_1968}, \citet{leamer_information_1979} previously motivated the use of the Kullback-Leibler (KL) divergence \citep{KullbackLeibler} to quantify the distance between models' predictive distributions, and \citet{martini_predictive_1984} constructed a Bayesian model-averaged reference model. \citet{bernardo_bayesian_2002} later used a symmetric divergence measure, and \citet{nott_bayesian_2010} and \citet{tran2012predictive} extended this to use different divergences. \citet{Vehtari:2021} discuss the interpretation implied by the different divergence measures used.

The idea of a reference model is also common in non-Bayesian literature, where, for example, \citet{harrell_regression_2001} referred to them as a ``gold standard'' model one then seeks to approximate. Such approximation becomes increasingly important as the model space grows. \citet{faraggi_understanding_2001} discussed this in the context of identifying risk groups within large neural networks. Indeed, as wide neural networks grow wider, it becomes more likely that one of their subnetworks is useful for prediction. This is closely related to the so-called ``student-teacher'' learning framework \citep{wang_knowledge_2022}. Under the terminology of ``pre-conditioning,'' \citet{paul_preconditioning_2008} and \citet{hahn2015decoupling} discuss fitting models to consistent estimates from the reference model. Across different fields and under different names, the notion of fitting to the fit of a reference model has become a popular approach to model selection.

Furthermore, projective inference is significantly cheaper than rerunning Markov chain Monte Carlo (MCMC) for submodel fitting (we will use the terms ``submodel'' and ``restricted model'' interchangeably throughout this paper), making it desirable also from a computational perspective. The projection has the advantage of being a deterministic transformation of the reference model's posterior and therefore requires a prior only for the reference model. 

\citet{Pavone2020} show that aside from the projection, other components of the projection predictive model selection framework contribute to the superior performance of projection predictive model selection. The use of cross-validation (CV) for selecting from the space of possible model sizes is one such component (see also Section~\ref{sec:cv}). When we perform forward search through the model space (see Section~\ref{sec:search}), the objective optimised at each of its steps reduces the risk of over-fitting by operating in terms of the difference between predictive distributions and not the submodels' predictive performance directly. Within the projection, \citet{Pavone2020} argue, the use of the reference model's complete posterior predictive distribution (or some sensibly clustered or thinned approximation of it) allows the submodels to better incorporate uncertainty in their own posteriors. 

\citet{piironen2016} presented projective inference in relation to other practical model selection techniques (including sparsifying priors). Later, \citet{piironen2018} introduced the so-called ``clustered projection'', combined projective model selection with Pareto-smoothed importance sampling leave-one-out cross-validation \citep[PSIS-LOO-CV;][]{Vehtari+etal:PSIS:2024, loo}, and investigated reasons for the good performance of projective inference. In more complex settings, \citet{projpred-gp} showed the potential for the procedure's use in Gaussian processes, and \citet{Afrabandpey+etal:2020} demonstrated its use to create more explainable tree models. \citet{mclatchie_identification_2022} motivated an application of the procedure to identify the order of auto-regressive moving-average models. Finally, \citet{catalina_latent_2021}, \citet{catalina_projection_2020}, and \citet{weber_projection_2023} demonstrated how the procedure can be elevated beyond the observation models used in generalised linear models (GLMs), allowing for even broader application. 
We limit our discussion and examples in this paper to GLMs.
\subsection{Motivating examples}\label{sec:examples}
To make the discussion of the properties of projection predictive inference easier to follow, we use a couple of examples throughout the paper. These examples use model selection to different ends, and have data that accentuate certain aspects of the workflow.

The first of these examples is a dataset used to predict lean body weight given weights, body measurements, and ages of $n=251$ individuals \citep{penrose_generalized_1985}. Since accurate measurements of body fat are costly, we wish to predict it using $p=13$ measurements. Some of these measurements require tape measurements of the patients, and we would like to minimise the human cost as much as possible. Thus our task with regards to model selection is purely predictive: identify the minimal set of these $13$ attributes required to accurately predict body fat. Namely, there is no causal interpretation to the results.

The second data provided by \citet{cortez_using_2008} relates Portuguese student performances in mathematics and Portuguese language to their demographics and various social and school factors ($n=395, p=30$). This is a purely observational study, making causal analysis difficult, but learning which predictors are relevant to prediction can help design an intervention study for causal analysis later. As such, model selection in this case is primarily used to identify which external factors best predict the requirement for teacher intervention.
\section{The role of a reference model}\label{sec:ref-model}
The rich reference model in question is the best-performing (in terms of its predictive performance) model we have at our disposal, and is one we would be happy to use as-is. Model selection then becomes applicable when we
\begin{enumerate*}[label=(\arabic*), ref=\arabic*, nosep]
    \item have a predictive rich model, but would like to reduce computational burden,
    \item would like to use a model more robust to changes in the data-generating distribution, or
    \item would like to gain a better understanding of important correlation structures.
\end{enumerate*}
Given the importance of this model, it is reasonable to begin by describing how one might construct one, and the other uses of a reference model outside of projection predictive inference.
\subsection{Building a reference model}
When diagnosing a reference model, there are three primary dimensions we recommend the statistician to investigate:
\begin{enumerate}[nosep]
    \item posterior sensitivity to the prior and likelihood; \label{diag:prior}
    \item posterior predictive checks; \label{diag:ppc}
    \item cross-validation and the influence of data on the posterior. \label{diag:cv}
\end{enumerate}
Corresponding discussion on model diagnosis and visualisation thereof has been covered by \citet{kallioinen_detecting_2022}, \citet{Gabry2019}, \citet{Aki:2020}, and \citet{loo}. The \pkg{priorsense} \citep{priorsense}, \pkg{bayesplot} \citep{bayesplot}, and \pkg{loo} \citep{loo-package} packages are free resources in \proglang{R} to such ends, and \pkg{ArviZ} \citep{arviz_2019} provides an analogue in \proglang{Python}.

Our reference model should be fit with priors that allow for model complexity without over-fitting the observed data. This can include sparsity-inducing priors such as the spike-and-slab \citep{spikeslab}, regularised horseshoe priors \citep{horseshoe}, or the $L_1$-ball prior \citep{l1ball}. More recently, priors model fit (namely $R^2$) such as the R2D2 \citep{r2d2,zhang_bayesian_2022} and R2D2M2 priors \citep{aguilar_intuitive_2023} have been found well-suited to this.
Naturally, subjective prior specification is perfectly inline with our view of building a reference model, and the projection of the reference posterior translates also this information to the restricted model. 
For more discussion on prior elicitation, see the review by \citet{mikkola_prior_2023}.
These priors can be verified with prior predictive checking as described by \citet[Section 2.4]{Aki:2020}, wherein we simulate from the prior distribution and investigate its implications for model predictions in the response space. 

Once we have defined some reasonable priors, we look to diagnose the posterior predictions of our reference model. These checks are in many ways related to the prior predictive checking of before. Here, we simulate from the posterior predictive distribution of our fitted model to check that the simulated predictions resemble the observed data sufficiently well. This can mean sufficient coverage of the posterior predictive simulations, sufficient concentration of the posterior predictive simulations, or sufficiently close moments of the posterior predictive simulations to the observed data. For more discussion surrounding posterior predictive checks, the reader is advised to consult the work of \citet[Section 6.1]{Aki:2020} and \citet[Section 5]{Gabry2019}.

Finally, we can investigate the effect of individual data to the posterior from our reference model. Doing so allows us to understand which regions of data are better represented by our model, and can also act as a sanity check for model misspecification. Since posterior predictive checks ``double-dip'' our data by using it both for fitting and diagnosis, it is liable to be overly optimistic in its conclusions. Cross-validation shows how the model would behave on unseen data. For example, investigating the leave-one-out (LOO) posterior predictive distributions by comparing the distribution of the LOO probability integral transform (LOO-PIT) values at the observed data to the uniform distribution which would hold under a calibrated model can reveal over- or under-dispersion of the LOO posterior predictive distributions \citep{Gabry2019,Aki:2020}. 
PSIS-LOO-CV \citep{Vehtari+etal:PSIS:2024, loo} can further automatically provide insight into the behaviour of our cross-validation approximation. 

In a word, should a reference model pass checks and were it not for the need for model selection, we would be happy to use it as-is.
\subsection{Using a reference model}
While we interest ourselves primarily in the use of reference models as a predictive benchmark for projections, they can be used more generally as approximations of the underlying data-generating distribution. We presently discuss other use cases of reference models before returning to their role specifically in a projective model selection workflow. 

Indeed, reference models can be combined with many other model selection procedures to increase selection stability. 
\citet{Pavone2020} indicate that the more complex the models considered and the fewer data there are, the more valuable a reference model is in filtering some of the noise present in data. For instance, reference models can be used in minimal subset selection as part of a Bayesian step-wise selection strategy \citep{Pavone2020}. This might involve beginning with the reference model and at each step, excluding the predictor with associated regression coefficient $\theta$ having the highest Bayesian $p$-value. The selection continues for as long as the reduced model has a better estimate for the expected log (pointwise) predictive density (for a new dataset) \citep[elpd;][]{loo} than the current model. \citet{Pavone2020} discuss how reference models can also be used in complete variable selection through either local false discovery rate minimisation \citep{efron_microarrays_2008,efron_large-scale_2010}, empirical Bayesian median analysis \citep{johnstone_needles_2004}, or based on posterior credible intervals \citep{Pavone2020}.
\section{Practical projection predictive inference}\label{sec:projpred}
Given a reference model, we wish to achieve a parsimonious restricted model whose posterior predictive distribution most closely resembles that of the reference model.\footnote{Unless stated otherwise, the term ``restricted model'' will always refer to a candidate model that the reference model has been projected onto. Thus, the associated ``posterior predictive distribution'' is in fact a predictive distribution based on the \emph{projected} posterior.} 
\subsection{Computational projections}
Given $\mathcal{S}$ posterior draws $\theta^{(s)},\,s \in \{1, \dotsc, \mathcal{S}\}$ from the reference model,
we can approximate Equation~\ref{eq:projpred-theory} as
\begin{multline}
    \argmin_{\theta^{(1)}_\perp,\,\dotsc,\,\theta^{(\mathcal{S})}_\perp} \sum_{i=1}^n D_\mathrm{KL}\bigg\{\frac{1}{\mathcal{S}} \sum_{s} p(\tilde{y}_i\mid\theta^{(s)}, \refmodel);\\
    \frac{1}{\mathcal{S}} \sum_{s} p(\tilde{y}_i\mid \theta^{(s)}_\perp, \modelp) \bigg\}.
\end{multline}
The minimisation over all of these posterior draws simultaneously, however, remains infeasible.
\citet{goutis1998} and \citet{dupuis2003} propose to swap the order of integration and minimisation, leading to a computationally efficient approximation to the original KL minimisation task.
Now the minimisation is done draw-by-draw:
\begin{multline}
    \argmin_{\theta^{(s)}_\perp} \sum_{i=1}^n D_\mathrm{KL}\bigg\{p(\tilde{y}_i\mid\theta^{(s)}, \refmodel); \\
    p(\tilde{y}_i\mid \theta^{(s)}_\perp, \modelp) \bigg\}. \label{eq:draw-wise}
\end{multline}
After finding each $\theta^{(s)}_\perp$ individually, we concatenate them to represent the projected posterior.
Empirically, this approximation has been found to perform well (see the references in the last paragraph of Section \ref{sec:literature}).

\citet{tran2012predictive} simplified the projection by finding a single value $\thetap$ that minimises the divergence 
\begin{multline}
    \argmin_{\thetap} \sum_{i=1}^n D_\mathrm{KL}\bigg\{\frac{1}{\mathcal{S}} \sum_{s} p(\tilde{y}_i\mid\theta^{(s)}, \refmodel); \\
    p(\tilde{y}_i\mid \thetap, \modelp) \bigg\}. \label{eq:single-point}
\end{multline}
This leads to a projected posterior predictive distribution which is less expressive, and has strictly the parametric shape of $p(\tilde{y_i}\mid\thetap, \modelp)$. 

\citet{piironen2018} suggested an approach which is between the draw-by-draw and single-point projections. 
The $\mathcal{S}$ posterior draws are clustered so that we may write $\{1, \dotsc, \mathcal{S}\} \supseteq \mathop{\bigcup}_{c = 1}^C \mathcal{I}^\ast_c$ with disjoint and non-empty index sets $\mathcal{I}^\ast_c$ (more discussion on how this clustering is performed can be found in their Section 3.3).
For each cluster $c \in \{1, \dotsc, C\}$, the single-point projection is performed, and we get $C$ projections $\theta^{(c)}_\perp$. The corresponding projected predictive distribution is 
\begin{equation*}
\sum_{c=1}^C w_c\, p(\tilde{y}\mid\theta^{(c)}_\perp, \modelp),
\end{equation*}
where $w_c$ is weight of a cluster proportional to the number of draws in cluster $c$.

Setting $C = \mathcal{S}$ recovers the draw-by-draw approach of Equation~\ref{eq:draw-wise}.
Similarly, setting $C = 1$ (and $\mathcal{I}^\ast_1 = \{1, \dotsc, \mathcal{S}\}$) is the single-point approach (Equation~\ref{eq:single-point}). 
This is computationally cheaper, and can be useful for searching the model space (see Section~\ref{sec:lasso}).
Choosing the number of clusters regulates the trade-off between computational convenience and projective accuracy: we prefer using small $C$ in the search phase, and larger $C$ for post-selection validation and inference.

In the case of exponential family models, and in particular (hierarchical) GLMs, KL-divergence can be efficiently minimized by equivalent (marginal) maximum likelihood using, for example, the \pkg{lme4} package \citep{bates_fitting_2015} in \proglang{R}. For implementation details, see the papers by \citet{catalina_latent_2021,catalina_projection_2022,weber_projection_2023,piironen2018}. In case of clustering, the only difference compared to traditional MCMC draws is that in any post-projection inference, the cluster weights need to be taken into account. 
The (marginal) maximum likelihood solution does not depend on the value of possible dispersion parameter. After computing the projected values for the model parameters $\theta$, the projections for exponential family dispersion parameters are obtained via Equation~17 in the paper by \citet{piironen2018}.
\subsection{An efficient projection predictive workflow}
Having provided a theoretical motivation of projections, we present the primary contribution of this paper: an efficient workflow for projection predictive inference. The steps inherent to the workflow are visualised in Figure~\ref{fig:flow}:
\begin{enumerate}[nosep]
    \item begin with a reference model, which we will assume to be well-considered and appropriate to the task (see Section~\ref{sec:ref-model} for more discussion);
    \item perform an initial search through the model space, or some heuristic subset of it, without cross-validation and using all available data to achieve a series of nested, increasingly complex submodels deemed the \emph{solution path} (Section~\ref{sec:search});
    \item diagnose this solution path to detect possible over-optimism using the cross-validated predictive performance of the submodels along this solution path (Section~\ref{sec:cv});
    \item if the initial solution path is deemed to be over-optimistic, then in a second stage include the search in the cross-validation, and then combine the resulting cross-validated performance evaluations of models along each fold-wise solution path (Section~\ref{sec:cv-paths});
    \item select the minimal submodel size capable of producing similar predictive performance to our reference model (Section~\ref{sec:select});
    \item finally, diagnose whether inference directly with our projected posterior (if such inference is desired) is reasonable with calibration tests (Section~\ref{sec:posteriors}).
\end{enumerate}
Before any of this, however, it is reasonable to ask whether model selection is required at all, and whether the projection predictive inference is the flavour best-suited to one's needs.
\section{Search heuristics}\label{sec:search}
Once we have built our reference model and identified that the projection predictive inference fits our use case, we begin by producing some set of models to compare our reference model to. It is this search stage which produces our solution path, and there are two primary search methods we consider: KL divergence-based forward search and Lasso ($L_1$) regularisation search. These two have useful speed-accuracy balance and afford the possibility of an efficient stopping rule, but other algorithms can also be used.

\subsection{Forward search}
Forward search is a systematic approach to model selection where we iteratively project the reference model onto submodels of growing size, starting from the ``empty'' (intercept-only) model. 
At each step, we choose the predictor that brings the model closest to the reference model in terms of KL divergence of their (projected) posterior predictive distributions. 
This process continues until all predictors are included or a pre-defined model size limit is reached. 

We favour forward search over backward search for the ability to perform early-stopping along the path (i.e. only search up to some number of predictors fewer than the size of the reference model) starting with the least computationally expensive models (while backward search begins with the most computationally expensive calculations).
Although forward search is a sensible default strategy and can be implemented with any number of clusters in projection, it can become computationally expensive when dealing with a large number of predictors.
\subsection{Lasso-type search}\label{sec:lasso}
\citet{piironen2018} discuss how in the case of single-cluster projections, one can reduce computational cost further by performing a Lasso-type search \citep{Tibshirani2011, Zou2006}, also called \emph{$L_1$ search}. Concretely, this involves fitting a model with Lasso penalty to the in-sample predictions made by the reference model \citep[thus, the $L_1$ search solves an $L_1$-penalised projection problem, see e.g.][whereas the forward search solves the original projection problem]{piironen2018} and investigating the order in which the predictors enter the Lasso model. 
Concretely, the single-point projection for the regression coefficients, denoted $\beta$, in GLMs is given by \citep{piironen2018}
\begin{multline}
    \argmin_{\beta, \phi} \Bigg[\sum_{i=1}^n D_\mathrm{KL}\bigg\{\frac{1}{\mathcal{S}} \sum_{s} p(\tilde{y}_i\mid\theta^{(s)}, \refmodel); \\
    p(\tilde{y}_i\mid \beta, \phi, \modelp) \bigg\} + \lambda \lVert \beta \rVert_1\Bigg].
\end{multline}
In GLMs, the model parameters $\theta$ comprise both the regression coefficients $\beta$, and the dispersion parameter $\phi$ if present: $\theta = \{\beta, \phi\}$.
In the case of the exponential family, the optimisation with respect to $\beta$ does not depend on the dispersion parameter $\phi$, and we can solve the two sequentially.

Optimising this objective over a range of $\lambda$ yields a sequence of submodels with different number of predictors, which we then use to order the predictors.
In our experience, Lasso-type search is computationally faster than forward search, but has higher variability in the search results (with that variability referring to variability across data realizations, as can be seen, e.g., by bootstrapping). 
\subsection{Output of the search heuristic}
After our projection predictive workflow has identified a predictor ordering (the solution path), either through forward or $L_1$ search, we store this ordering so as to avoid repeated computations. The predictor ordering alone can also be interesting to the practitioner, who might identify some intuition about their problem from this alone. In a forward search, we can store (and later access) the KL divergence from the reference model to each submodel, so it might be the case that we can observe an elbow when plotting the KL divergence along the solution path, indicating some saturation of information being provided by predictors. In our experience, these elbows are usually very soft and difficult to infer from. In Section~\ref{sec:select}, predictive performance \emph{utilities} are reasoned on instead of KL divergence for picking a submodel size.
\subsection{Modularity of the search heuristic}
\begin{figure}[!t]
    \centering
    \input{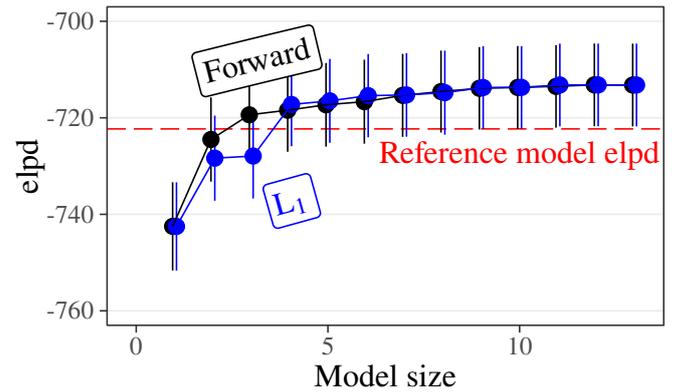}
    \caption{Body fat example. The elpd point estimate and one standard error bars of submodels along the full-data solution path under forward search (in black) and $L_1$ search (in \mybox[fill=blue!20]{blue}) relative to the reference model (shown in \mybox[fill=red!20]{red}). The two methods identify different predictor orderings, leading to different elpd values along the solution path. In particular, forward search tends to produce smoother such elpd curves than $L_1$ search. We have omitted the submodel of size zero (the intercept-only model) as it is much worse than the submodel of size one, and is the same under both search methods trivially.}
    \label{fig:method-paths}
\end{figure}
\begin{figure*}[!t]
    \centering
    \input{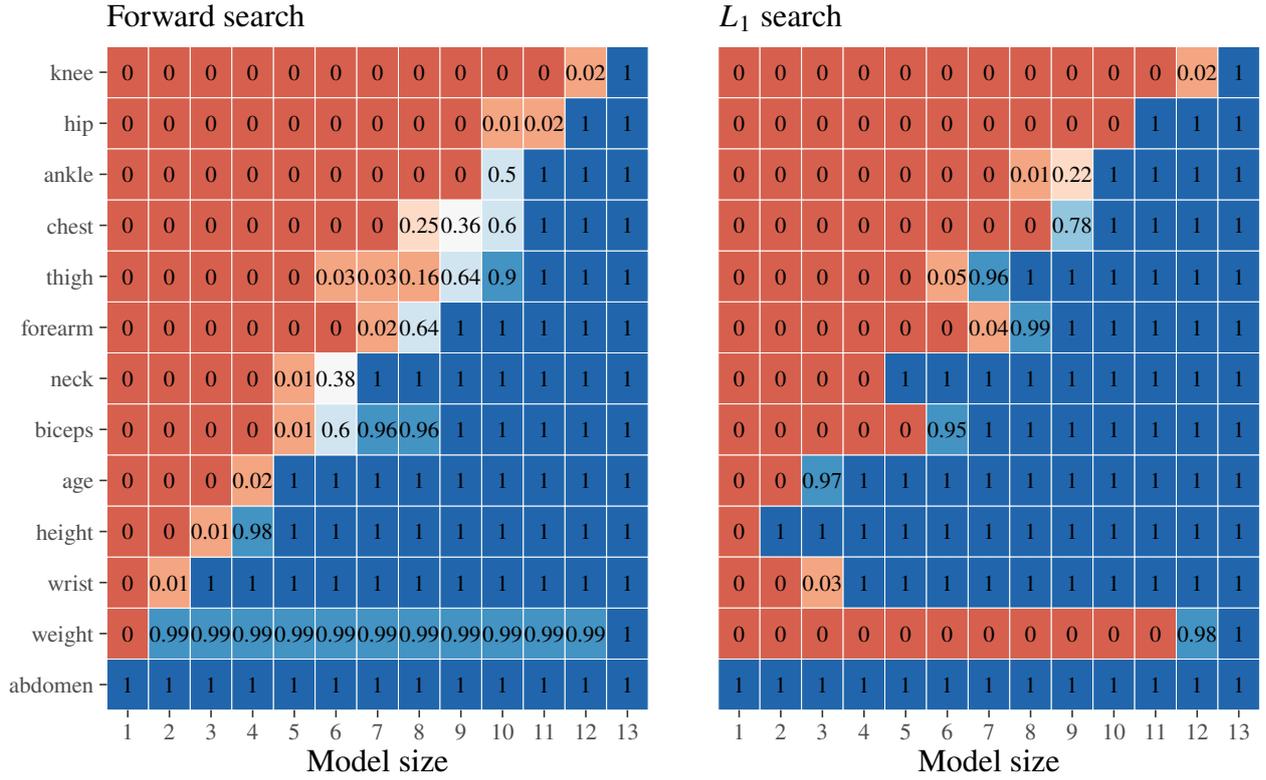}
    \caption{Body fat example. A comparison of the occurrence of predictors over multiple LOO-CV selection folds when using forward search and $L_1$ search heuristics. Shown are \emph{cumulative} rates, i.e., the proportions of CV folds that include a given predictor from the $y$-axis at model sizes smaller than or equal to a given model size from the $x$-axis. In this case, the search paths are quite stable across the CV folds as most of the non-zero cumulative rates are 1. In case of collinearity of predictors, the used search method can have a large effect on the search order, but it is unlikely to have a big change in the predictive performance of the selected model.}
    \label{fig:lasso-v-forward}
\end{figure*}
Our workflow remains a modular initiative. Indeed, we do not require or assume that any specific search heuristic is implemented to continue.

Other search heuristics can be just as easily implemented, including exhaustive search \citep{galatenko_highly_2015}, backward search \citep{nilsson_artificial_1998}, and stochastic \citep{george_variable_1993,ntzoufras_stochastic_2000} and shotgun stochastic \citep{jones_experiments_2005} search. Our general workflow can admit any of these without modification, although we have chosen the above two given they remain efficient and stable in selection.

In particular, while Lasso search may achieve more stable paths over CV folds, forward search often achieves more predictive submodels and converges to the performance of the reference model faster. We have noted empirically that the Lasso path is also liable to fail in situations of high block-wise correlation. In general, it is possible for the two methods to identify different predictor orderings; in Figure~\ref{fig:method-paths} we visualise how these different ordering are reflected by the different predictive performance of submodels along the two paths in our body fat example. 
We use the elpd introduced previously as our predictive metric, which we will define more formally in Section~\ref{sec:loo-cv}.

We can investigate exactly which predictors differ in the ordering in Figure~\ref{fig:lasso-v-forward}. Indeed, we find that in the body fat example, forward search is slightly less stable across cross-validation folds compared to lasso search. It is also clear from this figure that the order with which predictors enter the solution path is very different, hence why the elpds of smaller submodels differ between methods. 
\subsection{Computational efficiency of search heuristics}\label{sec:search-efficiency}
Since at each step in forward search, we project the reference model onto a subset of predictors, the complete forward search requires $\frac{p (p + 1)}{2} + 1$ projections if there are $p$ predictor terms (excluding the intercept).

We can perform the search using $C$ draws (or clusters), and then re-project the submodels along the solution path using more draws (or clusters), say $C^\prime$ (where $C' \gg C$) when evaluating the predictive performance of the submodels. If it takes $t_{\text{proj}}$ seconds to perform the projection for one model in the forward search using $C$ projected draws (or clusters of posterior draws), then a second projection using $C'$ projected draws will take approximately $\frac{C'}{C} \cdot t_{\text{proj}}$ seconds. In our experience, it is usually sufficient to perform the search with $20$ clusters, to perform the performance evaluation along the solution path with $400$ thinned posterior draws (\textit{not} clusters), and finally to use all posterior draws when projecting onto the final selected model. This way, we optimise the balance between stability and efficiency.

We recommend one use forward search over Lasso search for final decision making. However, this is not to say that the latter is not a useful tool: when the number of predictors in the search space is large (greater than $40$, say), $L_1$ search is significantly faster to compute; one can use it to ascertain a heuristic upper bound on the submodel size, and then re-run forward search only up to this bound.

Naturally, performing the projections using more draws or clusters can improve the stability of the solution path at an increased computational cost. With extreme thinning, variability in model selection can increase. This can be reduced by smarter thinning \citep[for example using Wasserstein distance-based thinning as proposed by][]{south_postprocessing_2022} or \textit{ad hoc} clustering. We note, however, that the cluster centroids may not be the best possible approximation to the posterior. As such, one may inadvertently introduce a small amount of bias into the model size selection by using clusters in cases of difficult (for example multi-modal, skewed, or thick-tailed) parameter posteriors. On the whole, this bias will likely remain small since we use the same draws over all projections (in a word, we repeat the same mistake at each step), so that the bias tends to be in the same direction intuitively.
\section{Selection methods}\label{sec:cv}
Having achieved a predictor ordering in the search phase, we turn our attention to the predictive performance of submodels along the path. While in principle, this performance evaluation can be run on the same data that the search was run with, it is more sensible to run the performance evaluation on new data, typically leading to cross-validation. In the ideal case, the whole procedure (search and performance evaluation) is included in the cross-validation, meaning that the search is performed with the training data of each CV fold separately and the performance evaluation is performed with the test data of the respective CV fold. As this is computationally expensive, it makes sense to run a full-data search first, and approximately cross-validate only for the evaluation part using PSIS-LOO-CV. However, it may be that this is not sufficient, and in this case we include the search in the cross-validation.

In this section, we first detail two methods of cross-validating the predictive performance of individual models, before discussing how one might conduct cross-validation for the entire procedure should we over-fit the data in the search path.
\subsection{K-fold cross-validation}
In the $K$-fold-CV paradigm, we fit the reference model and repeat the model selection $K$ times \citep{piironen2018}. We then estimate a pre-specified utility (in practice we interest ourselves in the elpd and thus it is used as an example here) of the model at position $k \in \{0, 1, \dotsc\}$ of the solution path and the reference model at observation $i$, denoted $u_k^{(i)}$ and $u_\ast^{(i)}$, respectively. The utility differential between a submodel of complexity $k$ and the reference model is then estimated with
\begin{equation}
    \Delta\bar{U}_k = \sum_{i=1}^n(u_k^{(i)} - u_\ast^{(i)}). 
\end{equation}
\subsection{Leave-one-out cross-validation}\label{sec:loo-cv}
The Bayesian LOO-CV estimate for a model's elpd on $n$ data observations is defined as
\begin{align}
    \elpd_{\loo} &= \sum_{i=1}^n\log p(y_i\mid y_{-i}) \nonumber\\
    &= \sum_{i=1}^n \log\int p(y_i\mid\theta)p(\theta\mid y_{-i})\,\mathrm{d}\theta ,
\end{align}
where $y_{-i}$ denotes all observations omitting the $i^\text{th}$. In PSIS-LOO-CV, rather than na\"ively computing this integral $n$ times (once for each fold, which requires refitting the reference model $n$ times), we instead estimate the posterior predictive distribution $p(y_i\mid y_{-i})$ based on the importance sampling estimate
\begin{equation}
     p(y_i\mid y_{-i}) \approx \frac{\sum_{s=1}^\mathcal{S}p(y_i\mid\theta^{(s)})r(\theta^{(s)})}{\sum_{s=1}^\mathcal{S}r(\theta^{(s)})}, 
\end{equation}
wherein the weights take the form
\begin{equation}
    r(\theta^{(s)}) \propto \frac{1}{p(y_i\mid\theta^{(s)})}. 
\end{equation}
In PSIS-LOO-CV, these importance weights are stabilised and diagnosed with Pareto smoothing as previously discussed in Section~\ref{sec:ref-model} \citep{Vehtari+etal:PSIS:2024}. The primary advantage of PSIS-LOO-CV is that it does not require repeatedly refitting the reference model (and, optionally, it also allows running only a full-data search instead of fold-wise searches). When the number of observations or the complexity of the model grows, even this estimate can be expensive to compute and may occasionally fail. In this case, we fall back on $K$-fold-CV instead for the predictive utility.
\subsection{Cross-validation over solution paths}\label{sec:cv-paths}
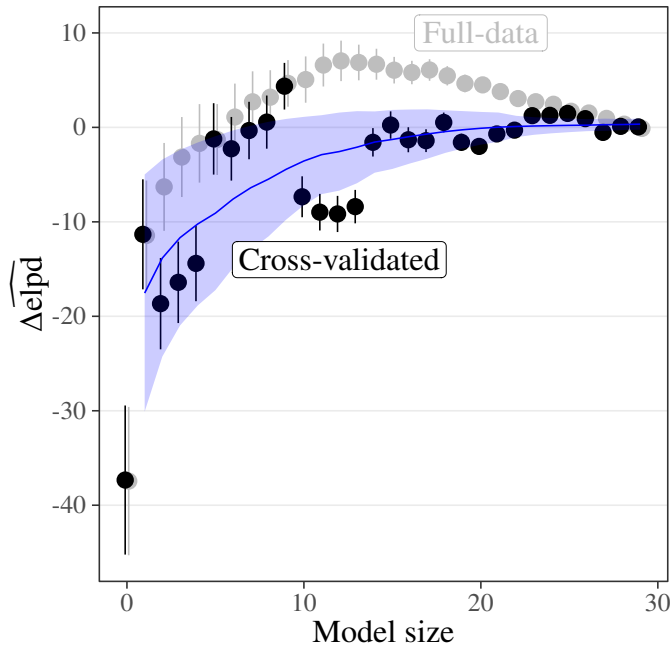
\begin{figure}[!t]
    \centering
    \input{tikz/schools-spline.tex}
    \caption{Portuguese student example. We show point estimates for the elpd difference to the reference model of both the full-data search in \mybox[fill=gray!20]{grey}, and the cross-validated search in black, along with one standard error bars. When we include the search in the cross-validation, we find that instability in the predictor ordering leads to ``jumpy'' behaviour in the elpd differences. We smooth the elpd difference estimates with a monotonic spline shown in the \mybox[fill=blue!20]{blue} line, where the enveloping ribbon represents the one standard error interval around the point estimate. We do not include the null model in the spline, since we understand that the increase in predictive performance is so great that the intercept-only model is of little importance to later analysis.}
    \label{fig:school-spline}
\end{figure}
Once we have cross-validated the predictive performance of the submodels along our initial full-data solution path, we can then immediately diagnose selection-induced over-optimism in the elpd estimates. This is apparent when we observe a ``bulge'' in the elpd of submodels along the solution path. Specifically, if a submodel smaller than the reference model has elpd much higher (better) than the latter, we can conclude that the cross-validation estimates for full data solution path are over-optimistic. Recall the predictive performance along the two solution paths from  Figure~\ref{fig:method-paths}, where we see that both initial solution paths have over-optimistic performance estimates to the point where models of more than roughly five predictors achieve better elpd than the reference model of 13 predictors. Moreover, the over-optimistic submodel elpds are in some cases ``significantly'' better than that of the reference model.\footnote{By ``significant'', we mean that the estimated elpd difference is at least one standard error larger than zero.}

In this case, understanding that we are witnessing over-optimism in the elpd estimates along the full data solution path, we should cross-validate the models along the solution path \textit{and} cross-validate the model search. Due to the vastly increased computational requirements, we suggest that the search be re-run only up to the model complexity inducing the most over-optimistic elpd estimate (for example in Figure~\ref{fig:method-paths}, we might re-run up to model size nine or ten to cut down some computational cost). Doing so achieves submodel performance evaluations that have smaller bias \citep{piironen2016}.

Consider now the Portuguese students motivating example. In Figure~\ref{fig:school-spline} we see how the elpd estimates along full-data solution path are over-optimistic, inducing a bulge at model size roughly 12 which gradually diminishes to recover the elpd of the reference model. By including the search in the cross-validation, the bias in the elpd estimates is reduced. However, in doing so, we introduce some ``jumpy'' behaviour in the elpd differences at some model sizes. This is caused by instability across cross-validation folds in the predictor ordering.
\subsection{Computational efficiency of cross-validation}
The initial full-data search lends itself naturally to an estimate of the time required to perform $K$-fold-CV over the whole procedure (both search and evaluation): suppose each projection uses $C$ projected draws and takes $t_{\text{proj}}$ seconds. Then running the search for each of the $K$ CV folds with the same number $C$ of projected draws will take approximately 
\begin{equation}
    t_{\text{search}} = K \cdot \left(\frac{p (p + 1)}{2} + 1\right) \cdot t_{\text{proj}}
\end{equation}
seconds if not run in parallel and not taking into account the time required for refitting the reference model in $K$-fold-CV. Since a CV including the search is always a CV for predictive performance evaluation on the test data of each CV fold, it makes sense to combine this estimated search time with the estimated time for the corresponding performance evaluation based on $C'$ projected draws, which is 
\begin{equation}
    t_{\text{eval}} = K \cdot (p + 1) \cdot \frac{C'}{C} \cdot t_{\text{proj}}
\end{equation}
seconds, giving an estimate of
\begin{equation}
    t_{\text{total}} = t_{\text{search}} + t_{\text{eval}}
\end{equation}
seconds in total. These heuristics assume that one wishes to perform an exhaustive search over the model space. That is, we search all model sizes up to the size of the full model, $p_{\text{max}} = p$.

As was previously discussed, we can use a smaller number of clusters in the search phase to achieve a solution path and then re-project the models along that path using more draws in order to achieve more stable cross-validation results. Typically, we have found that using approximately $20$ clusters in the search phase, and $400$ thinned posterior draws for cross-validation provides a good balance of stability and efficiency. We do not recommend using clustered projections for performance evaluation due to the risk of introducing bias as previously discussed in Section~\ref{sec:search-efficiency} \citep[although in the search phase it is likely that the bias induced is in the same direction across all submodels so that the solution path is not as affected;][]{piironen2018}.

Additional speed-up in cross-validation computation can be obtained by combining the fast PSIS-LOO results along the full data search path with running the cross-validation over solution paths only for subset of cross-validation folds using the subsampling LOO with difference estimator proposed by \citep{Magnusson+etal:2020:bigloocomparison}.

Having then attained a cross-validated performance estimate for the model sizes along the solution path, we look to select the minimal set of predictors replicating the performance of the reference model to a sufficient degree.
\subsection{Choosing a submodel size}\label{sec:select}
We are now faced with the final decision of identifying the smallest submodel (minimal predictor set) whose predictive performance most closely resembles that of the reference model. For this task, we propose two primary heuristics: one based on differential utility intervals, and another on the mean elpd difference to the reference model.

The former is most common, and has been previously considered by \citet{piironen2016, piironen2018, catalina_latent_2021, piironen_projpred_2022}, and \citet{weber_projection_2023}. In this case, we denote the utility of the reference model as $u_\ast$, and the utility of the submodel with $k$ predictors as $u_k$. Having cross-validated the predictive performance of the submodels, we then select the smallest submodel whose utility is one standard error less than the reference model's utility, with the standard error being that of the utility \emph{difference}, denoted $s_k$. Formally, we select the smallest $k$ such that
\begin{equation}
    \hat{u}_\ast \leq \hat{u}_k + s_k,\tag{SE}\label{eq:utility-ci}
\end{equation}

The second heuristic, motivated by \citet{sivula_uncertainty_2022}, is to choose the smallest $k$ such that the estimated elpd of the submodel is at most four less than the estimated elpd of the reference model. That is, we choose the smallest $k$ such that
\begin{equation}
    \hat{u}_k - \hat{u}_\ast \geq -4. \tag{$\Delta$utility}\label{eq:delta-elpd}
\end{equation}
We have found through our own experiments that the \ref{eq:delta-elpd} heuristic results in more stable selection, and selects better-performing submodels (results not shown in this paper). We have also observed that it may select larger models than the \ref{eq:utility-ci} heuristic. In general, however, these have been noted to behave similarly.

In certain ``large $n$, small $p$'' regimes, it may be that even following a cross-validation including the search, the elpds at the model sizes along the path can be erratic and uncertain. This will be seen later in Figure~\ref{fig:elpd_plot_case_study_4.2} in our second case study of Section~\ref{sec:weakly-relevant}. In order to stabilise selection of submodel size in these cases, and reduce the noise in the solution path, we smooth the submodels' estimated elpds. One well-established tool for smoothing a collection of points $\{(x_i,y_i)\}_{i=1}^n$ are B-splines \citep{ramsay_monotone_1988,eilers_flexible_1996,eilers_perfect_2003,eilers_unimodal_2005}. These are smoothly joined polynomial segments, knotted together at some (usually regular) intervals that allow us to model some functional form. For instance, the true elpd differences along model sizes are typically monotonically increasing and negative. In practice, we use the \pkg{scam} package \citep{pya_scam_2022} to compute these splines. 

\begin{figure}[!t]
    \centering
    \input{tikz/schools-dendro.tex}
    \caption{Portuguese student example. Dendrogram of the predictors in the reference model, using the correlation of single-predictor models (Equation~\ref{eq:cov-dist}) as the distance metric. Such dendrograms can reveal predictive similarities between predictors and aid intuition in variable subset selection.}
    \label{fig:school-dendro}
\end{figure}
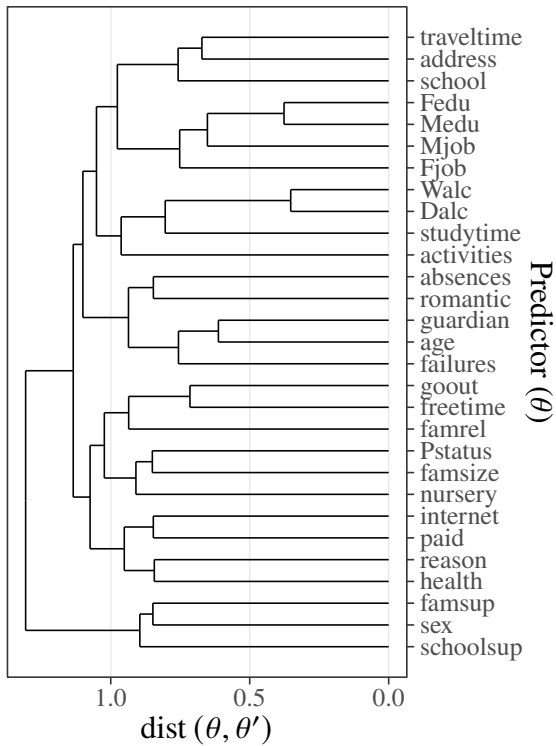

Usually the intercept-only model has so much worse predictive performance  than the model with just one additional predictor, that it is usually best to omit the intercept-only model (and even occasionally the size-one model) when fitting the spline. Further, we fit these splines to the normalised elpd difference values (elpd difference divided by its standard error), such that accuracy of each elpd difference estimate is taken into account, too. The smoothing of the elpd values is illustrated in Figure~\ref{fig:school-spline}.
Once a minimal predictor set is identified, we may find that one or more predictors could be substituted with others to achieve very similarly performing submodels of the same complexity but at a lower measurement cost. Such instances can be diagnosed by inducing the distance metric between predictors with regression coefficients $\beta, \beta^\prime$,
\begin{multline}
    \dist{\beta}{\beta^\prime} = 1 - |\correlation\big(\E{}{\tilde{y}\mid\mathcal{M}_{\beta}, \data}, \\
    \E{}{\tilde{y}\mid\mathcal{M}_{\beta^\prime},\data}\big)|, \label{eq:cov-dist}
\end{multline}
where $\E{}{\tilde{y}\mid\mathcal{M}_{\beta}, \data}$ are the in-sample predictions made by the single-predictor model, $\mathcal{M}_\beta$, containing only the intercept term and $\beta$ \citep{paasiniemi_methods_2018}. Computing then the pair-wise distances as defined in Equation~\ref{eq:cov-dist} between all predictors, we can achieve such dendrograms as in Figure~\ref{fig:school-dendro}. Specifically, we understand that should we identify a submodel admitting an individual's weekend alcohol consumption in the predictor set (named ``walc'' in the plot), but not their weekday alcohol consumption (``dalc''), that similar results might be achieved by including the latter instead of the former. This can be useful for building predictive models without explicit costs for the measurements. For instance, should it be found that similar predictions can be achieved in a medical setting through non-invasive measurements instead of invasive measurements, then the less intrusive measurement could be preferred. 

And so concludes projection predictive model selection: we have identified the minimal predictor set and identified correlation structures between them. What remains to be discussed is what can be done with this information.
\section{Numerical demonstrations}\label{sec:case-studies}
Presently we manifest the recommendations laid out in this paper through several simulated and real-data examples. The below case studies were carried out using the \pkg{projpred} package in \proglang{R} \citep{piironen_projpred_2022}.\footnote{The code is freely available at \url{https://github.com/yannmclatchie/projpred-workflow/}.}
\subsection{Highly correlated predictors}\label{sec:high-corr}
In the first case study, we employ projection predictive inference on simulated data where predictors are highly correlated. The setup is the same as in \citet[Section 4.2]{piironen2016}, with data generated according to 
\begin{align}
    x &\sim \mathrm{multivariate\,}\normal(0_p,R) \nonumber \\
    y &\sim \normal(w^\top x,\sigma^2),  \nonumber
\end{align}
where the number of predictors is set to $p = 100$.
The matrix $R \in \mathbb{R}^{p\times p}$ is block diagonal, each block being of dimension $5 \times 5$. Each predictor has mean zero and unit variance and is correlated with the other four predictors in its block with coefficient $\rho=0.9$, and uncorrelated with the predictors in all other blocks. Further, the weights $w$ are such that only the first $15$ predictors influence the target $y$ with weights $(w^{1:5},w^{6:10},w^{11:15}) = (\xi,0.5 \xi,0.25 \xi)$ and zero otherwise. We set $\xi = 0.59$ to fix $R^2 = 0.7$, and $\sigma^2=1$. We simulate $n=500$ data points according to this data-generating process (DGP).
\begin{figure}[!t]
    \centering
    \input{tikz/dendrogram_case_study_4.1.tex}
    \caption{Highly correlated predictor simulated case study. A dendrogram of the predictors (with associated regression coefficients $\theta$) present in the reference model, where the distance recorded on the $x$-axis is computed according to Equation~\ref{eq:cov-dist}, and where clusters containing chosen predictors have been coloured \mybox[fill=blue!20]{blue}.}
    \label{fig:dendrogram_case_study_4.1}
\end{figure}
\begin{figure*}[!t]
    \centering
    \input{tikz/elpd_plot_case_study_4.2.tex}
    \caption{Many weakly-revelant predictors case study. Over-optimistic predictive performance estimates based on a full-data search with many weakly-relevant predictors (in \mybox[fill=gray!20]{grey}), along with predictive performance estimates from PSIS-LOO-CV where the search has been included (in black) up to the lowest model size inducing the maximum degree of over-confidence (judged by elpd point estimate) in the original search. Since the cross-validated predictive performance estimates are very ``jumpy'', we smooth them with a monotonic spline (shown in \mybox[fill=blue!20]{blue}) and perform our heuristic selection on that instead. Once more, we show the reference model's elpd by the horizontal dashed \mybox[fill=red!20]{red} line, and the selected model size in the vertical dashed \mybox[fill=gray!20]{grey} line. }
    \label{fig:elpd_plot_case_study_4.2}
\end{figure*}

We fit to these data a linear regression on all predictors and with an R2D2 prior (see Appendix \ref{appendix:R2D2} for more information on the construction of the R2D2 prior).
Our reference model is the linear regression on all available predictors.
As a result, the model passes basic computational diagnostics as well as basic model diagnostics, and we are comfortable using it as a reference model.\footnote{In this case, $\widehat{R}$ values are all smaller than 1.01 \citep[following the recommendations of][]{gelman_inference_1992,vehtari_rank-normalization_2021}, and bulk and tail effective sample sizes are high. All Pareto $\hat{k}$-values are satisfactory ($\hat{k}<0.7$) meaning  the second-order moment of the elpd estimate is likely finite, and the central limit theorem holds as a result \citep{Vehtari+etal:PSIS:2024}.}

We begin by performing an initial forward search on the full data, where we find that there is an elpd "bulge" (not shown here, but similar to the full-data search path of Figure~\ref{fig:school-spline}). That is, for some model sizes, the elpd point estimate and its one standard error interval lie completely above the point estimate of the reference model elpd, indicating potentially over-optimistic elpd estimates. We therefore include the search in the CV (and terminate the search at the model size inducing the highest over-optimistic elpd estimate inline with previous discussion), yielding more reliable elpd estimates and uncertainties for the different model sizes. 

Since we have set up our experiment to have highly-correlated predictors, it is likely that our procedure was able to find a very small subset of predictors by choosing only one or two from each of the relevant correlated blocks, since they should repeat some predictive information. As such, we might ask whether any of the predictors in the minimal predictor set chosen could have been substituted with another predictor. To do so, we calculate the pairwise distances as previously described in Section~\ref{sec:select} and represent them with the dendrogram in Figure \ref{fig:dendrogram_case_study_4.1}. We find that the predictors fall nicely into clusters of five, in line with the DGP. Further, it is likely that we can replace any of the selected predictors with any other predictor from the same cluster to achieve similar predictive performance.

This case study demonstrates that in the case of correlated predictors, projection predictive model selection is able to choose parsimonious submodels, and we can identify when predictors repeat or share predictive information with the distance metric we have previously motivated.
\subsection{Many weakly-relevant predictors}\label{sec:weakly-relevant}
Another situation not uncommon in statistical analyses is the presence of many predictors, each one contributing only a marginal amount of predictive information to the final model. Indeed, in situations where the data are already difficult to explain, having predictors that are only weakly-relevant can pose issues in model selection, and in particular in identifying a minimal predictor subset.

In this case study, we sample $n = 500$ observations from $p = 50$ predictors that are distributed according to a multivariate Gaussian, where all predictors are correlated with $\rho = 0.1$. As before, $y \sim \normal(w^\top x,\sigma^2)$, with $\sigma^2 = 1$ and $w$ computed and fixed such that $R^2 = 0.5$.

In these regimes, the predictor ordering in the solution path can be extremely erratic: since each predictor contributes so little predictive information, the search heuristic can be highly unstable and different runs may result in different orderings. As a remedy to this noise, we can smooth the difference of submodel elpds to the reference model elpd over the model sizes with a monotonic spline as motivated in Section~\ref{sec:select}.

This is seen in Figure~\ref{fig:elpd_plot_case_study_4.2}, wherein the noise in the cross-validated predictive performance estimates for the submodels becomes immediately obvious. Indeed, we find that following the inclusion of the search in the cross-validation, we observe instability in the predictive performance estimates, likely caused by different predictor orderings, and instances where adding a new predictor can reduce the predictive performance estimate of the submodel. As previously discussed, this is merely an artifact of including the search in the cross-validation, and in general the addition of a new predictor should not worsen predictive performance on expectation. We find that smoothing these noisy elpd estimates produces more consistent interpretation, and we can select a model size by applying our \ref{eq:delta-elpd} heuristic to the spline point estimate -- leading to model size $30$. 

As such, even in regimes of weakly-relevant predictors, projection predictive inference is able to identify a saturation point in model size after which the addition of subsequent predictors do not add significant predictive benefit. Our smoothing technique also mitigates the noise induced by the predictors' predictive weakness.
\subsection{Supervised principal component reference models}\label{sec:nutrimouse}
Although not often discussed, our reference model needs not be the largest of the set of nested models we search through. Indeed, it is entirely possible to search through a completely different model space than the one obtained by building increasingly complex models from the collection of the reference model's predictors. One such example is the use of principal components of the original predictor data as predictors in the reference model, whose fitted values are then used in the projection onto the original predictor space.

For example, we consider presently the task of predicting the presence of prostate cancer.\footnote{These data are freely available from \url{https://jundongl.github.io/scikit-feature}.} We observe $n = 102$ patients, each having $p = 5966$ highly collinear predictor measurements. To reduce the dimensionality of the data we compute $n-1 = 101$ iterative supervised principal components \citep{pmlr-v84-piironen18a}, and use them as predictors in our reference model. Having fit this reference model using the principal components and appropriate priors, we then project the predictive information back onto the original predictors so as to achieve an intuitive understanding of which measurements were most relevant in terms of predictive information. Using forward search in this case represents a large computational cost, so we first run $L_1$ search. This is found to be sufficient, as we did not observe a ``bulge'' in the predictive performance along the solution path (figure not illustrated here).

The use of principal components is then only to achieve a dimension reduction for the predictors in the reference model. This illustrates that we can project a reference model having a hard-to-interpret predictor space onto an interpretable predictor space. Naturally, one can equally use an additive Gaussian process \citep{longp}, Bayesian additive regression trees \citep{Chipman_2010}, or any other non-parametric model to produce the reference model.
And indeed, any (approximate or exact) inferential procedure to achieve posterior draws from the reference model.
\subsection{Interpreting the projected posterior}\label{sec:posteriors}
\begin{figure*}[!t]
    \centering
    \input{tikz/overconfidence.tex}
    \caption{Simulated linear regression example. An example of over-confidence in the projected posterior of a linear regression on $95$ independent predictors of which only $15$ are relevant. The first three plots show a subset of marginal posteriors (namely $\beta_1, \beta_2$, and $\sigma$) from a reference model fit with priors that are liable to over-fit the data (in black), and a projected submodel including only the relevant predictors (in \mybox[fill=blue!20]{blue}). The dashed vertical lines indicate the true value of the parameters. The fourth plot depicts a kernel density estimate over the posterior predictive mean at each observation for each model, denoted $\tilde{y}$, with the shaded region indicating the observed data density. The $\beta_1$ and $\beta_2$ marginals of the projected posterior concentrate more sharply than in the reference model, and closer to the true parameter values. The additional variance in the $\beta_1$ and $\beta_2$ marginals of the reference model's posterior (compared to the $\beta_1$ and $\beta_2$ marginals of the projected posterior) is projected onto $\sigma$, inflating it.}
    \label{fig:overconfident-posteriors}
\end{figure*}
The primary output of projection predictive inference is a collection of projected parameter draws for a submodel consisting of a minimal set of predictors that induces a predictive performance close to that of the reference model. However, it is interesting to understand under which circumstances valid post-selection inference can be done for these predictors using the marginals of the projected posterior directly.

When the reference model is projected onto a submodel, any additional structured variation captured by the reference model is projected onto the unstructured dispersion parameter of the submodel, thereby inflating it \citep{piironen2018}. Furthermore, in the case of an over-fitting reference model, we will demonstrate that the dispersion parameter may also absorb some of the posterior variance from the regression coefficients, leading to their over-confidence. 

Consider a linear regression on $n=100$ observations of $95$ predictors of which only $15$ are relevant. We fit a reference model using independent Gaussian priors, which we know to be prone to over-fitting in these regimes. In Figure \ref{fig:overconfident-posteriors}, we then project this reference model 
onto only the set of truly relevant predictors. What we find is that the marginal posteriors of the regression coefficients (here we only show the first two) concentrate more sharply than in the reference model, and their means approach the true parameter values.\footnote{This convergence towards the true values is likely because we have projected onto the known true model, and is not the focus of our procedure.} The variance in the reference model's marginal posteriors of the regression coefficients is then transferred to the magnitude of $\sigma$. We note that this behaviour is to be expected, and is not a failure of the procedure. Indeed, this arises directly from the projection objective being focused on predictive inference with respect to the fit of the reference model. And in this case, because we have chosen an over-fitting reference model to emphasise this behaviour.

Better reference models result in well calibrated projected posteriors, as we will see presently.
\subsection{Understanding the effect of the reference model on projected calibration}\label{sec:over-confidence}
\begin{figure*}[!t]
    \centering
    \input{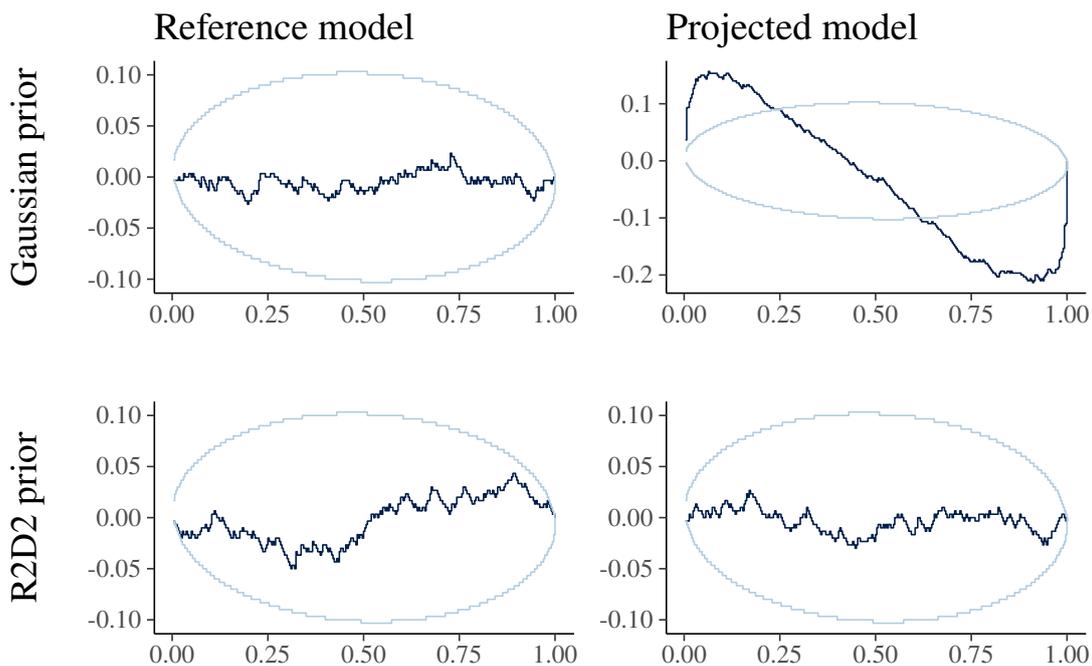}
    \caption{Simulated data case study. An example of miscalibration in the projected posterior in case of linear regression with independent predictors. The upper and lower rows compare the calibration of the treatment effect under the same model with two different priors: one using independent Gaussian priors for the regression coefficients, and a more predictive reference model using the R2D2 prior. The two columns then refer to the treatment effect in the reference model (on the left) and to the treatment effect after a projection of the reference model onto the selected minimal predictor set (on the right). A model is interpreted to be well-calibrated if the \mybox[fill=blue!20]{navy} line stays within the \mybox[fill=cyan!20]{cyan} envelope.}
    \label{fig:calibration}
\end{figure*}
We presently show how one might assess the calibration of a projected posterior after variable selection using simulation-based calibration \citep[SBC;][]{talts_validating_2020,Sailynoja+etal:2022:PIT-ECDF,modrak2023simulation} in the case of a linear regression. We generate data with $n=100$ observations on $p=70$ independent predictors from a standard normal distribution and a treatment predictor from a Bernoulli distribution. $10$ of the standard normal predictors and the treatment predictor were used to generate the noisy target, $y$, at each SBC iteration. We perform projection predictive variable selection using a $10$-fold cross-validation over the entire procedure (including forward search up to $p_{\text{max}} = 30$, and using the \ref{eq:utility-ci} selection heuristic). We forced the treatment predictor to be included in the final submodel across all SBC iterations, and compared the calibration for its regression coefficient in the reference model posterior and in the projected posterior of the submodel selected by projection predictive inference. These calibration checks are visualised in Figure \ref{fig:calibration}, where the top row belongs to a reference model using independent standard normal priors over the regression coefficients, and the bottom row belongs to a more predictive reference model (fitted using the R2D2 prior).

At least three aspects of the projection affect the calibration of the marginals in the projected posterior:
\begin{enumerate*}[label=(\arabic*), ref=\arabic*]
    \item the projection aims to minimise the KL divergence in the predictive space, and thus does not guarantee what happens in the parameter space; 
    \item the draw-wise projection swaps the order of minimisation and integration; and,
\item the search over the model space and the used selection heuristic.
\end{enumerate*}
If there is a substantial amount of structural uncertainty in the reference model (induced, e.g., by vague priors on coefficients) that needs to be projected on the non-structural variation part of the submodel, the projected
posterior of the remaining parameters can be over-confident. 
This is seen in the independent Gaussian case by the left tail of the probability integral transform (PIT) values exceeding the 95\% simultaneous confidence band in the projected model, and the right tail of the PIT values dropping far below it. 

On the other hand, if the opposite is true (as in the R2D2 case) then the projected posterior is well calibrated. 
In this latter case, the draw-wise projection and the search are not themselves causing substantial miscalibration.

In the Gaussian prior case, there is a lot of uncertainty about the irrelevant coefficients, while in R2D2 case the prior reduces the uncertainty so that it doesn't cause problems in the projection.
Should we wish to perform post-selection inference using the projected posterior, SBC checks can appropriately diagnose over-confident projected posteriors.
\section{Discussion}\label{sec:discussion}
It is worth reiterating the motivations for model selection. Namely, we do not consider model selection a remedy to over-fitting. Within a Bayesian framework, the statistician has the ability to mitigate over-fitting through the sensible use of priors. While we don't wish to distract from the core aim of this paper with a comprehensive discussion on sparsifying priors, we note that the spike-and-slab prior \citep{spikeslab}, regularised horseshoe prior \citep{horseshoe}, R2D2 prior \citep{zhang_bayesian_2022,r2d2}, and more recently the R2D2M2 \citep{aguilar_intuitive_2023} and the $L_1$-ball \citep{l1ball} priors have found success in explicating the statistician's desire for sparsity and consistent behavior with increasing dimensionality. Instead, we advertise model selection as a method to reduce measurement cost in future predictions, as a robustification technique against unrepresentative data, and to gain an understanding of predictor correlations.
\subsection{Limitations}
Projection predictive approach has been shown to perform well in many experiments, but it also has its limitations:
\begin{enumerate}[nosep]
    \item we need to have a full reference model, which may be difficult to construct in some cases;
    \item computational and memory cost when creating the reference model and performing the projections can render the approach infeasible for big data;
    \item efficient projection for new model types (e.g. non-log-concave likelihoods, or conditional auto-regressive models) is non-trivial;
    \item there seems to be no clear advantage to using projection predictive inference when the task involves finding \textit{all} relevant predictors \citep{Pavone2020}.
\end{enumerate}
\subsection{Calibration}
Common folklore states that when using step-wise selection, credible intervals are not well-calibrated. However, using forward search method based on KL divergence, and cross-validation thereof, there is less selection-induced over-fitting. Furthermore, for the inference for the treatment effect to be well-calibrated, it is important to have a sensible reference model (which we show empirically in Section~\ref{sec:over-confidence}).
\subsection{Causal settings}
In a causal setting, it might be tempting to perform model selection in order to increase the precision of a treatment effect posterior of interest \citep{cinelli_crash_2022}. However, to achieve valid causal inference, one must account for the causal relationship between the response and measured predictors. This is done by including in the model what \citet{pearl_causal_2009} calls the admissible set of predictors. When searching in the model space, all submodels must include this admissible set. If our main concern is to accurately infer a causally valid treatment effect, then no model selection is needed and we can use the reference model directly, provided it includes the admissible set.
\subsection{When not to use projection predictive}
It may also be unnecessary to involve projection predictive inference in one's model building workflow if one is dealing with relatively few (fewer than ten, say) models, especially if these models are not nested. In this case, it may be easier to compare the models either by estimating the difference in their predictive performance directly \citep{loo,loo-package}, or by investigating their Bayesian model stacking or averaging weights \citep{yao_using_2018,yao_bayesian_2021,yao_locking_2023}. 

If one is unable to achieve a suitable reference model, then projection predictive inference can be immediately discounted. And if the reference model can be constructed but is not very complex, then further simplification of it may not be worth the effort.
\subsection{Software}
Alongside \pkg{projpred} in \proglang{R}, the \pkg{kulprit} package \citep{kulprit} provides a projection predictive inference implementation in \proglang{Python}.
\subsection{Future directions}
While our proposed workflow mitigates many of the dangers associated with cross-validation and model selection in a Bayesian workflow, there remain some open pits one might fall into. For instance, in the case of forward search, adding a predictor to a model that does not improve its theoretical predictive performance can lead to overly-optimistic model selection in finite data regimes, resulting in selection-induced bias along the solution path. Future work may task to quantify this risk, and diagnose when the user is at risk with safe and robust stopping criteria.

The models we have considered thus far have remained relatively simple and all with univariate target. Indeed, an extension to the multivariate case presents itself as an interesting avenue to explore in time.
Taking into account information about future decision task fits naturally into the projection predictive approach, and is related to wider literature on loss-calibrated variational inference \citep{lacoste_julien_approximate_2011,kusmierczyk_variational_2019,morais_loss-calibrated_2022}. Such a variant of projection predictive inference would be beneficial when, for example, we are interested in predicting summary functions of the models.

%% file: tikz/flowchart.tex
\resizebox{\columnwidth}{!}{%
        \begin{tikzpicture}
            \node [ref, aspect=1] (ref_model){Fit the reference model};
            \node [ref, aspect=1, right=of ref_model] (diagnose_ref) {Diagnose the reference model};
            \node [search, aspect=1, right=of diagnose_ref] (search) {Full-data search};
            \node [cv, aspect=1, below=of search] (diagnose_search) {Diagnose the search path};
            \node [cv, aspect=1, left=of diagnose_search] (cv_search) {Cross-validate over search paths};
            \node [cv, aspect=1, left=of cv_search] (select) {Select a submodel size};
            \node [select, aspect=1, below=of cv_search] (posterior) {Interpret the projected posterior};

            \node [c, label={[font=\Large]above left:Section~\ref{sec:ref-model}}, fit=(ref_model) (diagnose_ref)] {};
            \node [c, label={[font=\Large]90:Section~\ref{sec:search}}, fit= (search)] {};
            \node [c, label={[font=\Large]335:Section~\ref{sec:cv}}, fit= (diagnose_search) (cv_search) (select)] {};
            \node [c, label={[font=\Large]below:Section~\ref{sec:posteriors}}, fit= (posterior)] {};

            \draw[-{Stealth[scale=2]}] (ref_model) edge [bend left = 1em] (diagnose_ref);
            \draw[-{Stealth[scale=2]}, dashed] (diagnose_ref) edge [bend left = 1em](ref_model);
            \draw[-{Stealth[scale=2]}] (diagnose_ref) edge (search);
            \draw[-{Stealth[scale=2]}] (search) edge (diagnose_search);
            \draw[-{Stealth[scale=2]}, dashed] (diagnose_search) edge (cv_search);
            \draw[-{Stealth[scale=2]}] (diagnose_search) edge [bend right = 3.4em] (select);
            \draw[-{Stealth[scale=2]}, dashed] (cv_search) edge (select);
            \draw[-{Stealth[scale=2]}, dashed] (select) edge [bend right = 3em](posterior);
        
        \end{tikzpicture}
    }

%% file: tikz/schools-spline.tex
\begin{tikzpicture}[x=1pt,y=1pt]
\definecolor{fillColor}{RGB}{255,255,255}
\begin{scope}
\definecolor{drawColor}{RGB}{255,255,255}
\definecolor{fillColor}{RGB}{255,255,255}

\path[draw=drawColor,line width= 0.6pt,line join=round,line cap=round,fill=fillColor] (  0.00,  0.00) rectangle (252.94,252.94);
\end{scope}
\begin{scope}
\definecolor{fillColor}{RGB}{255,255,255}

\path[fill=fillColor] ( 34.64, 30.69) rectangle (247.44,247.45);
\definecolor{drawColor}{gray}{0.92}

\path[draw=drawColor,line width= 0.6pt,line join=round] ( 34.64, 59.42) --
	(247.44, 59.42);

\path[draw=drawColor,line width= 0.6pt,line join=round] ( 34.64, 95.05) --
	(247.44, 95.05);

\path[draw=drawColor,line width= 0.6pt,line join=round] ( 34.64,130.69) --
	(247.44,130.69);

\path[draw=drawColor,line width= 0.6pt,line join=round] ( 34.64,166.32) --
	(247.44,166.32);

\path[draw=drawColor,line width= 0.6pt,line join=round] ( 34.64,201.96) --
	(247.44,201.96);

\path[draw=drawColor,line width= 0.6pt,line join=round] ( 34.64,237.59) --
	(247.44,237.59);
\definecolor{drawColor}{RGB}{190,190,190}

\path[draw=drawColor,line width= 0.6pt,line join=round] ( 45.64, 40.54) -- ( 45.64, 96.49);

\path[draw=drawColor,line width= 0.6pt,line join=round] ( 52.27,140.44) -- ( 52.27,181.89);

\path[draw=drawColor,line width= 0.6pt,line join=round] ( 58.89,162.93) -- ( 58.89,196.08);

\path[draw=drawColor,line width= 0.6pt,line join=round] ( 65.52,175.69) -- ( 65.52,205.86);

\path[draw=drawColor,line width= 0.6pt,line join=round] ( 72.14,181.13) -- ( 72.14,210.73);

\path[draw=drawColor,line width= 0.6pt,line join=round] ( 78.77,184.02) -- ( 78.77,210.74);

\path[draw=drawColor,line width= 0.6pt,line join=round] ( 85.39,193.26) -- ( 85.39,218.47);

\path[draw=drawColor,line width= 0.6pt,line join=round] ( 92.02,200.12) -- ( 92.02,223.10);

\path[draw=drawColor,line width= 0.6pt,line join=round] ( 98.64,203.15) -- ( 98.64,223.52);

\path[draw=drawColor,line width= 0.6pt,line join=round] (105.27,209.75) -- (105.27,227.34);

\path[draw=drawColor,line width= 0.6pt,line join=round] (111.89,211.26) -- (111.89,228.72);

\path[draw=drawColor,line width= 0.6pt,line join=round] (118.52,217.38) -- (118.52,233.57);

\path[draw=drawColor,line width= 0.6pt,line join=round] (125.14,219.41) -- (125.14,234.68);

\path[draw=drawColor,line width= 0.6pt,line join=round] (131.77,219.71) -- (131.77,233.16);

\path[draw=drawColor,line width= 0.6pt,line join=round] (138.39,219.95) -- (138.39,231.62);

\path[draw=drawColor,line width= 0.6pt,line join=round] (145.02,218.52) -- (145.02,228.55);

\path[draw=drawColor,line width= 0.6pt,line join=round] (151.65,218.23) -- (151.65,227.07);

\path[draw=drawColor,line width= 0.6pt,line join=round] (158.27,219.54) -- (158.27,227.68);

\path[draw=drawColor,line width= 0.6pt,line join=round] (164.90,217.78) -- (164.90,225.07);

\path[draw=drawColor,line width= 0.6pt,line join=round] (171.52,215.18) -- (171.52,221.87);

\path[draw=drawColor,line width= 0.6pt,line join=round] (178.15,215.31) -- (178.15,220.67);

\path[draw=drawColor,line width= 0.6pt,line join=round] (184.77,213.32) -- (184.77,217.60);

\path[draw=drawColor,line width= 0.6pt,line join=round] (191.40,210.80) -- (191.40,214.86);

\path[draw=drawColor,line width= 0.6pt,line join=round] (198.02,209.95) -- (198.02,213.33);

\path[draw=drawColor,line width= 0.6pt,line join=round] (204.65,209.32) -- (204.65,212.03);

\path[draw=drawColor,line width= 0.6pt,line join=round] (211.27,206.79) -- (211.27,209.29);

\path[draw=drawColor,line width= 0.6pt,line join=round] (217.90,206.13) -- (217.90,208.63);

\path[draw=drawColor,line width= 0.6pt,line join=round] (224.52,204.09) -- (224.52,206.61);

\path[draw=drawColor,line width= 0.6pt,line join=round] (231.15,202.03) -- (231.15,204.31);

\path[draw=drawColor,line width= 0.6pt,line join=round] (237.77,200.51) -- (237.77,202.63);
\definecolor{fillColor}{RGB}{190,190,190}

\path[draw=drawColor,line width= 0.8pt,line join=round,line cap=round,fill=fillColor] ( 45.64, 68.52) circle (  2.85);

\path[draw=drawColor,line width= 0.8pt,line join=round,line cap=round,fill=fillColor] ( 52.27,161.17) circle (  2.85);

\path[draw=drawColor,line width= 0.8pt,line join=round,line cap=round,fill=fillColor] ( 58.89,179.51) circle (  2.85);

\path[draw=drawColor,line width= 0.8pt,line join=round,line cap=round,fill=fillColor] ( 65.52,190.77) circle (  2.85);

\path[draw=drawColor,line width= 0.8pt,line join=round,line cap=round,fill=fillColor] ( 72.14,195.93) circle (  2.85);

\path[draw=drawColor,line width= 0.8pt,line join=round,line cap=round,fill=fillColor] ( 78.77,197.38) circle (  2.85);

\path[draw=drawColor,line width= 0.8pt,line join=round,line cap=round,fill=fillColor] ( 85.39,205.86) circle (  2.85);

\path[draw=drawColor,line width= 0.8pt,line join=round,line cap=round,fill=fillColor] ( 92.02,211.61) circle (  2.85);

\path[draw=drawColor,line width= 0.8pt,line join=round,line cap=round,fill=fillColor] ( 98.64,213.34) circle (  2.85);

\path[draw=drawColor,line width= 0.8pt,line join=round,line cap=round,fill=fillColor] (105.27,218.55) circle (  2.85);

\path[draw=drawColor,line width= 0.8pt,line join=round,line cap=round,fill=fillColor] (111.89,219.99) circle (  2.85);

\path[draw=drawColor,line width= 0.8pt,line join=round,line cap=round,fill=fillColor] (118.52,225.47) circle (  2.85);

\path[draw=drawColor,line width= 0.8pt,line join=round,line cap=round,fill=fillColor] (125.14,227.04) circle (  2.85);

\path[draw=drawColor,line width= 0.8pt,line join=round,line cap=round,fill=fillColor] (131.77,226.43) circle (  2.85);

\path[draw=drawColor,line width= 0.8pt,line join=round,line cap=round,fill=fillColor] (138.39,225.78) circle (  2.85);

\path[draw=drawColor,line width= 0.8pt,line join=round,line cap=round,fill=fillColor] (145.02,223.53) circle (  2.85);

\path[draw=drawColor,line width= 0.8pt,line join=round,line cap=round,fill=fillColor] (151.65,222.65) circle (  2.85);

\path[draw=drawColor,line width= 0.8pt,line join=round,line cap=round,fill=fillColor] (158.27,223.61) circle (  2.85);

\path[draw=drawColor,line width= 0.8pt,line join=round,line cap=round,fill=fillColor] (164.90,221.43) circle (  2.85);

\path[draw=drawColor,line width= 0.8pt,line join=round,line cap=round,fill=fillColor] (171.52,218.53) circle (  2.85);

\path[draw=drawColor,line width= 0.8pt,line join=round,line cap=round,fill=fillColor] (178.15,217.99) circle (  2.85);

\path[draw=drawColor,line width= 0.8pt,line join=round,line cap=round,fill=fillColor] (184.77,215.46) circle (  2.85);

\path[draw=drawColor,line width= 0.8pt,line join=round,line cap=round,fill=fillColor] (191.40,212.83) circle (  2.85);

\path[draw=drawColor,line width= 0.8pt,line join=round,line cap=round,fill=fillColor] (198.02,211.64) circle (  2.85);

\path[draw=drawColor,line width= 0.8pt,line join=round,line cap=round,fill=fillColor] (204.65,210.68) circle (  2.85);

\path[draw=drawColor,line width= 0.8pt,line join=round,line cap=round,fill=fillColor] (211.27,208.04) circle (  2.85);

\path[draw=drawColor,line width= 0.8pt,line join=round,line cap=round,fill=fillColor] (217.90,207.38) circle (  2.85);

\path[draw=drawColor,line width= 0.8pt,line join=round,line cap=round,fill=fillColor] (224.52,205.35) circle (  2.85);

\path[draw=drawColor,line width= 0.8pt,line join=round,line cap=round,fill=fillColor] (231.15,203.17) circle (  2.85);

\path[draw=drawColor,line width= 0.8pt,line join=round,line cap=round,fill=fillColor] (237.77,201.57) circle (  2.85);
\definecolor{drawColor}{RGB}{0,0,0}

\path[draw=drawColor,line width= 0.6pt,line join=round] ( 44.32, 40.80) -- ( 44.32, 97.04);

\path[draw=drawColor,line width= 0.6pt,line join=round] ( 50.94,140.86) -- ( 50.94,182.36);

\path[draw=drawColor,line width= 0.6pt,line join=round] ( 57.57,118.23) -- ( 57.57,152.69);

\path[draw=drawColor,line width= 0.6pt,line join=round] ( 64.19,128.14) -- ( 64.19,158.86);

\path[draw=drawColor,line width= 0.6pt,line join=round] ( 70.82,136.38) -- ( 70.82,164.89);

\path[draw=drawColor,line width= 0.6pt,line join=round] ( 77.44,184.15) -- ( 77.44,211.07);

\path[draw=drawColor,line width= 0.6pt,line join=round] ( 84.07,181.88) -- ( 84.07,205.89);

\path[draw=drawColor,line width= 0.6pt,line join=round] ( 90.69,189.93) -- ( 90.69,211.61);

\path[draw=drawColor,line width= 0.6pt,line join=round] ( 97.32,193.92) -- ( 97.32,213.97);

\path[draw=drawColor,line width= 0.6pt,line join=round] (103.94,208.69) -- (103.94,226.29);

\path[draw=drawColor,line width= 0.6pt,line join=round] (110.57,168.06) -- (110.57,183.50);

\path[draw=drawColor,line width= 0.6pt,line join=round] (117.19,163.06) -- (117.19,176.87);

\path[draw=drawColor,line width= 0.6pt,line join=round] (123.82,162.47) -- (123.82,176.12);

\path[draw=drawColor,line width= 0.6pt,line join=round] (130.44,165.76) -- (130.44,178.37);

\path[draw=drawColor,line width= 0.6pt,line join=round] (137.07,190.96) -- (137.07,201.68);

\path[draw=drawColor,line width= 0.6pt,line join=round] (143.69,197.67) -- (143.69,208.00);

\path[draw=drawColor,line width= 0.6pt,line join=round] (150.32,192.57) -- (150.32,202.01);

\path[draw=drawColor,line width= 0.6pt,line join=round] (156.95,192.64) -- (156.95,201.25);

\path[draw=drawColor,line width= 0.6pt,line join=round] (163.57,200.09) -- (163.57,207.54);

\path[draw=drawColor,line width= 0.6pt,line join=round] (170.20,193.07) -- (170.20,199.58);

\path[draw=drawColor,line width= 0.6pt,line join=round] (176.82,191.91) -- (176.82,197.64);

\path[draw=drawColor,line width= 0.6pt,line join=round] (183.45,196.93) -- (183.45,202.04);

\path[draw=drawColor,line width= 0.6pt,line join=round] (190.07,199.05) -- (190.07,202.78);

\path[draw=drawColor,line width= 0.6pt,line join=round] (196.70,204.84) -- (196.70,207.92);

\path[draw=drawColor,line width= 0.6pt,line join=round] (203.32,205.16) -- (203.32,207.84);

\path[draw=drawColor,line width= 0.6pt,line join=round] (209.95,206.04) -- (209.95,208.41);

\path[draw=drawColor,line width= 0.6pt,line join=round] (216.57,204.21) -- (216.57,206.30);

\path[draw=drawColor,line width= 0.6pt,line join=round] (223.20,199.00) -- (223.20,201.12);

\path[draw=drawColor,line width= 0.6pt,line join=round] (229.82,201.37) -- (229.82,203.31);

\path[draw=drawColor,line width= 0.6pt,line join=round] (236.45,201.17) -- (236.45,203.12);
\definecolor{fillColor}{RGB}{0,0,0}

\path[draw=drawColor,line width= 0.8pt,line join=round,line cap=round,fill=fillColor] ( 44.32, 68.92) circle (  2.85);

\path[draw=drawColor,line width= 0.8pt,line join=round,line cap=round,fill=fillColor] ( 50.94,161.61) circle (  2.85);

\path[draw=drawColor,line width= 0.8pt,line join=round,line cap=round,fill=fillColor] ( 57.57,135.46) circle (  2.85);

\path[draw=drawColor,line width= 0.8pt,line join=round,line cap=round,fill=fillColor] ( 64.19,143.50) circle (  2.85);

\path[draw=drawColor,line width= 0.8pt,line join=round,line cap=round,fill=fillColor] ( 70.82,150.64) circle (  2.85);

\path[draw=drawColor,line width= 0.8pt,line join=round,line cap=round,fill=fillColor] ( 77.44,197.61) circle (  2.85);

\path[draw=drawColor,line width= 0.8pt,line join=round,line cap=round,fill=fillColor] ( 84.07,193.89) circle (  2.85);

\path[draw=drawColor,line width= 0.8pt,line join=round,line cap=round,fill=fillColor] ( 90.69,200.77) circle (  2.85);

\path[draw=drawColor,line width= 0.8pt,line join=round,line cap=round,fill=fillColor] ( 97.32,203.95) circle (  2.85);

\path[draw=drawColor,line width= 0.8pt,line join=round,line cap=round,fill=fillColor] (103.94,217.49) circle (  2.85);

\path[draw=drawColor,line width= 0.8pt,line join=round,line cap=round,fill=fillColor] (110.57,175.78) circle (  2.85);

\path[draw=drawColor,line width= 0.8pt,line join=round,line cap=round,fill=fillColor] (117.19,169.97) circle (  2.85);

\path[draw=drawColor,line width= 0.8pt,line join=round,line cap=round,fill=fillColor] (123.82,169.30) circle (  2.85);

\path[draw=drawColor,line width= 0.8pt,line join=round,line cap=round,fill=fillColor] (130.44,172.06) circle (  2.85);

\path[draw=drawColor,line width= 0.8pt,line join=round,line cap=round,fill=fillColor] (137.07,196.32) circle (  2.85);

\path[draw=drawColor,line width= 0.8pt,line join=round,line cap=round,fill=fillColor] (143.69,202.84) circle (  2.85);

\path[draw=drawColor,line width= 0.8pt,line join=round,line cap=round,fill=fillColor] (150.32,197.29) circle (  2.85);

\path[draw=drawColor,line width= 0.8pt,line join=round,line cap=round,fill=fillColor] (156.95,196.94) circle (  2.85);

\path[draw=drawColor,line width= 0.8pt,line join=round,line cap=round,fill=fillColor] (163.57,203.82) circle (  2.85);

\path[draw=drawColor,line width= 0.8pt,line join=round,line cap=round,fill=fillColor] (170.20,196.32) circle (  2.85);

\path[draw=drawColor,line width= 0.8pt,line join=round,line cap=round,fill=fillColor] (176.82,194.78) circle (  2.85);

\path[draw=drawColor,line width= 0.8pt,line join=round,line cap=round,fill=fillColor] (183.45,199.48) circle (  2.85);

\path[draw=drawColor,line width= 0.8pt,line join=round,line cap=round,fill=fillColor] (190.07,200.91) circle (  2.85);

\path[draw=drawColor,line width= 0.8pt,line join=round,line cap=round,fill=fillColor] (196.70,206.38) circle (  2.85);

\path[draw=drawColor,line width= 0.8pt,line join=round,line cap=round,fill=fillColor] (203.32,206.50) circle (  2.85);

\path[draw=drawColor,line width= 0.8pt,line join=round,line cap=round,fill=fillColor] (209.95,207.22) circle (  2.85);

\path[draw=drawColor,line width= 0.8pt,line join=round,line cap=round,fill=fillColor] (216.57,205.25) circle (  2.85);

\path[draw=drawColor,line width= 0.8pt,line join=round,line cap=round,fill=fillColor] (223.20,200.06) circle (  2.85);

\path[draw=drawColor,line width= 0.8pt,line join=round,line cap=round,fill=fillColor] (229.82,202.34) circle (  2.85);

\path[draw=drawColor,line width= 0.8pt,line join=round,line cap=round,fill=fillColor] (236.45,202.14) circle (  2.85);
\definecolor{fillColor}{RGB}{0,0,255}

\path[fill=fillColor,fill opacity=0.20] ( 51.61,184.19) --
	( 58.23,189.83) --
	( 64.86,193.48) --
	( 71.48,196.26) --
	( 78.11,198.63) --
	( 84.73,200.82) --
	( 91.36,202.58) --
	( 97.98,204.06) --
	(104.61,205.15) --
	(111.23,205.93) --
	(117.86,206.56) --
	(124.48,207.55) --
	(131.11,208.08) --
	(137.73,207.98) --
	(144.36,208.55) --
	(150.98,208.70) --
	(157.61,208.76) --
	(164.23,208.41) --
	(170.86,208.10) --
	(177.48,207.79) --
	(184.11,207.56) --
	(190.73,206.32) --
	(197.36,205.80) --
	(203.98,205.50) --
	(210.61,205.28) --
	(217.23,205.05) --
	(223.86,205.26) --
	(230.48,205.12) --
	(237.11,205.27) --
	(237.11,201.08) --
	(230.48,200.93) --
	(223.86,200.67) --
	(217.23,200.53) --
	(210.61,200.16) --
	(203.98,199.72) --
	(197.36,199.16) --
	(190.73,198.28) --
	(184.11,196.53) --
	(177.48,195.44) --
	(170.86,194.04) --
	(164.23,192.35) --
	(157.61,190.20) --
	(150.98,188.35) --
	(144.36,186.27) --
	(137.73,184.86) --
	(131.11,180.89) --
	(124.48,178.10) --
	(117.86,176.77) --
	(111.23,172.62) --
	(104.61,167.17) --
	( 97.98,160.82) --
	( 91.36,155.81) --
	( 84.73,149.03) --
	( 78.11,140.56) --
	( 71.48,134.76) --
	( 64.86,127.24) --
	( 58.23,115.50) --
	( 51.61, 94.69) --
	cycle;

\path[] ( 51.61,184.19) --
	( 58.23,189.83) --
	( 64.86,193.48) --
	( 71.48,196.26) --
	( 78.11,198.63) --
	( 84.73,200.82) --
	( 91.36,202.58) --
	( 97.98,204.06) --
	(104.61,205.15) --
	(111.23,205.93) --
	(117.86,206.56) --
	(124.48,207.55) --
	(131.11,208.08) --
	(137.73,207.98) --
	(144.36,208.55) --
	(150.98,208.70) --
	(157.61,208.76) --
	(164.23,208.41) --
	(170.86,208.10) --
	(177.48,207.79) --
	(184.11,207.56) --
	(190.73,206.32) --
	(197.36,205.80) --
	(203.98,205.50) --
	(210.61,205.28) --
	(217.23,205.05) --
	(223.86,205.26) --
	(230.48,205.12) --
	(237.11,205.27);

\path[] (237.11,201.08) --
	(230.48,200.93) --
	(223.86,200.67) --
	(217.23,200.53) --
	(210.61,200.16) --
	(203.98,199.72) --
	(197.36,199.16) --
	(190.73,198.28) --
	(184.11,196.53) --
	(177.48,195.44) --
	(170.86,194.04) --
	(164.23,192.35) --
	(157.61,190.20) --
	(150.98,188.35) --
	(144.36,186.27) --
	(137.73,184.86) --
	(131.11,180.89) --
	(124.48,178.10) --
	(117.86,176.77) --
	(111.23,172.62) --
	(104.61,167.17) --
	( 97.98,160.82) --
	( 91.36,155.81) --
	( 84.73,149.03) --
	( 78.11,140.56) --
	( 71.48,134.76) --
	( 64.86,127.24) --
	( 58.23,115.50) --
	( 51.61, 94.69);
\definecolor{drawColor}{RGB}{0,0,255}

\path[draw=drawColor,line width= 0.6pt,line join=round] ( 51.61,139.44) --
	( 58.23,152.67) --
	( 64.86,160.36) --
	( 71.48,165.51) --
	( 78.11,169.60) --
	( 84.73,174.93) --
	( 91.36,179.20) --
	( 97.98,182.44) --
	(104.61,186.16) --
	(111.23,189.28) --
	(117.86,191.67) --
	(124.48,192.83) --
	(131.11,194.49) --
	(137.73,196.42) --
	(144.36,197.41) --
	(150.98,198.52) --
	(157.61,199.48) --
	(164.23,200.38) --
	(170.86,201.07) --
	(177.48,201.62) --
	(184.11,202.04) --
	(190.73,202.30) --
	(197.36,202.48) --
	(203.98,202.61) --
	(210.61,202.72) --
	(217.23,202.79) --
	(223.86,202.97) --
	(230.48,203.03) --
	(237.11,203.18);
\definecolor{drawColor}{RGB}{0,0,0}
\definecolor{fillColor}{RGB}{255,255,255}

\path[draw=drawColor,line width= 0.3pt,line join=round,line cap=round,fill=fillColor] ( 86.26,145.25) --
	(162.71,145.25) --
	(162.63,145.26) --
	(162.93,145.27) --
	(163.21,145.33) --
	(163.48,145.43) --
	(163.73,145.57) --
	(163.96,145.76) --
	(164.15,145.98) --
	(164.31,146.22) --
	(164.42,146.49) --
	(164.49,146.77) --
	(164.51,147.06) --
	(164.51,147.06) --
	(164.51,157.07) --
	(164.51,157.07) --
	(164.49,157.36) --
	(164.42,157.65) --
	(164.31,157.91) --
	(164.15,158.16) --
	(163.96,158.38) --
	(163.73,158.56) --
	(163.48,158.71) --
	(163.21,158.81) --
	(162.93,158.87) --
	(162.71,158.88) --
	( 86.26,158.88) --
	( 86.47,158.87) --
	( 86.18,158.88) --
	( 85.90,158.84) --
	( 85.62,158.76) --
	( 85.35,158.64) --
	( 85.11,158.47) --
	( 84.90,158.27) --
	( 84.73,158.04) --
	( 84.59,157.78) --
	( 84.50,157.51) --
	( 84.46,157.22) --
	( 84.45,157.07) --
	( 84.45,147.06) --
	( 84.46,147.21) --
	( 84.46,146.92) --
	( 84.50,146.63) --
	( 84.59,146.35) --
	( 84.73,146.10) --
	( 84.90,145.86) --
	( 85.11,145.66) --
	( 85.35,145.50) --
	( 85.62,145.37) --
	( 85.90,145.29) --
	( 86.18,145.26) --
	cycle;
\end{scope}
\begin{scope}
\definecolor{drawColor}{RGB}{0,0,0}

\node[text=drawColor,anchor=base,inner sep=0pt, outer sep=0pt, scale=  1.10] at (124.48,148.27) {Cross-validated};
\definecolor{drawColor}{RGB}{190,190,190}
\definecolor{fillColor}{RGB}{255,255,255}

\path[draw=drawColor,line width= 0.3pt,line join=round,line cap=round,fill=fillColor] (154.43,230.78) --
	(200.53,230.78) --
	(200.46,230.78) --
	(200.75,230.79) --
	(201.04,230.85) --
	(201.31,230.95) --
	(201.56,231.10) --
	(201.78,231.28) --
	(201.98,231.50) --
	(202.13,231.75) --
	(202.25,232.01) --
	(202.32,232.30) --
	(202.34,232.59) --
	(202.34,232.59) --
	(202.34,242.60) --
	(202.34,242.60) --
	(202.32,242.89) --
	(202.25,243.17) --
	(202.13,243.44) --
	(201.98,243.68) --
	(201.78,243.90) --
	(201.56,244.09) --
	(201.31,244.23) --
	(201.04,244.33) --
	(200.75,244.39) --
	(200.53,244.41) --
	(154.43,244.41) --
	(154.65,244.39) --
	(154.36,244.40) --
	(154.07,244.37) --
	(153.79,244.29) --
	(153.53,244.16) --
	(153.29,244.00) --
	(153.08,243.80) --
	(152.91,243.56) --
	(152.77,243.31) --
	(152.68,243.03) --
	(152.63,242.74) --
	(152.63,242.60) --
	(152.63,232.59) --
	(152.63,232.73) --
	(152.63,232.44) --
	(152.68,232.15) --
	(152.77,231.88) --
	(152.91,231.62) --
	(153.08,231.39) --
	(153.29,231.19) --
	(153.53,231.02) --
	(153.79,230.90) --
	(154.07,230.82) --
	(154.36,230.78) --
	cycle;
\end{scope}
\begin{scope}
\definecolor{drawColor}{RGB}{190,190,190}

\node[text=drawColor,anchor=base,inner sep=0pt, outer sep=0pt, scale=  1.10] at (177.48,233.79) {Full-data};
\definecolor{drawColor}{gray}{0.20}

\path[draw=drawColor,line width= 0.6pt,line join=round,line cap=round] ( 34.64, 30.69) rectangle (247.44,247.45);
\end{scope}
\begin{scope}
\definecolor{drawColor}{gray}{0.30}

\node[text=drawColor,anchor=base east,inner sep=0pt, outer sep=0pt, scale=  0.88] at ( 29.69, 56.38) {-40};

\node[text=drawColor,anchor=base east,inner sep=0pt, outer sep=0pt, scale=  0.88] at ( 29.69, 92.02) {-30};

\node[text=drawColor,anchor=base east,inner sep=0pt, outer sep=0pt, scale=  0.88] at ( 29.69,127.66) {-20};

\node[text=drawColor,anchor=base east,inner sep=0pt, outer sep=0pt, scale=  0.88] at ( 29.69,163.29) {-10};

\node[text=drawColor,anchor=base east,inner sep=0pt, outer sep=0pt, scale=  0.88] at ( 29.69,198.93) {0};

\node[text=drawColor,anchor=base east,inner sep=0pt, outer sep=0pt, scale=  0.88] at ( 29.69,234.56) {10};
\end{scope}
\begin{scope}
\definecolor{drawColor}{gray}{0.20}

\path[draw=drawColor,line width= 0.6pt,line join=round] ( 31.89, 59.42) --
	( 34.64, 59.42);

\path[draw=drawColor,line width= 0.6pt,line join=round] ( 31.89, 95.05) --
	( 34.64, 95.05);

\path[draw=drawColor,line width= 0.6pt,line join=round] ( 31.89,130.69) --
	( 34.64,130.69);

\path[draw=drawColor,line width= 0.6pt,line join=round] ( 31.89,166.32) --
	( 34.64,166.32);

\path[draw=drawColor,line width= 0.6pt,line join=round] ( 31.89,201.96) --
	( 34.64,201.96);

\path[draw=drawColor,line width= 0.6pt,line join=round] ( 31.89,237.59) --
	( 34.64,237.59);
\end{scope}
\begin{scope}
\definecolor{drawColor}{gray}{0.20}

\path[draw=drawColor,line width= 0.6pt,line join=round] ( 44.98, 27.94) --
	( 44.98, 30.69);

\path[draw=drawColor,line width= 0.6pt,line join=round] (111.23, 27.94) --
	(111.23, 30.69);

\path[draw=drawColor,line width= 0.6pt,line join=round] (177.48, 27.94) --
	(177.48, 30.69);

\path[draw=drawColor,line width= 0.6pt,line join=round] (243.73, 27.94) --
	(243.73, 30.69);
\end{scope}
\begin{scope}
\definecolor{drawColor}{gray}{0.30}

\node[text=drawColor,anchor=base,inner sep=0pt, outer sep=0pt, scale=  0.88] at ( 44.98, 19.68) {0};

\node[text=drawColor,anchor=base,inner sep=0pt, outer sep=0pt, scale=  0.88] at (111.23, 19.68) {10};

\node[text=drawColor,anchor=base,inner sep=0pt, outer sep=0pt, scale=  0.88] at (177.48, 19.68) {20};

\node[text=drawColor,anchor=base,inner sep=0pt, outer sep=0pt, scale=  0.88] at (243.73, 19.68) {30};
\end{scope}
\begin{scope}
\definecolor{drawColor}{RGB}{0,0,0}

\node[text=drawColor,anchor=base,inner sep=0pt, outer sep=0pt, scale=  1.10] at (141.04,  7.64) {Model size};
\end{scope}
\begin{scope}
\definecolor{drawColor}{RGB}{0,0,0}

\node[text=drawColor,rotate= 90.00,anchor=base,inner sep=0pt, outer sep=0pt, scale=  1.10] at ( 13.08,139.07) {$\Delta \elpdHatPlain$};
\end{scope}
\end{tikzpicture}

%% file: tikz/schools-dendro.tex
\begin{tikzpicture}[x=1pt,y=1pt]
\definecolor{fillColor}{RGB}{255,255,255}
\begin{scope}
\definecolor{drawColor}{RGB}{255,255,255}
\definecolor{fillColor}{RGB}{255,255,255}

\path[draw=drawColor,line width= 0.6pt,line join=round,line cap=round,fill=fillColor] (  0.00,  0.00) rectangle (216.81,289.08);
\end{scope}
\begin{scope}
\definecolor{fillColor}{RGB}{255,255,255}

\path[fill=fillColor] (  5.50, 30.69) rectangle (155.01,283.58);
\definecolor{drawColor}{gray}{0.92}

\path[draw=drawColor,line width= 0.6pt,line join=round] (148.22, 30.69) --
	(148.22,283.58);

\path[draw=drawColor,line width= 0.6pt,line join=round] ( 96.19, 30.69) --
	( 96.19,283.58);

\path[draw=drawColor,line width= 0.6pt,line join=round] ( 44.16, 30.69) --
	( 44.16,283.58);
\definecolor{drawColor}{RGB}{0,0,0}

\path[draw=drawColor,line width= 0.6pt,line join=round] ( 12.30, 97.32) -- ( 12.30, 48.34);

\path[draw=drawColor,line width= 0.6pt,line join=round] ( 12.30, 48.34) -- ( 55.06, 48.34);

\path[draw=drawColor,line width= 0.6pt,line join=round] ( 55.06, 48.34) -- ( 55.06, 42.18);

\path[draw=drawColor,line width= 0.6pt,line join=round] ( 55.06, 42.18) -- (148.22, 42.18);

\path[draw=drawColor,line width= 0.6pt,line join=round] ( 55.06, 48.34) -- ( 55.06, 54.50);

\path[draw=drawColor,line width= 0.6pt,line join=round] ( 55.06, 54.50) -- ( 59.94, 54.50);

\path[draw=drawColor,line width= 0.6pt,line join=round] ( 59.94, 54.50) -- ( 59.94, 50.39);

\path[draw=drawColor,line width= 0.6pt,line join=round] ( 59.94, 50.39) -- (148.22, 50.39);

\path[draw=drawColor,line width= 0.6pt,line join=round] ( 59.94, 54.50) -- ( 59.94, 58.60);

\path[draw=drawColor,line width= 0.6pt,line join=round] ( 59.94, 58.60) -- (148.22, 58.60);

\path[draw=drawColor,line width= 0.6pt,line join=round] ( 12.30, 97.32) -- ( 12.30,146.29);

\path[draw=drawColor,line width= 0.6pt,line join=round] ( 12.30,146.29) -- ( 30.12,146.29);

\path[draw=drawColor,line width= 0.6pt,line join=round] ( 30.12,146.29) -- ( 30.12, 98.63);

\path[draw=drawColor,line width= 0.6pt,line join=round] ( 30.12, 98.63) -- ( 36.43, 98.63);

\path[draw=drawColor,line width= 0.6pt,line join=round] ( 36.43, 98.63) -- ( 36.43, 79.13);

\path[draw=drawColor,line width= 0.6pt,line join=round] ( 36.43, 79.13) -- ( 49.23, 79.13);

\path[draw=drawColor,line width= 0.6pt,line join=round] ( 49.23, 79.13) -- ( 49.23, 70.92);

\path[draw=drawColor,line width= 0.6pt,line join=round] ( 49.23, 70.92) -- ( 60.47, 70.92);

\path[draw=drawColor,line width= 0.6pt,line join=round] ( 60.47, 70.92) -- ( 60.47, 66.81);

\path[draw=drawColor,line width= 0.6pt,line join=round] ( 60.47, 66.81) -- (148.22, 66.81);

\path[draw=drawColor,line width= 0.6pt,line join=round] ( 60.47, 70.92) -- ( 60.47, 75.02);

\path[draw=drawColor,line width= 0.6pt,line join=round] ( 60.47, 75.02) -- (148.22, 75.02);

\path[draw=drawColor,line width= 0.6pt,line join=round] ( 49.23, 79.13) -- ( 49.23, 87.34);

\path[draw=drawColor,line width= 0.6pt,line join=round] ( 49.23, 87.34) -- ( 60.10, 87.34);

\path[draw=drawColor,line width= 0.6pt,line join=round] ( 60.10, 87.34) -- ( 60.10, 83.24);

\path[draw=drawColor,line width= 0.6pt,line join=round] ( 60.10, 83.24) -- (148.22, 83.24);

\path[draw=drawColor,line width= 0.6pt,line join=round] ( 60.10, 87.34) -- ( 60.10, 91.45);

\path[draw=drawColor,line width= 0.6pt,line join=round] ( 60.10, 91.45) -- (148.22, 91.45);

\path[draw=drawColor,line width= 0.6pt,line join=round] ( 36.43, 98.63) -- ( 36.43,118.13);

\path[draw=drawColor,line width= 0.6pt,line join=round] ( 36.43,118.13) -- ( 41.76,118.13);

\path[draw=drawColor,line width= 0.6pt,line join=round] ( 41.76,118.13) -- ( 41.76,105.82);

\path[draw=drawColor,line width= 0.6pt,line join=round] ( 41.76,105.82) -- ( 53.59,105.82);

\path[draw=drawColor,line width= 0.6pt,line join=round] ( 53.59,105.82) -- ( 53.59, 99.66);

\path[draw=drawColor,line width= 0.6pt,line join=round] ( 53.59, 99.66) -- (148.22, 99.66);

\path[draw=drawColor,line width= 0.6pt,line join=round] ( 53.59,105.82) -- ( 53.59,111.97);

\path[draw=drawColor,line width= 0.6pt,line join=round] ( 53.59,111.97) -- ( 59.73,111.97);

\path[draw=drawColor,line width= 0.6pt,line join=round] ( 59.73,111.97) -- ( 59.73,107.87);

\path[draw=drawColor,line width= 0.6pt,line join=round] ( 59.73,107.87) -- (148.22,107.87);

\path[draw=drawColor,line width= 0.6pt,line join=round] ( 59.73,111.97) -- ( 59.73,116.08);

\path[draw=drawColor,line width= 0.6pt,line join=round] ( 59.73,116.08) -- (148.22,116.08);

\path[draw=drawColor,line width= 0.6pt,line join=round] ( 41.76,118.13) -- ( 41.76,130.45);

\path[draw=drawColor,line width= 0.6pt,line join=round] ( 41.76,130.45) -- ( 50.88,130.45);

\path[draw=drawColor,line width= 0.6pt,line join=round] ( 50.88,130.45) -- ( 50.88,124.29);

\path[draw=drawColor,line width= 0.6pt,line join=round] ( 50.88,124.29) -- (148.22,124.29);

\path[draw=drawColor,line width= 0.6pt,line join=round] ( 50.88,130.45) -- ( 50.88,136.61);

\path[draw=drawColor,line width= 0.6pt,line join=round] ( 50.88,136.61) -- ( 73.82,136.61);

\path[draw=drawColor,line width= 0.6pt,line join=round] ( 73.82,136.61) -- ( 73.82,132.50);

\path[draw=drawColor,line width= 0.6pt,line join=round] ( 73.82,132.50) -- (148.22,132.50);

\path[draw=drawColor,line width= 0.6pt,line join=round] ( 73.82,136.61) -- ( 73.82,140.71);

\path[draw=drawColor,line width= 0.6pt,line join=round] ( 73.82,140.71) -- (148.22,140.71);

\path[draw=drawColor,line width= 0.6pt,line join=round] ( 30.12,146.29) -- ( 30.12,193.95);

\path[draw=drawColor,line width= 0.6pt,line join=round] ( 30.12,193.95) -- ( 33.73,193.95);

\path[draw=drawColor,line width= 0.6pt,line join=round] ( 33.73,193.95) -- ( 33.73,166.37);

\path[draw=drawColor,line width= 0.6pt,line join=round] ( 33.73,166.37) -- ( 50.80,166.37);

\path[draw=drawColor,line width= 0.6pt,line join=round] ( 50.80,166.37) -- ( 50.80,155.08);

\path[draw=drawColor,line width= 0.6pt,line join=round] ( 50.80,155.08) -- ( 69.52,155.08);

\path[draw=drawColor,line width= 0.6pt,line join=round] ( 69.52,155.08) -- ( 69.52,148.92);

\path[draw=drawColor,line width= 0.6pt,line join=round] ( 69.52,148.92) -- (148.22,148.92);

\path[draw=drawColor,line width= 0.6pt,line join=round] ( 69.52,155.08) -- ( 69.52,161.24);

\path[draw=drawColor,line width= 0.6pt,line join=round] ( 69.52,161.24) -- ( 84.47,161.24);

\path[draw=drawColor,line width= 0.6pt,line join=round] ( 84.47,161.24) -- ( 84.47,157.13);

\path[draw=drawColor,line width= 0.6pt,line join=round] ( 84.47,157.13) -- (148.22,157.13);

\path[draw=drawColor,line width= 0.6pt,line join=round] ( 84.47,161.24) -- ( 84.47,165.34);

\path[draw=drawColor,line width= 0.6pt,line join=round] ( 84.47,165.34) -- (148.22,165.34);

\path[draw=drawColor,line width= 0.6pt,line join=round] ( 50.80,166.37) -- ( 50.80,177.66);

\path[draw=drawColor,line width= 0.6pt,line join=round] ( 50.80,177.66) -- ( 60.12,177.66);

\path[draw=drawColor,line width= 0.6pt,line join=round] ( 60.12,177.66) -- ( 60.12,173.55);

\path[draw=drawColor,line width= 0.6pt,line join=round] ( 60.12,173.55) -- (148.22,173.55);

\path[draw=drawColor,line width= 0.6pt,line join=round] ( 60.12,177.66) -- ( 60.12,181.77);

\path[draw=drawColor,line width= 0.6pt,line join=round] ( 60.12,181.77) -- (148.22,181.77);

\path[draw=drawColor,line width= 0.6pt,line join=round] ( 33.73,193.95) -- ( 33.73,221.54);

\path[draw=drawColor,line width= 0.6pt,line join=round] ( 33.73,221.54) -- ( 38.82,221.54);

\path[draw=drawColor,line width= 0.6pt,line join=round] ( 38.82,221.54) -- ( 38.82,197.16);

\path[draw=drawColor,line width= 0.6pt,line join=round] ( 38.82,197.16) -- ( 48.06,197.16);

\path[draw=drawColor,line width= 0.6pt,line join=round] ( 48.06,197.16) -- ( 48.06,189.98);

\path[draw=drawColor,line width= 0.6pt,line join=round] ( 48.06,189.98) -- (148.22,189.98);

\path[draw=drawColor,line width= 0.6pt,line join=round] ( 48.06,197.16) -- ( 48.06,204.35);

\path[draw=drawColor,line width= 0.6pt,line join=round] ( 48.06,204.35) -- ( 64.56,204.35);

\path[draw=drawColor,line width= 0.6pt,line join=round] ( 64.56,204.35) -- ( 64.56,198.19);

\path[draw=drawColor,line width= 0.6pt,line join=round] ( 64.56,198.19) -- (148.22,198.19);

\path[draw=drawColor,line width= 0.6pt,line join=round] ( 64.56,204.35) -- ( 64.56,210.50);

\path[draw=drawColor,line width= 0.6pt,line join=round] ( 64.56,210.50) -- (111.54,210.50);

\path[draw=drawColor,line width= 0.6pt,line join=round] (111.54,210.50) -- (111.54,206.40);

\path[draw=drawColor,line width= 0.6pt,line join=round] (111.54,206.40) -- (148.22,206.40);

\path[draw=drawColor,line width= 0.6pt,line join=round] (111.54,210.50) -- (111.54,214.61);

\path[draw=drawColor,line width= 0.6pt,line join=round] (111.54,214.61) -- (148.22,214.61);

\path[draw=drawColor,line width= 0.6pt,line join=round] ( 38.82,221.54) -- ( 38.82,245.91);

\path[draw=drawColor,line width= 0.6pt,line join=round] ( 38.82,245.91) -- ( 46.64,245.91);

\path[draw=drawColor,line width= 0.6pt,line join=round] ( 46.64,245.91) -- ( 46.64,230.00);

\path[draw=drawColor,line width= 0.6pt,line join=round] ( 46.64,230.00) -- ( 69.97,230.00);

\path[draw=drawColor,line width= 0.6pt,line join=round] ( 69.97,230.00) -- ( 69.97,222.82);

\path[draw=drawColor,line width= 0.6pt,line join=round] ( 69.97,222.82) -- (148.22,222.82);

\path[draw=drawColor,line width= 0.6pt,line join=round] ( 69.97,230.00) -- ( 69.97,237.19);

\path[draw=drawColor,line width= 0.6pt,line join=round] ( 69.97,237.19) -- ( 80.35,237.19);

\path[draw=drawColor,line width= 0.6pt,line join=round] ( 80.35,237.19) -- ( 80.35,231.03);

\path[draw=drawColor,line width= 0.6pt,line join=round] ( 80.35,231.03) -- (148.22,231.03);

\path[draw=drawColor,line width= 0.6pt,line join=round] ( 80.35,237.19) -- ( 80.35,243.35);

\path[draw=drawColor,line width= 0.6pt,line join=round] ( 80.35,243.35) -- (109.04,243.35);

\path[draw=drawColor,line width= 0.6pt,line join=round] (109.04,243.35) -- (109.04,239.24);

\path[draw=drawColor,line width= 0.6pt,line join=round] (109.04,239.24) -- (148.22,239.24);

\path[draw=drawColor,line width= 0.6pt,line join=round] (109.04,243.35) -- (109.04,247.45);

\path[draw=drawColor,line width= 0.6pt,line join=round] (109.04,247.45) -- (148.22,247.45);

\path[draw=drawColor,line width= 0.6pt,line join=round] ( 46.64,245.91) -- ( 46.64,261.82);

\path[draw=drawColor,line width= 0.6pt,line join=round] ( 46.64,261.82) -- ( 69.38,261.82);

\path[draw=drawColor,line width= 0.6pt,line join=round] ( 69.38,261.82) -- ( 69.38,255.66);

\path[draw=drawColor,line width= 0.6pt,line join=round] ( 69.38,255.66) -- (148.22,255.66);

\path[draw=drawColor,line width= 0.6pt,line join=round] ( 69.38,261.82) -- ( 69.38,267.98);

\path[draw=drawColor,line width= 0.6pt,line join=round] ( 69.38,267.98) -- ( 78.30,267.98);

\path[draw=drawColor,line width= 0.6pt,line join=round] ( 78.30,267.98) -- ( 78.30,263.87);

\path[draw=drawColor,line width= 0.6pt,line join=round] ( 78.30,263.87) -- (148.22,263.87);

\path[draw=drawColor,line width= 0.6pt,line join=round] ( 78.30,267.98) -- ( 78.30,272.08);

\path[draw=drawColor,line width= 0.6pt,line join=round] ( 78.30,272.08) -- (148.22,272.08);
\definecolor{drawColor}{gray}{0.20}

\path[draw=drawColor,line width= 0.6pt,line join=round,line cap=round] (  5.50, 30.69) rectangle (155.01,283.58);
\end{scope}
\begin{scope}
\definecolor{drawColor}{gray}{0.20}

\path[draw=drawColor,line width= 0.6pt,line join=round] (155.01, 42.18) --
	(157.76, 42.18);

\path[draw=drawColor,line width= 0.6pt,line join=round] (155.01, 50.39) --
	(157.76, 50.39);

\path[draw=drawColor,line width= 0.6pt,line join=round] (155.01, 58.60) --
	(157.76, 58.60);

\path[draw=drawColor,line width= 0.6pt,line join=round] (155.01, 66.81) --
	(157.76, 66.81);

\path[draw=drawColor,line width= 0.6pt,line join=round] (155.01, 75.02) --
	(157.76, 75.02);

\path[draw=drawColor,line width= 0.6pt,line join=round] (155.01, 83.24) --
	(157.76, 83.24);

\path[draw=drawColor,line width= 0.6pt,line join=round] (155.01, 91.45) --
	(157.76, 91.45);

\path[draw=drawColor,line width= 0.6pt,line join=round] (155.01, 99.66) --
	(157.76, 99.66);

\path[draw=drawColor,line width= 0.6pt,line join=round] (155.01,107.87) --
	(157.76,107.87);

\path[draw=drawColor,line width= 0.6pt,line join=round] (155.01,116.08) --
	(157.76,116.08);

\path[draw=drawColor,line width= 0.6pt,line join=round] (155.01,124.29) --
	(157.76,124.29);

\path[draw=drawColor,line width= 0.6pt,line join=round] (155.01,132.50) --
	(157.76,132.50);

\path[draw=drawColor,line width= 0.6pt,line join=round] (155.01,140.71) --
	(157.76,140.71);

\path[draw=drawColor,line width= 0.6pt,line join=round] (155.01,148.92) --
	(157.76,148.92);

\path[draw=drawColor,line width= 0.6pt,line join=round] (155.01,157.13) --
	(157.76,157.13);

\path[draw=drawColor,line width= 0.6pt,line join=round] (155.01,165.34) --
	(157.76,165.34);

\path[draw=drawColor,line width= 0.6pt,line join=round] (155.01,173.55) --
	(157.76,173.55);

\path[draw=drawColor,line width= 0.6pt,line join=round] (155.01,181.77) --
	(157.76,181.77);

\path[draw=drawColor,line width= 0.6pt,line join=round] (155.01,189.98) --
	(157.76,189.98);

\path[draw=drawColor,line width= 0.6pt,line join=round] (155.01,198.19) --
	(157.76,198.19);

\path[draw=drawColor,line width= 0.6pt,line join=round] (155.01,206.40) --
	(157.76,206.40);

\path[draw=drawColor,line width= 0.6pt,line join=round] (155.01,214.61) --
	(157.76,214.61);

\path[draw=drawColor,line width= 0.6pt,line join=round] (155.01,222.82) --
	(157.76,222.82);

\path[draw=drawColor,line width= 0.6pt,line join=round] (155.01,231.03) --
	(157.76,231.03);

\path[draw=drawColor,line width= 0.6pt,line join=round] (155.01,239.24) --
	(157.76,239.24);

\path[draw=drawColor,line width= 0.6pt,line join=round] (155.01,247.45) --
	(157.76,247.45);

\path[draw=drawColor,line width= 0.6pt,line join=round] (155.01,255.66) --
	(157.76,255.66);

\path[draw=drawColor,line width= 0.6pt,line join=round] (155.01,263.87) --
	(157.76,263.87);

\path[draw=drawColor,line width= 0.6pt,line join=round] (155.01,272.08) --
	(157.76,272.08);
\end{scope}
\begin{scope}
\definecolor{drawColor}{gray}{0.30}

\node[text=drawColor,anchor=base west,inner sep=0pt, outer sep=0pt, scale=  0.88] at (159.96, 39.15) {schoolsup};

\node[text=drawColor,anchor=base west,inner sep=0pt, outer sep=0pt, scale=  0.88] at (159.96, 47.36) {sex};

\node[text=drawColor,anchor=base west,inner sep=0pt, outer sep=0pt, scale=  0.88] at (159.96, 55.57) {famsup};

\node[text=drawColor,anchor=base west,inner sep=0pt, outer sep=0pt, scale=  0.88] at (159.96, 63.78) {health};

\node[text=drawColor,anchor=base west,inner sep=0pt, outer sep=0pt, scale=  0.88] at (159.96, 71.99) {reason};

\node[text=drawColor,anchor=base west,inner sep=0pt, outer sep=0pt, scale=  0.88] at (159.96, 80.20) {paid};

\node[text=drawColor,anchor=base west,inner sep=0pt, outer sep=0pt, scale=  0.88] at (159.96, 88.42) {internet};

\node[text=drawColor,anchor=base west,inner sep=0pt, outer sep=0pt, scale=  0.88] at (159.96, 96.63) {nursery};

\node[text=drawColor,anchor=base west,inner sep=0pt, outer sep=0pt, scale=  0.88] at (159.96,104.84) {famsize};

\node[text=drawColor,anchor=base west,inner sep=0pt, outer sep=0pt, scale=  0.88] at (159.96,113.05) {Pstatus};

\node[text=drawColor,anchor=base west,inner sep=0pt, outer sep=0pt, scale=  0.88] at (159.96,121.26) {famrel};

\node[text=drawColor,anchor=base west,inner sep=0pt, outer sep=0pt, scale=  0.88] at (159.96,129.47) {freetime};

\node[text=drawColor,anchor=base west,inner sep=0pt, outer sep=0pt, scale=  0.88] at (159.96,137.68) {goout};

\node[text=drawColor,anchor=base west,inner sep=0pt, outer sep=0pt, scale=  0.88] at (159.96,145.89) {failures};

\node[text=drawColor,anchor=base west,inner sep=0pt, outer sep=0pt, scale=  0.88] at (159.96,154.10) {age};

\node[text=drawColor,anchor=base west,inner sep=0pt, outer sep=0pt, scale=  0.88] at (159.96,162.31) {guardian};

\node[text=drawColor,anchor=base west,inner sep=0pt, outer sep=0pt, scale=  0.88] at (159.96,170.52) {romantic};

\node[text=drawColor,anchor=base west,inner sep=0pt, outer sep=0pt, scale=  0.88] at (159.96,178.74) {absences};

\node[text=drawColor,anchor=base west,inner sep=0pt, outer sep=0pt, scale=  0.88] at (159.96,186.95) {activities};

\node[text=drawColor,anchor=base west,inner sep=0pt, outer sep=0pt, scale=  0.88] at (159.96,195.16) {studytime};

\node[text=drawColor,anchor=base west,inner sep=0pt, outer sep=0pt, scale=  0.88] at (159.96,203.37) {Dalc};

\node[text=drawColor,anchor=base west,inner sep=0pt, outer sep=0pt, scale=  0.88] at (159.96,211.58) {Walc};

\node[text=drawColor,anchor=base west,inner sep=0pt, outer sep=0pt, scale=  0.88] at (159.96,219.79) {Fjob};

\node[text=drawColor,anchor=base west,inner sep=0pt, outer sep=0pt, scale=  0.88] at (159.96,228.00) {Mjob};

\node[text=drawColor,anchor=base west,inner sep=0pt, outer sep=0pt, scale=  0.88] at (159.96,236.21) {Medu};

\node[text=drawColor,anchor=base west,inner sep=0pt, outer sep=0pt, scale=  0.88] at (159.96,244.42) {Fedu};

\node[text=drawColor,anchor=base west,inner sep=0pt, outer sep=0pt, scale=  0.88] at (159.96,252.63) {school};

\node[text=drawColor,anchor=base west,inner sep=0pt, outer sep=0pt, scale=  0.88] at (159.96,260.84) {address};

\node[text=drawColor,anchor=base west,inner sep=0pt, outer sep=0pt, scale=  0.88] at (159.96,269.05) {traveltime};
\end{scope}
\begin{scope}
\definecolor{drawColor}{gray}{0.20}

\path[draw=drawColor,line width= 0.6pt,line join=round] (148.22, 27.94) --
	(148.22, 30.69);

\path[draw=drawColor,line width= 0.6pt,line join=round] ( 96.19, 27.94) --
	( 96.19, 30.69);

\path[draw=drawColor,line width= 0.6pt,line join=round] ( 44.16, 27.94) --
	( 44.16, 30.69);
\end{scope}
\begin{scope}
\definecolor{drawColor}{gray}{0.30}

\node[text=drawColor,anchor=base,inner sep=0pt, outer sep=0pt, scale=  0.88] at (148.22, 19.68) {0.0};

\node[text=drawColor,anchor=base,inner sep=0pt, outer sep=0pt, scale=  0.88] at ( 96.19, 19.68) {0.5};

\node[text=drawColor,anchor=base,inner sep=0pt, outer sep=0pt, scale=  0.88] at ( 44.16, 19.68) {1.0};
\end{scope}
\begin{scope}
\definecolor{drawColor}{RGB}{0,0,0}

\node[text=drawColor,anchor=base,inner sep=0pt, outer sep=0pt, scale=  1.10] at ( 80.26,  7.64) {$\dist{\theta}{\theta^\prime}$};
\end{scope}
\begin{scope}
\definecolor{drawColor}{RGB}{0,0,0}

\node[text=drawColor,rotate=-90.00,anchor=base,inner sep=0pt, outer sep=0pt, scale=  1.10] at (201.60,157.13) {Predictor ($\theta$)};
\end{scope}
\end{tikzpicture}

%% file: tikz/dendrogram_case_study_4.1.tex
\begin{tikzpicture}[x=1pt,y=1pt]
\definecolor{fillColor}{RGB}{255,255,255}
\begin{scope}
\definecolor{drawColor}{RGB}{255,255,255}
\definecolor{fillColor}{RGB}{255,255,255}

\path[draw=drawColor,line width= 0.6pt,line join=round,line cap=round,fill=fillColor] (  0.00,  0.00) rectangle (216.81,289.08);
\end{scope}
\begin{scope}
\definecolor{fillColor}{RGB}{255,255,255}

\path[fill=fillColor] (  5.50, 30.69) rectangle (176.06,283.58);
\definecolor{drawColor}{gray}{0.92}

\path[draw=drawColor,line width= 0.6pt,line join=round] (168.30, 30.69) --
	(168.30,283.58);

\path[draw=drawColor,line width= 0.6pt,line join=round] (127.75, 30.69) --
	(127.75,283.58);

\path[draw=drawColor,line width= 0.6pt,line join=round] ( 87.19, 30.69) --
	( 87.19,283.58);

\path[draw=drawColor,line width= 0.6pt,line join=round] ( 46.63, 30.69) --
	( 46.63,283.58);

\path[draw=drawColor,line width= 0.6pt,line join=round] (  6.07, 30.69) --
	(  6.07,283.58);
\definecolor{drawColor}{RGB}{0,0,0}

\path[draw=drawColor,line width= 0.6pt,line join=round] ( 13.25,146.04) -- ( 13.25, 82.36);

\path[draw=drawColor,line width= 0.6pt,line join=round] ( 13.25, 82.36) -- ( 77.76, 82.36);

\path[draw=drawColor,line width= 0.6pt,line join=round] ( 77.76, 82.36) -- ( 77.76, 54.74);

\path[draw=drawColor,line width= 0.6pt,line join=round] ( 77.76, 54.74) -- ( 84.82, 54.74);

\path[draw=drawColor,line width= 0.6pt,line join=round] ( 84.82, 54.74) -- ( 84.82, 45.95);

\path[draw=drawColor,line width= 0.6pt,line join=round] ( 84.82, 45.95) -- (158.70, 45.95);

\path[draw=drawColor,line width= 0.6pt,line join=round] (158.70, 45.95) -- (158.70, 43.34);

\path[draw=drawColor,line width= 0.6pt,line join=round] (158.70, 43.34) -- (159.64, 43.34);

\path[draw=drawColor,line width= 0.6pt,line join=round] (159.64, 43.34) -- (159.64, 42.18);

\path[draw=drawColor,line width= 0.6pt,line join=round] (159.64, 42.18) -- (168.30, 42.18);

\path[draw=drawColor,line width= 0.6pt,line join=round] (159.64, 43.34) -- (159.64, 44.50);

\path[draw=drawColor,line width= 0.6pt,line join=round] (159.64, 44.50) -- (168.30, 44.50);

\path[draw=drawColor,line width= 0.6pt,line join=round] (158.70, 45.95) -- (158.70, 48.57);

\path[draw=drawColor,line width= 0.6pt,line join=round] (158.70, 48.57) -- (159.02, 48.57);

\path[draw=drawColor,line width= 0.6pt,line join=round] (159.02, 48.57) -- (159.02, 46.83);

\path[draw=drawColor,line width= 0.6pt,line join=round] (159.02, 46.83) -- (168.30, 46.83);

\path[draw=drawColor,line width= 0.6pt,line join=round] (159.02, 48.57) -- (159.02, 50.31);

\path[draw=drawColor,line width= 0.6pt,line join=round] (159.02, 50.31) -- (159.52, 50.31);

\path[draw=drawColor,line width= 0.6pt,line join=round] (159.52, 50.31) -- (159.52, 49.15);

\path[draw=drawColor,line width= 0.6pt,line join=round] (159.52, 49.15) -- (168.30, 49.15);

\path[draw=drawColor,line width= 0.6pt,line join=round] (159.52, 50.31) -- (159.52, 51.47);

\path[draw=drawColor,line width= 0.6pt,line join=round] (159.52, 51.47) -- (168.30, 51.47);

\path[draw=drawColor,line width= 0.6pt,line join=round] ( 84.82, 54.74) -- ( 84.82, 63.52);

\path[draw=drawColor,line width= 0.6pt,line join=round] ( 84.82, 63.52) -- ( 88.61, 63.52);
\definecolor{drawColor}{RGB}{0,0,255}

\path[draw=drawColor,line width= 0.6pt,line join=round] ( 88.61, 63.52) -- ( 88.61, 58.73);

\path[draw=drawColor,line width= 0.6pt,line join=round] ( 88.61, 58.73) -- (159.46, 58.73);

\path[draw=drawColor,line width= 0.6pt,line join=round] (159.46, 58.73) -- (159.46, 55.53);

\path[draw=drawColor,line width= 0.6pt,line join=round] (159.46, 55.53) -- (160.62, 55.53);

\path[draw=drawColor,line width= 0.6pt,line join=round] (160.62, 55.53) -- (160.62, 53.79);

\path[draw=drawColor,line width= 0.6pt,line join=round] (160.62, 53.79) -- (168.30, 53.79);

\path[draw=drawColor,line width= 0.6pt,line join=round] (160.62, 55.53) -- (160.62, 57.28);

\path[draw=drawColor,line width= 0.6pt,line join=round] (160.62, 57.28) -- (161.04, 57.28);

\path[draw=drawColor,line width= 0.6pt,line join=round] (161.04, 57.28) -- (161.04, 56.11);

\path[draw=drawColor,line width= 0.6pt,line join=round] (161.04, 56.11) -- (168.30, 56.11);

\path[draw=drawColor,line width= 0.6pt,line join=round] (161.04, 57.28) -- (161.04, 58.44);

\path[draw=drawColor,line width= 0.6pt,line join=round] (161.04, 58.44) -- (168.30, 58.44);

\path[draw=drawColor,line width= 0.6pt,line join=round] (159.46, 58.73) -- (159.46, 61.92);

\path[draw=drawColor,line width= 0.6pt,line join=round] (159.46, 61.92) -- (159.69, 61.92);

\path[draw=drawColor,line width= 0.6pt,line join=round] (159.69, 61.92) -- (159.69, 60.76);

\path[draw=drawColor,line width= 0.6pt,line join=round] (159.69, 60.76) -- (168.30, 60.76);

\path[draw=drawColor,line width= 0.6pt,line join=round] (159.69, 61.92) -- (159.69, 63.08);

\path[draw=drawColor,line width= 0.6pt,line join=round] (159.69, 63.08) -- (168.30, 63.08);
\definecolor{drawColor}{RGB}{0,0,0}

\path[draw=drawColor,line width= 0.6pt,line join=round] ( 88.61, 63.52) -- ( 88.61, 68.31);

\path[draw=drawColor,line width= 0.6pt,line join=round] ( 88.61, 68.31) -- ( 91.54, 68.31);

\path[draw=drawColor,line width= 0.6pt,line join=round] ( 91.54, 68.31) -- ( 91.54, 65.40);

\path[draw=drawColor,line width= 0.6pt,line join=round] ( 91.54, 65.40) -- (168.30, 65.40);

\path[draw=drawColor,line width= 0.6pt,line join=round] ( 91.54, 68.31) -- ( 91.54, 71.21);

\path[draw=drawColor,line width= 0.6pt,line join=round] ( 91.54, 71.21) -- (159.42, 71.21);

\path[draw=drawColor,line width= 0.6pt,line join=round] (159.42, 71.21) -- (159.42, 68.89);

\path[draw=drawColor,line width= 0.6pt,line join=round] (159.42, 68.89) -- (160.50, 68.89);

\path[draw=drawColor,line width= 0.6pt,line join=round] (160.50, 68.89) -- (160.50, 67.73);

\path[draw=drawColor,line width= 0.6pt,line join=round] (160.50, 67.73) -- (168.30, 67.73);

\path[draw=drawColor,line width= 0.6pt,line join=round] (160.50, 68.89) -- (160.50, 70.05);

\path[draw=drawColor,line width= 0.6pt,line join=round] (160.50, 70.05) -- (168.30, 70.05);

\path[draw=drawColor,line width= 0.6pt,line join=round] (159.42, 71.21) -- (159.42, 73.53);

\path[draw=drawColor,line width= 0.6pt,line join=round] (159.42, 73.53) -- (159.75, 73.53);

\path[draw=drawColor,line width= 0.6pt,line join=round] (159.75, 73.53) -- (159.75, 72.37);

\path[draw=drawColor,line width= 0.6pt,line join=round] (159.75, 72.37) -- (168.30, 72.37);

\path[draw=drawColor,line width= 0.6pt,line join=round] (159.75, 73.53) -- (159.75, 74.69);

\path[draw=drawColor,line width= 0.6pt,line join=round] (159.75, 74.69) -- (168.30, 74.69);

\path[draw=drawColor,line width= 0.6pt,line join=round] ( 77.76, 82.36) -- ( 77.76,109.98);

\path[draw=drawColor,line width= 0.6pt,line join=round] ( 77.76,109.98) -- ( 81.07,109.98);

\path[draw=drawColor,line width= 0.6pt,line join=round] ( 81.07,109.98) -- ( 81.07, 88.66);

\path[draw=drawColor,line width= 0.6pt,line join=round] ( 81.07, 88.66) -- ( 89.14, 88.66);

\path[draw=drawColor,line width= 0.6pt,line join=round] ( 89.14, 88.66) -- ( 89.14, 79.92);

\path[draw=drawColor,line width= 0.6pt,line join=round] ( 89.14, 79.92) -- (158.38, 79.92);

\path[draw=drawColor,line width= 0.6pt,line join=round] (158.38, 79.92) -- (158.38, 77.01);

\path[draw=drawColor,line width= 0.6pt,line join=round] (158.38, 77.01) -- (168.30, 77.01);

\path[draw=drawColor,line width= 0.6pt,line join=round] (158.38, 79.92) -- (158.38, 82.82);

\path[draw=drawColor,line width= 0.6pt,line join=round] (158.38, 82.82) -- (159.14, 82.82);

\path[draw=drawColor,line width= 0.6pt,line join=round] (159.14, 82.82) -- (159.14, 80.50);

\path[draw=drawColor,line width= 0.6pt,line join=round] (159.14, 80.50) -- (160.13, 80.50);

\path[draw=drawColor,line width= 0.6pt,line join=round] (160.13, 80.50) -- (160.13, 79.34);

\path[draw=drawColor,line width= 0.6pt,line join=round] (160.13, 79.34) -- (168.30, 79.34);

\path[draw=drawColor,line width= 0.6pt,line join=round] (160.13, 80.50) -- (160.13, 81.66);

\path[draw=drawColor,line width= 0.6pt,line join=round] (160.13, 81.66) -- (168.30, 81.66);

\path[draw=drawColor,line width= 0.6pt,line join=round] (159.14, 82.82) -- (159.14, 85.14);

\path[draw=drawColor,line width= 0.6pt,line join=round] (159.14, 85.14) -- (159.75, 85.14);

\path[draw=drawColor,line width= 0.6pt,line join=round] (159.75, 85.14) -- (159.75, 83.98);

\path[draw=drawColor,line width= 0.6pt,line join=round] (159.75, 83.98) -- (168.30, 83.98);

\path[draw=drawColor,line width= 0.6pt,line join=round] (159.75, 85.14) -- (159.75, 86.30);

\path[draw=drawColor,line width= 0.6pt,line join=round] (159.75, 86.30) -- (168.30, 86.30);

\path[draw=drawColor,line width= 0.6pt,line join=round] ( 89.14, 88.66) -- ( 89.14, 97.41);

\path[draw=drawColor,line width= 0.6pt,line join=round] ( 89.14, 97.41) -- ( 93.53, 97.41);

\path[draw=drawColor,line width= 0.6pt,line join=round] ( 93.53, 97.41) -- ( 93.53, 92.40);

\path[draw=drawColor,line width= 0.6pt,line join=round] ( 93.53, 92.40) -- (158.65, 92.40);

\path[draw=drawColor,line width= 0.6pt,line join=round] (158.65, 92.40) -- (158.65, 89.79);

\path[draw=drawColor,line width= 0.6pt,line join=round] (158.65, 89.79) -- (159.75, 89.79);

\path[draw=drawColor,line width= 0.6pt,line join=round] (159.75, 89.79) -- (159.75, 88.63);

\path[draw=drawColor,line width= 0.6pt,line join=round] (159.75, 88.63) -- (168.30, 88.63);

\path[draw=drawColor,line width= 0.6pt,line join=round] (159.75, 89.79) -- (159.75, 90.95);

\path[draw=drawColor,line width= 0.6pt,line join=round] (159.75, 90.95) -- (168.30, 90.95);

\path[draw=drawColor,line width= 0.6pt,line join=round] (158.65, 92.40) -- (158.65, 95.01);

\path[draw=drawColor,line width= 0.6pt,line join=round] (158.65, 95.01) -- (159.04, 95.01);

\path[draw=drawColor,line width= 0.6pt,line join=round] (159.04, 95.01) -- (159.04, 93.27);

\path[draw=drawColor,line width= 0.6pt,line join=round] (159.04, 93.27) -- (168.30, 93.27);

\path[draw=drawColor,line width= 0.6pt,line join=round] (159.04, 95.01) -- (159.04, 96.75);

\path[draw=drawColor,line width= 0.6pt,line join=round] (159.04, 96.75) -- (160.33, 96.75);

\path[draw=drawColor,line width= 0.6pt,line join=round] (160.33, 96.75) -- (160.33, 95.59);

\path[draw=drawColor,line width= 0.6pt,line join=round] (160.33, 95.59) -- (168.30, 95.59);

\path[draw=drawColor,line width= 0.6pt,line join=round] (160.33, 96.75) -- (160.33, 97.92);

\path[draw=drawColor,line width= 0.6pt,line join=round] (160.33, 97.92) -- (168.30, 97.92);

\path[draw=drawColor,line width= 0.6pt,line join=round] ( 93.53, 97.41) -- ( 93.53,102.41);

\path[draw=drawColor,line width= 0.6pt,line join=round] ( 93.53,102.41) -- (158.60,102.41);

\path[draw=drawColor,line width= 0.6pt,line join=round] (158.60,102.41) -- (158.60,100.24);

\path[draw=drawColor,line width= 0.6pt,line join=round] (158.60,100.24) -- (168.30,100.24);

\path[draw=drawColor,line width= 0.6pt,line join=round] (158.60,102.41) -- (158.60,104.59);

\path[draw=drawColor,line width= 0.6pt,line join=round] (158.60,104.59) -- (159.23,104.59);

\path[draw=drawColor,line width= 0.6pt,line join=round] (159.23,104.59) -- (159.23,102.56);

\path[draw=drawColor,line width= 0.6pt,line join=round] (159.23,102.56) -- (168.30,102.56);

\path[draw=drawColor,line width= 0.6pt,line join=round] (159.23,104.59) -- (159.23,106.62);

\path[draw=drawColor,line width= 0.6pt,line join=round] (159.23,106.62) -- (159.85,106.62);

\path[draw=drawColor,line width= 0.6pt,line join=round] (159.85,106.62) -- (159.85,104.88);

\path[draw=drawColor,line width= 0.6pt,line join=round] (159.85,104.88) -- (168.30,104.88);

\path[draw=drawColor,line width= 0.6pt,line join=round] (159.85,106.62) -- (159.85,108.37);

\path[draw=drawColor,line width= 0.6pt,line join=round] (159.85,108.37) -- (160.16,108.37);

\path[draw=drawColor,line width= 0.6pt,line join=round] (160.16,108.37) -- (160.16,107.20);

\path[draw=drawColor,line width= 0.6pt,line join=round] (160.16,107.20) -- (168.30,107.20);

\path[draw=drawColor,line width= 0.6pt,line join=round] (160.16,108.37) -- (160.16,109.53);

\path[draw=drawColor,line width= 0.6pt,line join=round] (160.16,109.53) -- (168.30,109.53);

\path[draw=drawColor,line width= 0.6pt,line join=round] ( 81.07,109.98) -- ( 81.07,131.30);

\path[draw=drawColor,line width= 0.6pt,line join=round] ( 81.07,131.30) -- ( 81.82,131.30);

\path[draw=drawColor,line width= 0.6pt,line join=round] ( 81.82,131.30) -- ( 81.82,120.99);

\path[draw=drawColor,line width= 0.6pt,line join=round] ( 81.82,120.99) -- ( 90.11,120.99);

\path[draw=drawColor,line width= 0.6pt,line join=round] ( 90.11,120.99) -- ( 90.11,115.62);

\path[draw=drawColor,line width= 0.6pt,line join=round] ( 90.11,115.62) -- (159.53,115.62);

\path[draw=drawColor,line width= 0.6pt,line join=round] (159.53,115.62) -- (159.53,113.01);

\path[draw=drawColor,line width= 0.6pt,line join=round] (159.53,113.01) -- (160.33,113.01);

\path[draw=drawColor,line width= 0.6pt,line join=round] (160.33,113.01) -- (160.33,111.85);

\path[draw=drawColor,line width= 0.6pt,line join=round] (160.33,111.85) -- (168.30,111.85);

\path[draw=drawColor,line width= 0.6pt,line join=round] (160.33,113.01) -- (160.33,114.17);

\path[draw=drawColor,line width= 0.6pt,line join=round] (160.33,114.17) -- (168.30,114.17);

\path[draw=drawColor,line width= 0.6pt,line join=round] (159.53,115.62) -- (159.53,118.24);

\path[draw=drawColor,line width= 0.6pt,line join=round] (159.53,118.24) -- (160.04,118.24);

\path[draw=drawColor,line width= 0.6pt,line join=round] (160.04,118.24) -- (160.04,116.49);

\path[draw=drawColor,line width= 0.6pt,line join=round] (160.04,116.49) -- (168.30,116.49);

\path[draw=drawColor,line width= 0.6pt,line join=round] (160.04,118.24) -- (160.04,119.98);

\path[draw=drawColor,line width= 0.6pt,line join=round] (160.04,119.98) -- (160.50,119.98);

\path[draw=drawColor,line width= 0.6pt,line join=round] (160.50,119.98) -- (160.50,118.82);

\path[draw=drawColor,line width= 0.6pt,line join=round] (160.50,118.82) -- (168.30,118.82);

\path[draw=drawColor,line width= 0.6pt,line join=round] (160.50,119.98) -- (160.50,121.14);

\path[draw=drawColor,line width= 0.6pt,line join=round] (160.50,121.14) -- (168.30,121.14);

\path[draw=drawColor,line width= 0.6pt,line join=round] ( 90.11,120.99) -- ( 90.11,126.36);

\path[draw=drawColor,line width= 0.6pt,line join=round] ( 90.11,126.36) -- ( 92.28,126.36);

\path[draw=drawColor,line width= 0.6pt,line join=round] ( 92.28,126.36) -- ( 92.28,123.46);

\path[draw=drawColor,line width= 0.6pt,line join=round] ( 92.28,123.46) -- (168.30,123.46);

\path[draw=drawColor,line width= 0.6pt,line join=round] ( 92.28,126.36) -- ( 92.28,129.27);

\path[draw=drawColor,line width= 0.6pt,line join=round] ( 92.28,129.27) -- (159.15,129.27);

\path[draw=drawColor,line width= 0.6pt,line join=round] (159.15,129.27) -- (159.15,126.94);

\path[draw=drawColor,line width= 0.6pt,line join=round] (159.15,126.94) -- (160.49,126.94);

\path[draw=drawColor,line width= 0.6pt,line join=round] (160.49,126.94) -- (160.49,125.78);

\path[draw=drawColor,line width= 0.6pt,line join=round] (160.49,125.78) -- (168.30,125.78);

\path[draw=drawColor,line width= 0.6pt,line join=round] (160.49,126.94) -- (160.49,128.10);

\path[draw=drawColor,line width= 0.6pt,line join=round] (160.49,128.10) -- (168.30,128.10);

\path[draw=drawColor,line width= 0.6pt,line join=round] (159.15,129.27) -- (159.15,131.59);

\path[draw=drawColor,line width= 0.6pt,line join=round] (159.15,131.59) -- (159.79,131.59);

\path[draw=drawColor,line width= 0.6pt,line join=round] (159.79,131.59) -- (159.79,130.43);

\path[draw=drawColor,line width= 0.6pt,line join=round] (159.79,130.43) -- (168.30,130.43);

\path[draw=drawColor,line width= 0.6pt,line join=round] (159.79,131.59) -- (159.79,132.75);

\path[draw=drawColor,line width= 0.6pt,line join=round] (159.79,132.75) -- (168.30,132.75);

\path[draw=drawColor,line width= 0.6pt,line join=round] ( 81.82,131.30) -- ( 81.82,141.60);

\path[draw=drawColor,line width= 0.6pt,line join=round] ( 81.82,141.60) -- ( 88.61,141.60);

\path[draw=drawColor,line width= 0.6pt,line join=round] ( 88.61,141.60) -- ( 88.61,137.10);

\path[draw=drawColor,line width= 0.6pt,line join=round] ( 88.61,137.10) -- (160.31,137.10);

\path[draw=drawColor,line width= 0.6pt,line join=round] (160.31,137.10) -- (160.31,135.07);

\path[draw=drawColor,line width= 0.6pt,line join=round] (160.31,135.07) -- (168.30,135.07);

\path[draw=drawColor,line width= 0.6pt,line join=round] (160.31,137.10) -- (160.31,139.14);

\path[draw=drawColor,line width= 0.6pt,line join=round] (160.31,139.14) -- (160.62,139.14);

\path[draw=drawColor,line width= 0.6pt,line join=round] (160.62,139.14) -- (160.62,137.39);

\path[draw=drawColor,line width= 0.6pt,line join=round] (160.62,137.39) -- (168.30,137.39);

\path[draw=drawColor,line width= 0.6pt,line join=round] (160.62,139.14) -- (160.62,140.88);

\path[draw=drawColor,line width= 0.6pt,line join=round] (160.62,140.88) -- (161.02,140.88);

\path[draw=drawColor,line width= 0.6pt,line join=round] (161.02,140.88) -- (161.02,139.72);

\path[draw=drawColor,line width= 0.6pt,line join=round] (161.02,139.72) -- (168.30,139.72);

\path[draw=drawColor,line width= 0.6pt,line join=round] (161.02,140.88) -- (161.02,142.04);

\path[draw=drawColor,line width= 0.6pt,line join=round] (161.02,142.04) -- (168.30,142.04);

\path[draw=drawColor,line width= 0.6pt,line join=round] ( 88.61,141.60) -- ( 88.61,146.10);

\path[draw=drawColor,line width= 0.6pt,line join=round] ( 88.61,146.10) -- (159.30,146.10);

\path[draw=drawColor,line width= 0.6pt,line join=round] (159.30,146.10) -- (159.30,144.36);

\path[draw=drawColor,line width= 0.6pt,line join=round] (159.30,144.36) -- (168.30,144.36);

\path[draw=drawColor,line width= 0.6pt,line join=round] (159.30,146.10) -- (159.30,147.84);

\path[draw=drawColor,line width= 0.6pt,line join=round] (159.30,147.84) -- (160.03,147.84);

\path[draw=drawColor,line width= 0.6pt,line join=round] (160.03,147.84) -- (160.03,146.68);

\path[draw=drawColor,line width= 0.6pt,line join=round] (160.03,146.68) -- (168.30,146.68);

\path[draw=drawColor,line width= 0.6pt,line join=round] (160.03,147.84) -- (160.03,149.00);

\path[draw=drawColor,line width= 0.6pt,line join=round] (160.03,149.00) -- (168.30,149.00);

\path[draw=drawColor,line width= 0.6pt,line join=round] ( 13.25,146.04) -- ( 13.25,209.73);

\path[draw=drawColor,line width= 0.6pt,line join=round] ( 13.25,209.73) -- ( 76.76,209.73);

\path[draw=drawColor,line width= 0.6pt,line join=round] ( 76.76,209.73) -- ( 76.76,182.51);

\path[draw=drawColor,line width= 0.6pt,line join=round] ( 76.76,182.51) -- ( 83.66,182.51);

\path[draw=drawColor,line width= 0.6pt,line join=round] ( 83.66,182.51) -- ( 83.66,168.78);

\path[draw=drawColor,line width= 0.6pt,line join=round] ( 83.66,168.78) -- ( 86.00,168.78);

\path[draw=drawColor,line width= 0.6pt,line join=round] ( 86.00,168.78) -- ( 86.00,160.11);

\path[draw=drawColor,line width= 0.6pt,line join=round] ( 86.00,160.11) -- ( 92.55,160.11);
\definecolor{drawColor}{RGB}{0,0,255}

\path[draw=drawColor,line width= 0.6pt,line join=round] ( 92.55,160.11) -- ( 92.55,155.10);

\path[draw=drawColor,line width= 0.6pt,line join=round] ( 92.55,155.10) -- (159.43,155.10);

\path[draw=drawColor,line width= 0.6pt,line join=round] (159.43,155.10) -- (159.43,152.49);

\path[draw=drawColor,line width= 0.6pt,line join=round] (159.43,152.49) -- (160.16,152.49);

\path[draw=drawColor,line width= 0.6pt,line join=round] (160.16,152.49) -- (160.16,151.33);

\path[draw=drawColor,line width= 0.6pt,line join=round] (160.16,151.33) -- (168.30,151.33);

\path[draw=drawColor,line width= 0.6pt,line join=round] (160.16,152.49) -- (160.16,153.65);

\path[draw=drawColor,line width= 0.6pt,line join=round] (160.16,153.65) -- (168.30,153.65);

\path[draw=drawColor,line width= 0.6pt,line join=round] (159.43,155.10) -- (159.43,157.71);

\path[draw=drawColor,line width= 0.6pt,line join=round] (159.43,157.71) -- (159.73,157.71);

\path[draw=drawColor,line width= 0.6pt,line join=round] (159.73,157.71) -- (159.73,155.97);

\path[draw=drawColor,line width= 0.6pt,line join=round] (159.73,155.97) -- (168.30,155.97);

\path[draw=drawColor,line width= 0.6pt,line join=round] (159.73,157.71) -- (159.73,159.46);

\path[draw=drawColor,line width= 0.6pt,line join=round] (159.73,159.46) -- (160.53,159.46);

\path[draw=drawColor,line width= 0.6pt,line join=round] (160.53,159.46) -- (160.53,158.29);

\path[draw=drawColor,line width= 0.6pt,line join=round] (160.53,158.29) -- (168.30,158.29);

\path[draw=drawColor,line width= 0.6pt,line join=round] (160.53,159.46) -- (160.53,160.62);

\path[draw=drawColor,line width= 0.6pt,line join=round] (160.53,160.62) -- (168.30,160.62);
\definecolor{drawColor}{RGB}{0,0,0}

\path[draw=drawColor,line width= 0.6pt,line join=round] ( 92.55,160.11) -- ( 92.55,165.12);

\path[draw=drawColor,line width= 0.6pt,line join=round] ( 92.55,165.12) -- (158.31,165.12);

\path[draw=drawColor,line width= 0.6pt,line join=round] (158.31,165.12) -- (158.31,162.94);

\path[draw=drawColor,line width= 0.6pt,line join=round] (158.31,162.94) -- (168.30,162.94);

\path[draw=drawColor,line width= 0.6pt,line join=round] (158.31,165.12) -- (158.31,167.29);

\path[draw=drawColor,line width= 0.6pt,line join=round] (158.31,167.29) -- (159.38,167.29);

\path[draw=drawColor,line width= 0.6pt,line join=round] (159.38,167.29) -- (159.38,165.26);

\path[draw=drawColor,line width= 0.6pt,line join=round] (159.38,165.26) -- (168.30,165.26);

\path[draw=drawColor,line width= 0.6pt,line join=round] (159.38,167.29) -- (159.38,169.32);

\path[draw=drawColor,line width= 0.6pt,line join=round] (159.38,169.32) -- (159.69,169.32);

\path[draw=drawColor,line width= 0.6pt,line join=round] (159.69,169.32) -- (159.69,167.58);

\path[draw=drawColor,line width= 0.6pt,line join=round] (159.69,167.58) -- (168.30,167.58);

\path[draw=drawColor,line width= 0.6pt,line join=round] (159.69,169.32) -- (159.69,171.07);

\path[draw=drawColor,line width= 0.6pt,line join=round] (159.69,171.07) -- (160.32,171.07);

\path[draw=drawColor,line width= 0.6pt,line join=round] (160.32,171.07) -- (160.32,169.91);

\path[draw=drawColor,line width= 0.6pt,line join=round] (160.32,169.91) -- (168.30,169.91);

\path[draw=drawColor,line width= 0.6pt,line join=round] (160.32,171.07) -- (160.32,172.23);

\path[draw=drawColor,line width= 0.6pt,line join=round] (160.32,172.23) -- (168.30,172.23);

\path[draw=drawColor,line width= 0.6pt,line join=round] ( 86.00,168.78) -- ( 86.00,177.45);

\path[draw=drawColor,line width= 0.6pt,line join=round] ( 86.00,177.45) -- ( 91.29,177.45);

\path[draw=drawColor,line width= 0.6pt,line join=round] ( 91.29,177.45) -- ( 91.29,174.55);

\path[draw=drawColor,line width= 0.6pt,line join=round] ( 91.29,174.55) -- (168.30,174.55);

\path[draw=drawColor,line width= 0.6pt,line join=round] ( 91.29,177.45) -- ( 91.29,180.36);

\path[draw=drawColor,line width= 0.6pt,line join=round] ( 91.29,180.36) -- (159.57,180.36);

\path[draw=drawColor,line width= 0.6pt,line join=round] (159.57,180.36) -- (159.57,178.03);

\path[draw=drawColor,line width= 0.6pt,line join=round] (159.57,178.03) -- (160.59,178.03);

\path[draw=drawColor,line width= 0.6pt,line join=round] (160.59,178.03) -- (160.59,176.87);

\path[draw=drawColor,line width= 0.6pt,line join=round] (160.59,176.87) -- (168.30,176.87);

\path[draw=drawColor,line width= 0.6pt,line join=round] (160.59,178.03) -- (160.59,179.19);

\path[draw=drawColor,line width= 0.6pt,line join=round] (160.59,179.19) -- (168.30,179.19);

\path[draw=drawColor,line width= 0.6pt,line join=round] (159.57,180.36) -- (159.57,182.68);

\path[draw=drawColor,line width= 0.6pt,line join=round] (159.57,182.68) -- (160.19,182.68);

\path[draw=drawColor,line width= 0.6pt,line join=round] (160.19,182.68) -- (160.19,181.52);

\path[draw=drawColor,line width= 0.6pt,line join=round] (160.19,181.52) -- (168.30,181.52);

\path[draw=drawColor,line width= 0.6pt,line join=round] (160.19,182.68) -- (160.19,183.84);

\path[draw=drawColor,line width= 0.6pt,line join=round] (160.19,183.84) -- (168.30,183.84);

\path[draw=drawColor,line width= 0.6pt,line join=round] ( 83.66,182.51) -- ( 83.66,196.25);

\path[draw=drawColor,line width= 0.6pt,line join=round] ( 83.66,196.25) -- ( 84.70,196.25);

\path[draw=drawColor,line width= 0.6pt,line join=round] ( 84.70,196.25) -- ( 84.70,189.79);

\path[draw=drawColor,line width= 0.6pt,line join=round] ( 84.70,189.79) -- ( 93.11,189.79);

\path[draw=drawColor,line width= 0.6pt,line join=round] ( 93.11,189.79) -- ( 93.11,186.16);

\path[draw=drawColor,line width= 0.6pt,line join=round] ( 93.11,186.16) -- (168.30,186.16);

\path[draw=drawColor,line width= 0.6pt,line join=round] ( 93.11,189.79) -- ( 93.11,193.42);

\path[draw=drawColor,line width= 0.6pt,line join=round] ( 93.11,193.42) -- (160.06,193.42);

\path[draw=drawColor,line width= 0.6pt,line join=round] (160.06,193.42) -- (160.06,190.23);

\path[draw=drawColor,line width= 0.6pt,line join=round] (160.06,190.23) -- (160.69,190.23);

\path[draw=drawColor,line width= 0.6pt,line join=round] (160.69,190.23) -- (160.69,188.48);

\path[draw=drawColor,line width= 0.6pt,line join=round] (160.69,188.48) -- (168.30,188.48);

\path[draw=drawColor,line width= 0.6pt,line join=round] (160.69,190.23) -- (160.69,191.97);

\path[draw=drawColor,line width= 0.6pt,line join=round] (160.69,191.97) -- (161.54,191.97);

\path[draw=drawColor,line width= 0.6pt,line join=round] (161.54,191.97) -- (161.54,190.81);

\path[draw=drawColor,line width= 0.6pt,line join=round] (161.54,190.81) -- (168.30,190.81);

\path[draw=drawColor,line width= 0.6pt,line join=round] (161.54,191.97) -- (161.54,193.13);

\path[draw=drawColor,line width= 0.6pt,line join=round] (161.54,193.13) -- (168.30,193.13);

\path[draw=drawColor,line width= 0.6pt,line join=round] (160.06,193.42) -- (160.06,196.61);

\path[draw=drawColor,line width= 0.6pt,line join=round] (160.06,196.61) -- (160.60,196.61);

\path[draw=drawColor,line width= 0.6pt,line join=round] (160.60,196.61) -- (160.60,195.45);

\path[draw=drawColor,line width= 0.6pt,line join=round] (160.60,195.45) -- (168.30,195.45);

\path[draw=drawColor,line width= 0.6pt,line join=round] (160.60,196.61) -- (160.60,197.77);

\path[draw=drawColor,line width= 0.6pt,line join=round] (160.60,197.77) -- (168.30,197.77);

\path[draw=drawColor,line width= 0.6pt,line join=round] ( 84.70,196.25) -- ( 84.70,202.71);

\path[draw=drawColor,line width= 0.6pt,line join=round] ( 84.70,202.71) -- ( 91.86,202.71);

\path[draw=drawColor,line width= 0.6pt,line join=round] ( 91.86,202.71) -- ( 91.86,200.09);

\path[draw=drawColor,line width= 0.6pt,line join=round] ( 91.86,200.09) -- (168.30,200.09);

\path[draw=drawColor,line width= 0.6pt,line join=round] ( 91.86,202.71) -- ( 91.86,205.32);

\path[draw=drawColor,line width= 0.6pt,line join=round] ( 91.86,205.32) -- (159.17,205.32);

\path[draw=drawColor,line width= 0.6pt,line join=round] (159.17,205.32) -- (159.17,202.42);

\path[draw=drawColor,line width= 0.6pt,line join=round] (159.17,202.42) -- (168.30,202.42);

\path[draw=drawColor,line width= 0.6pt,line join=round] (159.17,205.32) -- (159.17,208.22);

\path[draw=drawColor,line width= 0.6pt,line join=round] (159.17,208.22) -- (159.41,208.22);

\path[draw=drawColor,line width= 0.6pt,line join=round] (159.41,208.22) -- (159.41,205.90);

\path[draw=drawColor,line width= 0.6pt,line join=round] (159.41,205.90) -- (160.42,205.90);

\path[draw=drawColor,line width= 0.6pt,line join=round] (160.42,205.90) -- (160.42,204.74);

\path[draw=drawColor,line width= 0.6pt,line join=round] (160.42,204.74) -- (168.30,204.74);

\path[draw=drawColor,line width= 0.6pt,line join=round] (160.42,205.90) -- (160.42,207.06);

\path[draw=drawColor,line width= 0.6pt,line join=round] (160.42,207.06) -- (168.30,207.06);

\path[draw=drawColor,line width= 0.6pt,line join=round] (159.41,208.22) -- (159.41,210.54);

\path[draw=drawColor,line width= 0.6pt,line join=round] (159.41,210.54) -- (159.71,210.54);

\path[draw=drawColor,line width= 0.6pt,line join=round] (159.71,210.54) -- (159.71,209.38);

\path[draw=drawColor,line width= 0.6pt,line join=round] (159.71,209.38) -- (168.30,209.38);

\path[draw=drawColor,line width= 0.6pt,line join=round] (159.71,210.54) -- (159.71,211.71);

\path[draw=drawColor,line width= 0.6pt,line join=round] (159.71,211.71) -- (168.30,211.71);

\path[draw=drawColor,line width= 0.6pt,line join=round] ( 76.76,209.73) -- ( 76.76,236.94);

\path[draw=drawColor,line width= 0.6pt,line join=round] ( 76.76,236.94) -- ( 82.44,236.94);

\path[draw=drawColor,line width= 0.6pt,line join=round] ( 82.44,236.94) -- ( 82.44,220.78);

\path[draw=drawColor,line width= 0.6pt,line join=round] ( 82.44,220.78) -- ( 92.49,220.78);

\path[draw=drawColor,line width= 0.6pt,line join=round] ( 92.49,220.78) -- ( 92.49,216.06);

\path[draw=drawColor,line width= 0.6pt,line join=round] ( 92.49,216.06) -- (159.30,216.06);

\path[draw=drawColor,line width= 0.6pt,line join=round] (159.30,216.06) -- (159.30,214.03);

\path[draw=drawColor,line width= 0.6pt,line join=round] (159.30,214.03) -- (168.30,214.03);

\path[draw=drawColor,line width= 0.6pt,line join=round] (159.30,216.06) -- (159.30,218.09);

\path[draw=drawColor,line width= 0.6pt,line join=round] (159.30,218.09) -- (159.82,218.09);

\path[draw=drawColor,line width= 0.6pt,line join=round] (159.82,218.09) -- (159.82,216.35);

\path[draw=drawColor,line width= 0.6pt,line join=round] (159.82,216.35) -- (168.30,216.35);

\path[draw=drawColor,line width= 0.6pt,line join=round] (159.82,218.09) -- (159.82,219.83);

\path[draw=drawColor,line width= 0.6pt,line join=round] (159.82,219.83) -- (160.32,219.83);

\path[draw=drawColor,line width= 0.6pt,line join=round] (160.32,219.83) -- (160.32,218.67);

\path[draw=drawColor,line width= 0.6pt,line join=round] (160.32,218.67) -- (168.30,218.67);

\path[draw=drawColor,line width= 0.6pt,line join=round] (160.32,219.83) -- (160.32,221.00);

\path[draw=drawColor,line width= 0.6pt,line join=round] (160.32,221.00) -- (168.30,221.00);

\path[draw=drawColor,line width= 0.6pt,line join=round] ( 92.49,220.78) -- ( 92.49,225.49);

\path[draw=drawColor,line width= 0.6pt,line join=round] ( 92.49,225.49) -- (157.97,225.49);

\path[draw=drawColor,line width= 0.6pt,line join=round] (157.97,225.49) -- (157.97,223.32);

\path[draw=drawColor,line width= 0.6pt,line join=round] (157.97,223.32) -- (168.30,223.32);

\path[draw=drawColor,line width= 0.6pt,line join=round] (157.97,225.49) -- (157.97,227.67);

\path[draw=drawColor,line width= 0.6pt,line join=round] (157.97,227.67) -- (158.58,227.67);

\path[draw=drawColor,line width= 0.6pt,line join=round] (158.58,227.67) -- (158.58,225.64);

\path[draw=drawColor,line width= 0.6pt,line join=round] (158.58,225.64) -- (168.30,225.64);

\path[draw=drawColor,line width= 0.6pt,line join=round] (158.58,227.67) -- (158.58,229.70);

\path[draw=drawColor,line width= 0.6pt,line join=round] (158.58,229.70) -- (159.34,229.70);

\path[draw=drawColor,line width= 0.6pt,line join=round] (159.34,229.70) -- (159.34,227.96);

\path[draw=drawColor,line width= 0.6pt,line join=round] (159.34,227.96) -- (168.30,227.96);

\path[draw=drawColor,line width= 0.6pt,line join=round] (159.34,229.70) -- (159.34,231.45);

\path[draw=drawColor,line width= 0.6pt,line join=round] (159.34,231.45) -- (159.44,231.45);

\path[draw=drawColor,line width= 0.6pt,line join=round] (159.44,231.45) -- (159.44,230.28);

\path[draw=drawColor,line width= 0.6pt,line join=round] (159.44,230.28) -- (168.30,230.28);

\path[draw=drawColor,line width= 0.6pt,line join=round] (159.44,231.45) -- (159.44,232.61);

\path[draw=drawColor,line width= 0.6pt,line join=round] (159.44,232.61) -- (168.30,232.61);

\path[draw=drawColor,line width= 0.6pt,line join=round] ( 82.44,236.94) -- ( 82.44,253.11);

\path[draw=drawColor,line width= 0.6pt,line join=round] ( 82.44,253.11) -- ( 83.80,253.11);

\path[draw=drawColor,line width= 0.6pt,line join=round] ( 83.80,253.11) -- ( 83.80,243.71);

\path[draw=drawColor,line width= 0.6pt,line join=round] ( 83.80,243.71) -- ( 93.15,243.71);

\path[draw=drawColor,line width= 0.6pt,line join=round] ( 93.15,243.71) -- ( 93.15,238.70);

\path[draw=drawColor,line width= 0.6pt,line join=round] ( 93.15,238.70) -- (160.19,238.70);

\path[draw=drawColor,line width= 0.6pt,line join=round] (160.19,238.70) -- (160.19,236.09);

\path[draw=drawColor,line width= 0.6pt,line join=round] (160.19,236.09) -- (161.45,236.09);

\path[draw=drawColor,line width= 0.6pt,line join=round] (161.45,236.09) -- (161.45,234.93);

\path[draw=drawColor,line width= 0.6pt,line join=round] (161.45,234.93) -- (168.30,234.93);

\path[draw=drawColor,line width= 0.6pt,line join=round] (161.45,236.09) -- (161.45,237.25);

\path[draw=drawColor,line width= 0.6pt,line join=round] (161.45,237.25) -- (168.30,237.25);

\path[draw=drawColor,line width= 0.6pt,line join=round] (160.19,238.70) -- (160.19,241.31);

\path[draw=drawColor,line width= 0.6pt,line join=round] (160.19,241.31) -- (160.86,241.31);

\path[draw=drawColor,line width= 0.6pt,line join=round] (160.86,241.31) -- (160.86,239.57);

\path[draw=drawColor,line width= 0.6pt,line join=round] (160.86,239.57) -- (168.30,239.57);

\path[draw=drawColor,line width= 0.6pt,line join=round] (160.86,241.31) -- (160.86,243.06);

\path[draw=drawColor,line width= 0.6pt,line join=round] (160.86,243.06) -- (161.14,243.06);

\path[draw=drawColor,line width= 0.6pt,line join=round] (161.14,243.06) -- (161.14,241.90);

\path[draw=drawColor,line width= 0.6pt,line join=round] (161.14,241.90) -- (168.30,241.90);

\path[draw=drawColor,line width= 0.6pt,line join=round] (161.14,243.06) -- (161.14,244.22);

\path[draw=drawColor,line width= 0.6pt,line join=round] (161.14,244.22) -- (168.30,244.22);
\definecolor{drawColor}{RGB}{0,0,255}

\path[draw=drawColor,line width= 0.6pt,line join=round] ( 93.15,243.71) -- ( 93.15,248.72);

\path[draw=drawColor,line width= 0.6pt,line join=round] ( 93.15,248.72) -- (159.54,248.72);

\path[draw=drawColor,line width= 0.6pt,line join=round] (159.54,248.72) -- (159.54,246.54);

\path[draw=drawColor,line width= 0.6pt,line join=round] (159.54,246.54) -- (168.30,246.54);

\path[draw=drawColor,line width= 0.6pt,line join=round] (159.54,248.72) -- (159.54,250.89);

\path[draw=drawColor,line width= 0.6pt,line join=round] (159.54,250.89) -- (160.77,250.89);

\path[draw=drawColor,line width= 0.6pt,line join=round] (160.77,250.89) -- (160.77,248.86);

\path[draw=drawColor,line width= 0.6pt,line join=round] (160.77,248.86) -- (168.30,248.86);

\path[draw=drawColor,line width= 0.6pt,line join=round] (160.77,250.89) -- (160.77,252.93);

\path[draw=drawColor,line width= 0.6pt,line join=round] (160.77,252.93) -- (161.01,252.93);

\path[draw=drawColor,line width= 0.6pt,line join=round] (161.01,252.93) -- (161.01,251.18);

\path[draw=drawColor,line width= 0.6pt,line join=round] (161.01,251.18) -- (168.30,251.18);

\path[draw=drawColor,line width= 0.6pt,line join=round] (161.01,252.93) -- (161.01,254.67);

\path[draw=drawColor,line width= 0.6pt,line join=round] (161.01,254.67) -- (161.70,254.67);

\path[draw=drawColor,line width= 0.6pt,line join=round] (161.70,254.67) -- (161.70,253.51);

\path[draw=drawColor,line width= 0.6pt,line join=round] (161.70,253.51) -- (168.30,253.51);

\path[draw=drawColor,line width= 0.6pt,line join=round] (161.70,254.67) -- (161.70,255.83);

\path[draw=drawColor,line width= 0.6pt,line join=round] (161.70,255.83) -- (168.30,255.83);
\definecolor{drawColor}{RGB}{0,0,0}

\path[draw=drawColor,line width= 0.6pt,line join=round] ( 83.80,253.11) -- ( 83.80,262.51);

\path[draw=drawColor,line width= 0.6pt,line join=round] ( 83.80,262.51) -- ( 88.95,262.51);

\path[draw=drawColor,line width= 0.6pt,line join=round] ( 88.95,262.51) -- ( 88.95,259.31);

\path[draw=drawColor,line width= 0.6pt,line join=round] ( 88.95,259.31) -- (158.50,259.31);

\path[draw=drawColor,line width= 0.6pt,line join=round] (158.50,259.31) -- (158.50,258.15);

\path[draw=drawColor,line width= 0.6pt,line join=round] (158.50,258.15) -- (168.30,258.15);

\path[draw=drawColor,line width= 0.6pt,line join=round] (158.50,259.31) -- (158.50,260.47);

\path[draw=drawColor,line width= 0.6pt,line join=round] (158.50,260.47) -- (168.30,260.47);

\path[draw=drawColor,line width= 0.6pt,line join=round] ( 88.95,262.51) -- ( 88.95,265.70);

\path[draw=drawColor,line width= 0.6pt,line join=round] ( 88.95,265.70) -- (158.29,265.70);

\path[draw=drawColor,line width= 0.6pt,line join=round] (158.29,265.70) -- (158.29,262.80);

\path[draw=drawColor,line width= 0.6pt,line join=round] (158.29,262.80) -- (168.30,262.80);

\path[draw=drawColor,line width= 0.6pt,line join=round] (158.29,265.70) -- (158.29,268.60);

\path[draw=drawColor,line width= 0.6pt,line join=round] (158.29,268.60) -- (159.19,268.60);

\path[draw=drawColor,line width= 0.6pt,line join=round] (159.19,268.60) -- (159.19,266.28);

\path[draw=drawColor,line width= 0.6pt,line join=round] (159.19,266.28) -- (160.04,266.28);

\path[draw=drawColor,line width= 0.6pt,line join=round] (160.04,266.28) -- (160.04,265.12);

\path[draw=drawColor,line width= 0.6pt,line join=round] (160.04,265.12) -- (168.30,265.12);

\path[draw=drawColor,line width= 0.6pt,line join=round] (160.04,266.28) -- (160.04,267.44);

\path[draw=drawColor,line width= 0.6pt,line join=round] (160.04,267.44) -- (168.30,267.44);

\path[draw=drawColor,line width= 0.6pt,line join=round] (159.19,268.60) -- (159.19,270.92);

\path[draw=drawColor,line width= 0.6pt,line join=round] (159.19,270.92) -- (159.34,270.92);

\path[draw=drawColor,line width= 0.6pt,line join=round] (159.34,270.92) -- (159.34,269.76);

\path[draw=drawColor,line width= 0.6pt,line join=round] (159.34,269.76) -- (168.30,269.76);

\path[draw=drawColor,line width= 0.6pt,line join=round] (159.34,270.92) -- (159.34,272.08);

\path[draw=drawColor,line width= 0.6pt,line join=round] (159.34,272.08) -- (168.30,272.08);
\definecolor{drawColor}{gray}{0.20}

\path[draw=drawColor,line width= 0.6pt,line join=round,line cap=round] (  5.50, 30.69) rectangle (176.06,283.58);
\end{scope}
\begin{scope}
\definecolor{drawColor}{gray}{0.20}

\path[draw=drawColor,line width= 0.6pt,line join=round] (176.06, 63.08) --
	(178.81, 63.08);

\path[draw=drawColor,line width= 0.6pt,line join=round] (176.06,151.33) --
	(178.81,151.33);

\path[draw=drawColor,line width= 0.6pt,line join=round] (176.06,155.97) --
	(178.81,155.97);

\path[draw=drawColor,line width= 0.6pt,line join=round] (176.06,160.62) --
	(178.81,160.62);

\path[draw=drawColor,line width= 0.6pt,line join=round] (176.06,246.54) --
	(178.81,246.54);

\path[draw=drawColor,line width= 0.6pt,line join=round] (176.06,248.86) --
	(178.81,248.86);
\end{scope}
\begin{scope}
\definecolor{drawColor}{gray}{0.30}

\node[text=drawColor,anchor=base west,inner sep=0pt, outer sep=0pt, scale=  0.88] at (181.01, 60.05) {$\theta_{14}$};

\node[text=drawColor,anchor=base west,inner sep=0pt, outer sep=0pt, scale=  0.88] at (181.01,146) {$\theta_1$};

\node[text=drawColor,anchor=base west,inner sep=0pt, outer sep=0pt, scale=  0.88] at (181.01,152.94) {$\theta_4$};

\node[text=drawColor,anchor=base west,inner sep=0pt, outer sep=0pt, scale=  0.88] at (181.01,160) {$\theta_5$};

\node[text=drawColor,anchor=base west,inner sep=0pt, outer sep=0pt, scale=  0.88] at (181.01,241) {$\theta_9$};

\node[text=drawColor,anchor=base west,inner sep=0pt, outer sep=0pt, scale=  0.88] at (181.01,248.3) {$\theta_7$};
\end{scope}
\begin{scope}
\definecolor{drawColor}{gray}{0.20}

\path[draw=drawColor,line width= 0.6pt,line join=round] (168.30, 27.94) --
	(168.30, 30.69);

\path[draw=drawColor,line width= 0.6pt,line join=round] (127.75, 27.94) --
	(127.75, 30.69);

\path[draw=drawColor,line width= 0.6pt,line join=round] ( 87.19, 27.94) --
	( 87.19, 30.69);

\path[draw=drawColor,line width= 0.6pt,line join=round] ( 46.63, 27.94) --
	( 46.63, 30.69);

\path[draw=drawColor,line width= 0.6pt,line join=round] (  6.07, 27.94) --
	(  6.07, 30.69);
\end{scope}
\begin{scope}
\definecolor{drawColor}{gray}{0.30}

\node[text=drawColor,anchor=base,inner sep=0pt, outer sep=0pt, scale=  0.88] at (168.30, 19.68) {0.0};

\node[text=drawColor,anchor=base,inner sep=0pt, outer sep=0pt, scale=  0.88] at (127.75, 19.68) {0.5};

\node[text=drawColor,anchor=base,inner sep=0pt, outer sep=0pt, scale=  0.88] at ( 87.19, 19.68) {1.0};

\node[text=drawColor,anchor=base,inner sep=0pt, outer sep=0pt, scale=  0.88] at ( 46.63, 19.68) {1.5};

\node[text=drawColor,anchor=base,inner sep=0pt, outer sep=0pt, scale=  0.88] at (  6.07, 19.68) {2.0};
\end{scope}
\begin{scope}
\definecolor{drawColor}{RGB}{0,0,0}

\node[text=drawColor,anchor=base,inner sep=0pt, outer sep=0pt, scale=  1.10] at ( 90.78,  6.64) {$\dist{\theta}{\theta^\prime}$};
\end{scope}
\begin{scope}
\definecolor{drawColor}{RGB}{0,0,0}

\node[text=drawColor,rotate=-90.00,anchor=base,inner sep=0pt, outer sep=0pt, scale=  1.10] at (201.60,157.13) {Predictor ($\theta$)};
\end{scope}
\end{tikzpicture}

%% file: tikz/elpd_plot_case_study_4.2.tex
\begin{tikzpicture}[x=1pt,y=1pt]
\definecolor{fillColor}{RGB}{255,255,255}
\begin{scope}
\definecolor{drawColor}{RGB}{255,255,255}
\definecolor{fillColor}{RGB}{255,255,255}

\path[draw=drawColor,line width= 0.6pt,line join=round,line cap=round,fill=fillColor] (  0.00,  0.00) rectangle (469.75,289.08);
\end{scope}
\begin{scope}
\definecolor{fillColor}{RGB}{255,255,255}

\path[fill=fillColor] ( 39.04, 30.69) rectangle (464.25,283.58);
\definecolor{drawColor}{gray}{0.92}

\path[draw=drawColor,line width= 0.6pt,line join=round] ( 39.04, 98.48) --
	(464.26, 98.48);

\path[draw=drawColor,line width= 0.6pt,line join=round] ( 39.04,176.50) --
	(464.26,176.50);

\path[draw=drawColor,line width= 0.6pt,line join=round] ( 39.04,254.51) --
	(464.26,254.51);
\definecolor{drawColor}{RGB}{255,0,0}

\path[draw=drawColor,line width= 0.6pt,dash pattern=on 7pt off 3pt ,line join=round] ( 39.04,254.51) -- (464.26,254.51);
\definecolor{drawColor}{RGB}{190,190,190}

\path[draw=drawColor,line width= 0.6pt,line join=round] ( 59.14, 63.13) --
	( 66.84,104.69) --
	( 74.54,122.05) --
	( 82.24,135.08) --
	( 89.94,147.69) --
	( 97.64,156.57) --
	(105.34,169.55) --
	(113.04,179.67) --
	(120.74,187.47) --
	(128.44,192.81) --
	(136.14,199.93) --
	(143.85,205.32) --
	(151.55,211.68) --
	(159.25,216.47) --
	(166.95,221.00) --
	(174.65,225.24) --
	(182.35,230.03) --
	(190.05,233.71) --
	(197.75,236.25) --
	(205.45,238.99) --
	(213.15,242.43) --
	(220.85,245.54) --
	(228.55,249.45) --
	(236.25,251.84) --
	(243.95,254.40) --
	(251.65,256.35) --
	(259.35,258.44) --
	(267.05,260.36) --
	(274.75,262.28) --
	(282.45,263.55) --
	(290.15,264.96) --
	(297.85,266.16) --
	(305.55,266.80) --
	(313.25,267.25) --
	(320.95,267.92) --
	(328.65,267.97) --
	(336.35,267.71) --
	(344.05,267.20) --
	(351.75,266.51) --
	(359.45,265.66) --
	(367.15,265.14) --
	(374.85,264.06) --
	(382.55,263.26) --
	(390.26,262.22) --
	(397.96,261.13) --
	(405.66,260.12) --
	(413.36,258.86) --
	(421.06,257.78) --
	(428.76,256.35) --
	(436.46,255.08) --
	(444.16,253.86);

\path[draw=drawColor,line width= 0.6pt,line join=round] ( 58.76, 42.18) -- ( 58.76, 84.08);

\path[draw=drawColor,line width= 0.6pt,line join=round] ( 66.46, 85.22) -- ( 66.46,124.16);

\path[draw=drawColor,line width= 0.6pt,line join=round] ( 74.16,103.46) -- ( 74.16,140.65);

\path[draw=drawColor,line width= 0.6pt,line join=round] ( 81.86,117.72) -- ( 81.86,152.44);

\path[draw=drawColor,line width= 0.6pt,line join=round] ( 89.56,130.73) -- ( 89.56,164.65);

\path[draw=drawColor,line width= 0.6pt,line join=round] ( 97.26,140.12) -- ( 97.26,173.02);

\path[draw=drawColor,line width= 0.6pt,line join=round] (104.96,154.16) -- (104.96,184.94);

\path[draw=drawColor,line width= 0.6pt,line join=round] (112.66,164.99) -- (112.66,194.34);

\path[draw=drawColor,line width= 0.6pt,line join=round] (120.36,173.13) -- (120.36,201.81);

\path[draw=drawColor,line width= 0.6pt,line join=round] (128.06,178.91) -- (128.06,206.70);

\path[draw=drawColor,line width= 0.6pt,line join=round] (135.76,186.41) -- (135.76,213.45);

\path[draw=drawColor,line width= 0.6pt,line join=round] (143.46,191.81) -- (143.46,218.83);

\path[draw=drawColor,line width= 0.6pt,line join=round] (151.16,198.71) -- (151.16,224.65);

\path[draw=drawColor,line width= 0.6pt,line join=round] (158.86,203.91) -- (158.86,229.02);

\path[draw=drawColor,line width= 0.6pt,line join=round] (166.56,208.65) -- (166.56,233.35);

\path[draw=drawColor,line width= 0.6pt,line join=round] (174.26,213.51) -- (174.26,236.97);

\path[draw=drawColor,line width= 0.6pt,line join=round] (181.96,218.64) -- (181.96,241.41);

\path[draw=drawColor,line width= 0.6pt,line join=round] (189.66,222.93) -- (189.66,244.50);

\path[draw=drawColor,line width= 0.6pt,line join=round] (197.36,225.62) -- (197.36,246.88);

\path[draw=drawColor,line width= 0.6pt,line join=round] (205.06,229.00) -- (205.06,248.99);

\path[draw=drawColor,line width= 0.6pt,line join=round] (212.76,232.60) -- (212.76,252.27);

\path[draw=drawColor,line width= 0.6pt,line join=round] (220.46,235.85) -- (220.46,255.22);

\path[draw=drawColor,line width= 0.6pt,line join=round] (228.16,240.35) -- (228.16,258.54);

\path[draw=drawColor,line width= 0.6pt,line join=round] (235.86,242.95) -- (235.86,260.74);

\path[draw=drawColor,line width= 0.6pt,line join=round] (243.56,246.05) -- (243.56,262.76);

\path[draw=drawColor,line width= 0.6pt,line join=round] (251.26,248.47) -- (251.26,264.24);

\path[draw=drawColor,line width= 0.6pt,line join=round] (258.96,250.98) -- (258.96,265.89);

\path[draw=drawColor,line width= 0.6pt,line join=round] (266.67,253.38) -- (266.67,267.33);

\path[draw=drawColor,line width= 0.6pt,line join=round] (274.37,255.66) -- (274.37,268.91);

\path[draw=drawColor,line width= 0.6pt,line join=round] (282.07,257.06) -- (282.07,270.04);

\path[draw=drawColor,line width= 0.6pt,line join=round] (289.77,258.92) -- (289.77,271.00);

\path[draw=drawColor,line width= 0.6pt,line join=round] (297.47,260.57) -- (297.47,271.74);

\path[draw=drawColor,line width= 0.6pt,line join=round] (305.17,261.66) -- (305.17,271.93);

\path[draw=drawColor,line width= 0.6pt,line join=round] (312.87,262.58) -- (312.87,271.92);

\path[draw=drawColor,line width= 0.6pt,line join=round] (320.57,263.75) -- (320.57,272.08);

\path[draw=drawColor,line width= 0.6pt,line join=round] (328.27,264.10) -- (328.27,271.84);

\path[draw=drawColor,line width= 0.6pt,line join=round] (336.35,264.38) -- (336.35,271.04);

\path[draw=drawColor,line width= 0.6pt,line join=round] (344.05,264.15) -- (344.05,270.25);

\path[draw=drawColor,line width= 0.6pt,line join=round] (351.75,263.82) -- (351.75,269.21);

\path[draw=drawColor,line width= 0.6pt,line join=round] (359.45,263.18) -- (359.45,268.14);

\path[draw=drawColor,line width= 0.6pt,line join=round] (367.15,262.96) -- (367.15,267.31);

\path[draw=drawColor,line width= 0.6pt,line join=round] (374.85,262.12) -- (374.85,266.00);

\path[draw=drawColor,line width= 0.6pt,line join=round] (382.55,261.64) -- (382.55,264.88);

\path[draw=drawColor,line width= 0.6pt,line join=round] (390.26,260.84) -- (390.26,263.59);

\path[draw=drawColor,line width= 0.6pt,line join=round] (397.96,259.90) -- (397.96,262.36);

\path[draw=drawColor,line width= 0.6pt,line join=round] (405.66,259.04) -- (405.66,261.19);

\path[draw=drawColor,line width= 0.6pt,line join=round] (413.36,257.96) -- (413.36,259.77);

\path[draw=drawColor,line width= 0.6pt,line join=round] (421.06,256.96) -- (421.06,258.60);

\path[draw=drawColor,line width= 0.6pt,line join=round] (428.76,255.62) -- (428.76,257.07);

\path[draw=drawColor,line width= 0.6pt,line join=round] (436.46,254.46) -- (436.46,255.70);

\path[draw=drawColor,line width= 0.6pt,line join=round] (444.16,253.29) -- (444.16,254.43);
\definecolor{drawColor}{RGB}{0,0,0}

\path[draw=drawColor,line width= 0.6pt,line join=round] ( 59.53, 42.19) -- ( 59.53, 84.08);

\path[draw=drawColor,line width= 0.6pt,line join=round] ( 67.23, 85.23) -- ( 67.23,124.16);

\path[draw=drawColor,line width= 0.6pt,line join=round] ( 74.93, 98.81) -- ( 74.93,135.70);

\path[draw=drawColor,line width= 0.6pt,line join=round] ( 82.63,117.74) -- ( 82.63,152.46);

\path[draw=drawColor,line width= 0.6pt,line join=round] ( 90.33,130.75) -- ( 90.33,164.67);

\path[draw=drawColor,line width= 0.6pt,line join=round] ( 98.03, 97.45) -- ( 98.03,130.03);

\path[draw=drawColor,line width= 0.6pt,line join=round] (105.73,150.37) -- (105.73,180.89);

\path[draw=drawColor,line width= 0.6pt,line join=round] (113.43,165.02) -- (113.43,194.37);

\path[draw=drawColor,line width= 0.6pt,line join=round] (121.13,173.18) -- (121.13,201.85);

\path[draw=drawColor,line width= 0.6pt,line join=round] (128.83,158.10) -- (128.83,186.02);

\path[draw=drawColor,line width= 0.6pt,line join=round] (136.53,169.13) -- (136.53,196.25);

\path[draw=drawColor,line width= 0.6pt,line join=round] (144.23,174.67) -- (144.23,201.18);

\path[draw=drawColor,line width= 0.6pt,line join=round] (151.93,177.66) -- (151.93,203.95);

\path[draw=drawColor,line width= 0.6pt,line join=round] (159.63,182.80) -- (159.63,208.18);

\path[draw=drawColor,line width= 0.6pt,line join=round] (167.33,180.64) -- (167.33,205.79);

\path[draw=drawColor,line width= 0.6pt,line join=round] (175.03,199.66) -- (175.03,223.82);

\path[draw=drawColor,line width= 0.6pt,line join=round] (182.73,210.92) -- (182.73,234.16);

\path[draw=drawColor,line width= 0.6pt,line join=round] (190.43,209.22) -- (190.43,230.82);

\path[draw=drawColor,line width= 0.6pt,line join=round] (198.13,184.27) -- (198.13,205.57);

\path[draw=drawColor,line width= 0.6pt,line join=round] (205.83,193.88) -- (205.83,214.26);

\path[draw=drawColor,line width= 0.6pt,line join=round] (213.53,205.70) -- (213.53,225.94);

\path[draw=drawColor,line width= 0.6pt,line join=round] (221.23,220.64) -- (221.23,239.66);

\path[draw=drawColor,line width= 0.6pt,line join=round] (228.93,219.18) -- (228.93,237.62);

\path[draw=drawColor,line width= 0.6pt,line join=round] (236.63,218.38) -- (236.63,235.66);

\path[draw=drawColor,line width= 0.6pt,line join=round] (244.33,238.94) -- (244.33,255.40);

\path[draw=drawColor,line width= 0.6pt,line join=round] (252.03,243.48) -- (252.03,258.98);

\path[draw=drawColor,line width= 0.6pt,line join=round] (259.73,247.54) -- (259.73,261.63);

\path[draw=drawColor,line width= 0.6pt,line join=round] (267.44,227.02) -- (267.44,240.30);

\path[draw=drawColor,line width= 0.6pt,line join=round] (275.14,245.86) -- (275.14,258.44);

\path[draw=drawColor,line width= 0.6pt,line join=round] (282.84,227.63) -- (282.84,239.30);

\path[draw=drawColor,line width= 0.6pt,line join=round] (290.54,230.16) -- (290.54,241.16);

\path[draw=drawColor,line width= 0.6pt,line join=round] (298.24,249.46) -- (298.24,259.35);

\path[draw=drawColor,line width= 0.6pt,line join=round] (305.94,243.10) -- (305.94,252.23);

\path[draw=drawColor,line width= 0.6pt,line join=round] (313.64,244.62) -- (313.64,253.20);

\path[draw=drawColor,line width= 0.6pt,line join=round] (321.34,260.61) -- (321.34,268.49);

\path[draw=drawColor,line width= 0.6pt,line join=round] (329.04,245.90) -- (329.04,253.17);
\definecolor{drawColor}{RGB}{190,190,190}
\definecolor{fillColor}{RGB}{190,190,190}

\path[draw=drawColor,line width= 0.8pt,line join=round,line cap=round,fill=fillColor] ( 58.76, 63.13) circle (  2.85);

\path[draw=drawColor,line width= 0.8pt,line join=round,line cap=round,fill=fillColor] ( 66.46,104.69) circle (  2.85);

\path[draw=drawColor,line width= 0.8pt,line join=round,line cap=round,fill=fillColor] ( 74.16,122.05) circle (  2.85);

\path[draw=drawColor,line width= 0.8pt,line join=round,line cap=round,fill=fillColor] ( 81.86,135.08) circle (  2.85);

\path[draw=drawColor,line width= 0.8pt,line join=round,line cap=round,fill=fillColor] ( 89.56,147.69) circle (  2.85);

\path[draw=drawColor,line width= 0.8pt,line join=round,line cap=round,fill=fillColor] ( 97.26,156.57) circle (  2.85);

\path[draw=drawColor,line width= 0.8pt,line join=round,line cap=round,fill=fillColor] (104.96,169.55) circle (  2.85);

\path[draw=drawColor,line width= 0.8pt,line join=round,line cap=round,fill=fillColor] (112.66,179.67) circle (  2.85);

\path[draw=drawColor,line width= 0.8pt,line join=round,line cap=round,fill=fillColor] (120.36,187.47) circle (  2.85);

\path[draw=drawColor,line width= 0.8pt,line join=round,line cap=round,fill=fillColor] (128.06,192.81) circle (  2.85);

\path[draw=drawColor,line width= 0.8pt,line join=round,line cap=round,fill=fillColor] (135.76,199.93) circle (  2.85);

\path[draw=drawColor,line width= 0.8pt,line join=round,line cap=round,fill=fillColor] (143.46,205.32) circle (  2.85);

\path[draw=drawColor,line width= 0.8pt,line join=round,line cap=round,fill=fillColor] (151.16,211.68) circle (  2.85);

\path[draw=drawColor,line width= 0.8pt,line join=round,line cap=round,fill=fillColor] (158.86,216.47) circle (  2.85);

\path[draw=drawColor,line width= 0.8pt,line join=round,line cap=round,fill=fillColor] (166.56,221.00) circle (  2.85);

\path[draw=drawColor,line width= 0.8pt,line join=round,line cap=round,fill=fillColor] (174.26,225.24) circle (  2.85);

\path[draw=drawColor,line width= 0.8pt,line join=round,line cap=round,fill=fillColor] (181.96,230.03) circle (  2.85);

\path[draw=drawColor,line width= 0.8pt,line join=round,line cap=round,fill=fillColor] (189.66,233.71) circle (  2.85);

\path[draw=drawColor,line width= 0.8pt,line join=round,line cap=round,fill=fillColor] (197.36,236.25) circle (  2.85);

\path[draw=drawColor,line width= 0.8pt,line join=round,line cap=round,fill=fillColor] (205.06,238.99) circle (  2.85);

\path[draw=drawColor,line width= 0.8pt,line join=round,line cap=round,fill=fillColor] (212.76,242.43) circle (  2.85);

\path[draw=drawColor,line width= 0.8pt,line join=round,line cap=round,fill=fillColor] (220.46,245.54) circle (  2.85);

\path[draw=drawColor,line width= 0.8pt,line join=round,line cap=round,fill=fillColor] (228.16,249.45) circle (  2.85);

\path[draw=drawColor,line width= 0.8pt,line join=round,line cap=round,fill=fillColor] (235.86,251.84) circle (  2.85);

\path[draw=drawColor,line width= 0.8pt,line join=round,line cap=round,fill=fillColor] (243.56,254.40) circle (  2.85);

\path[draw=drawColor,line width= 0.8pt,line join=round,line cap=round,fill=fillColor] (251.26,256.35) circle (  2.85);

\path[draw=drawColor,line width= 0.8pt,line join=round,line cap=round,fill=fillColor] (258.96,258.44) circle (  2.85);

\path[draw=drawColor,line width= 0.8pt,line join=round,line cap=round,fill=fillColor] (266.67,260.36) circle (  2.85);

\path[draw=drawColor,line width= 0.8pt,line join=round,line cap=round,fill=fillColor] (274.37,262.28) circle (  2.85);

\path[draw=drawColor,line width= 0.8pt,line join=round,line cap=round,fill=fillColor] (282.07,263.55) circle (  2.85);

\path[draw=drawColor,line width= 0.8pt,line join=round,line cap=round,fill=fillColor] (289.77,264.96) circle (  2.85);

\path[draw=drawColor,line width= 0.8pt,line join=round,line cap=round,fill=fillColor] (297.47,266.16) circle (  2.85);

\path[draw=drawColor,line width= 0.8pt,line join=round,line cap=round,fill=fillColor] (305.17,266.80) circle (  2.85);

\path[draw=drawColor,line width= 0.8pt,line join=round,line cap=round,fill=fillColor] (312.87,267.25) circle (  2.85);

\path[draw=drawColor,line width= 0.8pt,line join=round,line cap=round,fill=fillColor] (320.57,267.92) circle (  2.85);

\path[draw=drawColor,line width= 0.8pt,line join=round,line cap=round,fill=fillColor] (328.27,267.97) circle (  2.85);

\path[draw=drawColor,line width= 0.8pt,line join=round,line cap=round,fill=fillColor] (336.35,267.71) circle (  2.85);

\path[draw=drawColor,line width= 0.8pt,line join=round,line cap=round,fill=fillColor] (344.05,267.20) circle (  2.85);

\path[draw=drawColor,line width= 0.8pt,line join=round,line cap=round,fill=fillColor] (351.75,266.51) circle (  2.85);

\path[draw=drawColor,line width= 0.8pt,line join=round,line cap=round,fill=fillColor] (359.45,265.66) circle (  2.85);

\path[draw=drawColor,line width= 0.8pt,line join=round,line cap=round,fill=fillColor] (367.15,265.14) circle (  2.85);

\path[draw=drawColor,line width= 0.8pt,line join=round,line cap=round,fill=fillColor] (374.85,264.06) circle (  2.85);

\path[draw=drawColor,line width= 0.8pt,line join=round,line cap=round,fill=fillColor] (382.55,263.26) circle (  2.85);

\path[draw=drawColor,line width= 0.8pt,line join=round,line cap=round,fill=fillColor] (390.26,262.22) circle (  2.85);

\path[draw=drawColor,line width= 0.8pt,line join=round,line cap=round,fill=fillColor] (397.96,261.13) circle (  2.85);

\path[draw=drawColor,line width= 0.8pt,line join=round,line cap=round,fill=fillColor] (405.66,260.12) circle (  2.85);

\path[draw=drawColor,line width= 0.8pt,line join=round,line cap=round,fill=fillColor] (413.36,258.86) circle (  2.85);

\path[draw=drawColor,line width= 0.8pt,line join=round,line cap=round,fill=fillColor] (421.06,257.78) circle (  2.85);

\path[draw=drawColor,line width= 0.8pt,line join=round,line cap=round,fill=fillColor] (428.76,256.35) circle (  2.85);

\path[draw=drawColor,line width= 0.8pt,line join=round,line cap=round,fill=fillColor] (436.46,255.08) circle (  2.85);

\path[draw=drawColor,line width= 0.8pt,line join=round,line cap=round,fill=fillColor] (444.16,253.86) circle (  2.85);
\definecolor{drawColor}{RGB}{0,0,0}
\definecolor{fillColor}{RGB}{0,0,0}

\path[draw=drawColor,line width= 0.8pt,line join=round,line cap=round,fill=fillColor] ( 59.53, 63.13) circle (  2.85);

\path[draw=drawColor,line width= 0.8pt,line join=round,line cap=round,fill=fillColor] ( 67.23,104.70) circle (  2.85);

\path[draw=drawColor,line width= 0.8pt,line join=round,line cap=round,fill=fillColor] ( 74.93,117.26) circle (  2.85);

\path[draw=drawColor,line width= 0.8pt,line join=round,line cap=round,fill=fillColor] ( 82.63,135.10) circle (  2.85);

\path[draw=drawColor,line width= 0.8pt,line join=round,line cap=round,fill=fillColor] ( 90.33,147.71) circle (  2.85);

\path[draw=drawColor,line width= 0.8pt,line join=round,line cap=round,fill=fillColor] ( 98.03,113.74) circle (  2.85);

\path[draw=drawColor,line width= 0.8pt,line join=round,line cap=round,fill=fillColor] (105.73,165.63) circle (  2.85);

\path[draw=drawColor,line width= 0.8pt,line join=round,line cap=round,fill=fillColor] (113.43,179.70) circle (  2.85);

\path[draw=drawColor,line width= 0.8pt,line join=round,line cap=round,fill=fillColor] (121.13,187.52) circle (  2.85);

\path[draw=drawColor,line width= 0.8pt,line join=round,line cap=round,fill=fillColor] (128.83,172.06) circle (  2.85);

\path[draw=drawColor,line width= 0.8pt,line join=round,line cap=round,fill=fillColor] (136.53,182.69) circle (  2.85);

\path[draw=drawColor,line width= 0.8pt,line join=round,line cap=round,fill=fillColor] (144.23,187.93) circle (  2.85);

\path[draw=drawColor,line width= 0.8pt,line join=round,line cap=round,fill=fillColor] (151.93,190.80) circle (  2.85);

\path[draw=drawColor,line width= 0.8pt,line join=round,line cap=round,fill=fillColor] (159.63,195.49) circle (  2.85);

\path[draw=drawColor,line width= 0.8pt,line join=round,line cap=round,fill=fillColor] (167.33,193.21) circle (  2.85);

\path[draw=drawColor,line width= 0.8pt,line join=round,line cap=round,fill=fillColor] (175.03,211.74) circle (  2.85);

\path[draw=drawColor,line width= 0.8pt,line join=round,line cap=round,fill=fillColor] (182.73,222.54) circle (  2.85);

\path[draw=drawColor,line width= 0.8pt,line join=round,line cap=round,fill=fillColor] (190.43,220.02) circle (  2.85);

\path[draw=drawColor,line width= 0.8pt,line join=round,line cap=round,fill=fillColor] (198.13,194.92) circle (  2.85);

\path[draw=drawColor,line width= 0.8pt,line join=round,line cap=round,fill=fillColor] (205.83,204.07) circle (  2.85);

\path[draw=drawColor,line width= 0.8pt,line join=round,line cap=round,fill=fillColor] (213.53,215.82) circle (  2.85);

\path[draw=drawColor,line width= 0.8pt,line join=round,line cap=round,fill=fillColor] (221.23,230.15) circle (  2.85);

\path[draw=drawColor,line width= 0.8pt,line join=round,line cap=round,fill=fillColor] (228.93,228.40) circle (  2.85);

\path[draw=drawColor,line width= 0.8pt,line join=round,line cap=round,fill=fillColor] (236.63,227.02) circle (  2.85);

\path[draw=drawColor,line width= 0.8pt,line join=round,line cap=round,fill=fillColor] (244.33,247.17) circle (  2.85);

\path[draw=drawColor,line width= 0.8pt,line join=round,line cap=round,fill=fillColor] (252.03,251.23) circle (  2.85);

\path[draw=drawColor,line width= 0.8pt,line join=round,line cap=round,fill=fillColor] (259.73,254.58) circle (  2.85);

\path[draw=drawColor,line width= 0.8pt,line join=round,line cap=round,fill=fillColor] (267.44,233.66) circle (  2.85);

\path[draw=drawColor,line width= 0.8pt,line join=round,line cap=round,fill=fillColor] (275.14,252.15) circle (  2.85);

\path[draw=drawColor,line width= 0.8pt,line join=round,line cap=round,fill=fillColor] (282.84,233.47) circle (  2.85);

\path[draw=drawColor,line width= 0.8pt,line join=round,line cap=round,fill=fillColor] (290.54,235.66) circle (  2.85);

\path[draw=drawColor,line width= 0.8pt,line join=round,line cap=round,fill=fillColor] (298.24,254.41) circle (  2.85);

\path[draw=drawColor,line width= 0.8pt,line join=round,line cap=round,fill=fillColor] (305.94,247.66) circle (  2.85);

\path[draw=drawColor,line width= 0.8pt,line join=round,line cap=round,fill=fillColor] (313.64,248.91) circle (  2.85);

\path[draw=drawColor,line width= 0.8pt,line join=round,line cap=round,fill=fillColor] (321.34,264.55) circle (  2.85);

\path[draw=drawColor,line width= 0.8pt,line join=round,line cap=round,fill=fillColor] (329.04,249.54) circle (  2.85);
\definecolor{fillColor}{RGB}{0,0,255}

\path[fill=fillColor,fill opacity=0.20] ( 66.84,137.13) --
	( 74.54,147.19) --
	( 82.24,157.18) --
	( 89.94,163.04) --
	( 97.64,170.09) --
	(105.34,178.67) --
	(113.04,184.67) --
	(120.74,189.31) --
	(128.44,194.01) --
	(136.14,198.59) --
	(143.85,202.66) --
	(151.55,205.89) --
	(159.25,210.25) --
	(166.95,213.31) --
	(174.65,217.50) --
	(182.35,221.37) --
	(190.05,225.99) --
	(197.75,228.64) --
	(205.45,231.92) --
	(213.15,234.22) --
	(220.85,237.45) --
	(228.55,239.93) --
	(236.25,242.68) --
	(243.95,244.98) --
	(251.65,247.18) --
	(259.35,249.34) --
	(267.05,251.04) --
	(274.75,252.56) --
	(282.45,253.94) --
	(290.15,255.13) --
	(297.85,256.12) --
	(305.55,256.96) --
	(313.25,257.72) --
	(320.95,258.29) --
	(328.65,258.77) --
	(328.65,249.62) --
	(320.95,248.38) --
	(313.25,246.92) --
	(305.55,245.47) --
	(297.85,243.66) --
	(290.15,241.28) --
	(282.45,239.24) --
	(274.75,236.73) --
	(267.05,234.31) --
	(259.35,231.60) --
	(251.65,227.66) --
	(243.95,224.26) --
	(236.25,220.93) --
	(228.55,216.71) --
	(220.85,213.50) --
	(213.15,208.73) --
	(205.45,206.25) --
	(197.75,201.80) --
	(190.05,198.78) --
	(182.35,192.11) --
	(174.65,187.08) --
	(166.95,181.65) --
	(159.25,178.29) --
	(151.55,172.80) --
	(143.85,169.27) --
	(136.14,164.44) --
	(128.44,158.86) --
	(120.74,153.20) --
	(113.04,147.71) --
	(105.34,140.24) --
	( 97.64,129.07) --
	( 89.94,120.33) --
	( 82.24,113.46) --
	( 74.54,100.73) --
	( 66.84, 88.11) --
	cycle;

\path[] ( 66.84,137.13) --
	( 74.54,147.19) --
	( 82.24,157.18) --
	( 89.94,163.04) --
	( 97.64,170.09) --
	(105.34,178.67) --
	(113.04,184.67) --
	(120.74,189.31) --
	(128.44,194.01) --
	(136.14,198.59) --
	(143.85,202.66) --
	(151.55,205.89) --
	(159.25,210.25) --
	(166.95,213.31) --
	(174.65,217.50) --
	(182.35,221.37) --
	(190.05,225.99) --
	(197.75,228.64) --
	(205.45,231.92) --
	(213.15,234.22) --
	(220.85,237.45) --
	(228.55,239.93) --
	(236.25,242.68) --
	(243.95,244.98) --
	(251.65,247.18) --
	(259.35,249.34) --
	(267.05,251.04) --
	(274.75,252.56) --
	(282.45,253.94) --
	(290.15,255.13) --
	(297.85,256.12) --
	(305.55,256.96) --
	(313.25,257.72) --
	(320.95,258.29) --
	(328.65,258.77);

\path[] (328.65,249.62) --
	(320.95,248.38) --
	(313.25,246.92) --
	(305.55,245.47) --
	(297.85,243.66) --
	(290.15,241.28) --
	(282.45,239.24) --
	(274.75,236.73) --
	(267.05,234.31) --
	(259.35,231.60) --
	(251.65,227.66) --
	(243.95,224.26) --
	(236.25,220.93) --
	(228.55,216.71) --
	(220.85,213.50) --
	(213.15,208.73) --
	(205.45,206.25) --
	(197.75,201.80) --
	(190.05,198.78) --
	(182.35,192.11) --
	(174.65,187.08) --
	(166.95,181.65) --
	(159.25,178.29) --
	(151.55,172.80) --
	(143.85,169.27) --
	(136.14,164.44) --
	(128.44,158.86) --
	(120.74,153.20) --
	(113.04,147.71) --
	(105.34,140.24) --
	( 97.64,129.07) --
	( 89.94,120.33) --
	( 82.24,113.46) --
	( 74.54,100.73) --
	( 66.84, 88.11);
\definecolor{drawColor}{RGB}{0,0,255}

\path[draw=drawColor,line width= 0.6pt,line join=round] ( 66.84,112.62) --
	( 74.54,123.96) --
	( 82.24,135.32) --
	( 89.94,141.68) --
	( 97.64,149.58) --
	(105.34,159.46) --
	(113.04,166.19) --
	(120.74,171.25) --
	(128.44,176.43) --
	(136.14,181.52) --
	(143.85,185.96) --
	(151.55,189.34) --
	(159.25,194.27) --
	(166.95,197.48) --
	(174.65,202.29) --
	(182.35,206.74) --
	(190.05,212.39) --
	(197.75,215.22) --
	(205.45,219.08) --
	(213.15,221.47) --
	(220.85,225.48) --
	(228.55,228.32) --
	(236.25,231.81) --
	(243.95,234.62) --
	(251.65,237.42) --
	(259.35,240.47) --
	(267.05,242.68) --
	(274.75,244.64) --
	(282.45,246.59) --
	(290.15,248.21) --
	(297.85,249.89) --
	(305.55,251.22) --
	(313.25,252.32) --
	(320.95,253.34) --
	(328.65,254.20);
\definecolor{drawColor}{RGB}{190,190,190}

\path[draw=drawColor,line width= 0.6pt,dash pattern=on 7pt off 3pt ,line join=round] (290.15, 30.69) -- (290.15,283.58);
\definecolor{drawColor}{RGB}{255,0,0}

\node[text=drawColor,anchor=base,inner sep=0pt, outer sep=0pt, scale=  0.85] at ( 82.24,256.25) {Reference model elpd};
\definecolor{drawColor}{gray}{0.20}

\path[draw=drawColor,line width= 0.6pt,line join=round,line cap=round] ( 39.04, 30.69) rectangle (464.25,283.58);
\end{scope}
\begin{scope}
\definecolor{drawColor}{gray}{0.30}

\node[text=drawColor,anchor=base east,inner sep=0pt, outer sep=0pt, scale=  0.88] at ( 34.09, 95.45) {-100};

\node[text=drawColor,anchor=base east,inner sep=0pt, outer sep=0pt, scale=  0.88] at ( 34.09,173.47) {-50};

\node[text=drawColor,anchor=base east,inner sep=0pt, outer sep=0pt, scale=  0.88] at ( 34.09,251.48) {0};
\end{scope}
\begin{scope}
\definecolor{drawColor}{gray}{0.20}

\path[draw=drawColor,line width= 0.6pt,line join=round] ( 36.29, 98.48) --
	( 39.04, 98.48);

\path[draw=drawColor,line width= 0.6pt,line join=round] ( 36.29,176.50) --
	( 39.04,176.50);

\path[draw=drawColor,line width= 0.6pt,line join=round] ( 36.29,254.51) --
	( 39.04,254.51);
\end{scope}
\begin{scope}
\definecolor{drawColor}{gray}{0.20}

\path[draw=drawColor,line width= 0.6pt,line join=round] ( 59.14, 27.94) --
	( 59.14, 30.69);

\path[draw=drawColor,line width= 0.6pt,line join=round] ( 74.54, 27.94) --
	( 74.54, 30.69);

\path[draw=drawColor,line width= 0.6pt,line join=round] ( 89.94, 27.94) --
	( 89.94, 30.69);

\path[draw=drawColor,line width= 0.6pt,line join=round] (105.34, 27.94) --
	(105.34, 30.69);

\path[draw=drawColor,line width= 0.6pt,line join=round] (120.74, 27.94) --
	(120.74, 30.69);

\path[draw=drawColor,line width= 0.6pt,line join=round] (136.14, 27.94) --
	(136.14, 30.69);

\path[draw=drawColor,line width= 0.6pt,line join=round] (151.55, 27.94) --
	(151.55, 30.69);

\path[draw=drawColor,line width= 0.6pt,line join=round] (166.95, 27.94) --
	(166.95, 30.69);

\path[draw=drawColor,line width= 0.6pt,line join=round] (182.35, 27.94) --
	(182.35, 30.69);

\path[draw=drawColor,line width= 0.6pt,line join=round] (197.75, 27.94) --
	(197.75, 30.69);

\path[draw=drawColor,line width= 0.6pt,line join=round] (213.15, 27.94) --
	(213.15, 30.69);

\path[draw=drawColor,line width= 0.6pt,line join=round] (228.55, 27.94) --
	(228.55, 30.69);

\path[draw=drawColor,line width= 0.6pt,line join=round] (243.95, 27.94) --
	(243.95, 30.69);

\path[draw=drawColor,line width= 0.6pt,line join=round] (259.35, 27.94) --
	(259.35, 30.69);

\path[draw=drawColor,line width= 0.6pt,line join=round] (274.75, 27.94) --
	(274.75, 30.69);

\path[draw=drawColor,line width= 0.6pt,line join=round] (290.15, 27.94) --
	(290.15, 30.69);

\path[draw=drawColor,line width= 0.6pt,line join=round] (305.55, 27.94) --
	(305.55, 30.69);

\path[draw=drawColor,line width= 0.6pt,line join=round] (320.95, 27.94) --
	(320.95, 30.69);

\path[draw=drawColor,line width= 0.6pt,line join=round] (336.35, 27.94) --
	(336.35, 30.69);

\path[draw=drawColor,line width= 0.6pt,line join=round] (351.75, 27.94) --
	(351.75, 30.69);

\path[draw=drawColor,line width= 0.6pt,line join=round] (367.15, 27.94) --
	(367.15, 30.69);

\path[draw=drawColor,line width= 0.6pt,line join=round] (382.55, 27.94) --
	(382.55, 30.69);

\path[draw=drawColor,line width= 0.6pt,line join=round] (397.96, 27.94) --
	(397.96, 30.69);

\path[draw=drawColor,line width= 0.6pt,line join=round] (413.36, 27.94) --
	(413.36, 30.69);

\path[draw=drawColor,line width= 0.6pt,line join=round] (428.76, 27.94) --
	(428.76, 30.69);

\path[draw=drawColor,line width= 0.6pt,line join=round] (444.16, 27.94) --
	(444.16, 30.69);
\end{scope}
\begin{scope}
\definecolor{drawColor}{gray}{0.30}

\node[text=drawColor,anchor=base,inner sep=0pt, outer sep=0pt, scale=  0.88] at ( 59.14, 19.68) {0};

\node[text=drawColor,anchor=base,inner sep=0pt, outer sep=0pt, scale=  0.88] at ( 74.54, 19.68) {2};

\node[text=drawColor,anchor=base,inner sep=0pt, outer sep=0pt, scale=  0.88] at ( 89.94, 19.68) {4};

\node[text=drawColor,anchor=base,inner sep=0pt, outer sep=0pt, scale=  0.88] at (105.34, 19.68) {6};

\node[text=drawColor,anchor=base,inner sep=0pt, outer sep=0pt, scale=  0.88] at (120.74, 19.68) {8};

\node[text=drawColor,anchor=base,inner sep=0pt, outer sep=0pt, scale=  0.88] at (136.14, 19.68) {10};

\node[text=drawColor,anchor=base,inner sep=0pt, outer sep=0pt, scale=  0.88] at (151.55, 19.68) {12};

\node[text=drawColor,anchor=base,inner sep=0pt, outer sep=0pt, scale=  0.88] at (166.95, 19.68) {14};

\node[text=drawColor,anchor=base,inner sep=0pt, outer sep=0pt, scale=  0.88] at (182.35, 19.68) {16};

\node[text=drawColor,anchor=base,inner sep=0pt, outer sep=0pt, scale=  0.88] at (197.75, 19.68) {18};

\node[text=drawColor,anchor=base,inner sep=0pt, outer sep=0pt, scale=  0.88] at (213.15, 19.68) {20};

\node[text=drawColor,anchor=base,inner sep=0pt, outer sep=0pt, scale=  0.88] at (228.55, 19.68) {22};

\node[text=drawColor,anchor=base,inner sep=0pt, outer sep=0pt, scale=  0.88] at (243.95, 19.68) {24};

\node[text=drawColor,anchor=base,inner sep=0pt, outer sep=0pt, scale=  0.88] at (259.35, 19.68) {26};

\node[text=drawColor,anchor=base,inner sep=0pt, outer sep=0pt, scale=  0.88] at (274.75, 19.68) {28};

\node[text=drawColor,anchor=base,inner sep=0pt, outer sep=0pt, scale=  0.88] at (290.15, 19.68) {30};

\node[text=drawColor,anchor=base,inner sep=0pt, outer sep=0pt, scale=  0.88] at (305.55, 19.68) {32};

\node[text=drawColor,anchor=base,inner sep=0pt, outer sep=0pt, scale=  0.88] at (320.95, 19.68) {34};

\node[text=drawColor,anchor=base,inner sep=0pt, outer sep=0pt, scale=  0.88] at (336.35, 19.68) {36};

\node[text=drawColor,anchor=base,inner sep=0pt, outer sep=0pt, scale=  0.88] at (351.75, 19.68) {38};

\node[text=drawColor,anchor=base,inner sep=0pt, outer sep=0pt, scale=  0.88] at (367.15, 19.68) {40};

\node[text=drawColor,anchor=base,inner sep=0pt, outer sep=0pt, scale=  0.88] at (382.55, 19.68) {42};

\node[text=drawColor,anchor=base,inner sep=0pt, outer sep=0pt, scale=  0.88] at (397.96, 19.68) {44};

\node[text=drawColor,anchor=base,inner sep=0pt, outer sep=0pt, scale=  0.88] at (413.36, 19.68) {46};

\node[text=drawColor,anchor=base,inner sep=0pt, outer sep=0pt, scale=  0.88] at (428.76, 19.68) {48};

\node[text=drawColor,anchor=base,inner sep=0pt, outer sep=0pt, scale=  0.88] at (444.16, 19.68) {50};
\end{scope}
\begin{scope}
\definecolor{drawColor}{RGB}{0,0,0}

\node[text=drawColor,anchor=base,inner sep=0pt, outer sep=0pt, scale=  1.10] at (251.65,  7.64) {Model size};
\end{scope}
\begin{scope}
\definecolor{drawColor}{RGB}{0,0,0}

\node[text=drawColor,rotate= 90.00,anchor=base,inner sep=0pt, outer sep=0pt, scale=  1.10] at ( 13.08,157.13) {$\Delta$ elpd};
\end{scope}
\begin{scope}
\definecolor{fillColor}{RGB}{255,255,255}

\path[fill=fillColor] (300.21,116.93) rectangle (458.22,172.05);
\end{scope}
\begin{scope}
\definecolor{fillColor}{RGB}{255,255,255}

\path[fill=fillColor] (305.71,136.88) rectangle (320.16,151.34);
\end{scope}
\begin{scope}
\definecolor{drawColor}{RGB}{190,190,190}

\path[draw=drawColor,line width= 0.6pt,line join=round] (312.93,138.33) -- (312.93,149.89);
\definecolor{fillColor}{RGB}{190,190,190}

\path[draw=drawColor,line width= 0.8pt,line join=round,line cap=round,fill=fillColor] (312.93,144.11) circle (  2.85);
\end{scope}
\begin{scope}
\definecolor{fillColor}{RGB}{255,255,255}

\path[fill=fillColor] (305.71,122.43) rectangle (320.16,136.88);
\end{scope}
\begin{scope}
\definecolor{drawColor}{RGB}{0,0,0}

\path[draw=drawColor,line width= 0.6pt,line join=round] (312.93,123.87) -- (312.93,135.44);
\definecolor{fillColor}{RGB}{0,0,0}

\path[draw=drawColor,line width= 0.8pt,line join=round,line cap=round,fill=fillColor] (312.93,129.65) circle (  2.85);
\end{scope}
\begin{scope}
\definecolor{drawColor}{RGB}{0,0,0}

\node[text=drawColor,anchor=base west,inner sep=0pt, outer sep=0pt, scale=  0.88] at (325.66,141.08) {Single-path forward selection};
\end{scope}
\begin{scope}
\definecolor{drawColor}{RGB}{0,0,0}

\node[text=drawColor,anchor=base west,inner sep=0pt, outer sep=0pt, scale=  0.88] at (325.66,126.62) {Cross-validated forward selection};
\end{scope}
\end{tikzpicture}

%% file: appendix.tex
\section{Applications of projection predictive inference}\label{sec:references}
Projection predictive inference has found application in many fields, including medicine, psychology, and life sciences \citep[e.g.][]{nordstrom_time_2019,schrittenlocher_impact_2019,teng_spatial_2020,ng_noise_2020,greenop_invertebrate_2020,amirkhiz_investigating_2021,bartonicek_value_2021,milne_effects_2021,goto_genetic_2021,zhang_predicting_2021,sullivan_conventional_2022,mori_genotype-by-environment_2022,soliman_urine_2022,bratulic_noninvasive_2022,zhou_mapping_2022,leung_soo_development_2023,vanarsa_comprehensive_2023,kuck_prolonged_2023,digby_hidden_2023,bohn_great_2023,trier_emotions_2023,dean_continuous_2023,wirthgen_identifying_2023,schiller_fecal_2024}, chemistry \citep[e.g.][]{rosso_uniting_2019}, hydrology \citep[e.g.][]{mercer_atmospheric_2020}, and cultural studies and linguistics \citep[e.g.][]{karsdorp_cultural_2019}. 

Alongside \pkg{projpred}, there exist other model selection packages in \proglang{R}: \pkg{BayesVarSel} \citep{garcia-donato_bayesian_2018}; \pkg{BAS} \citep{clyde_bas_2022}; \pkg{varbvs} \citep{carbonetto_scalable_2012}; \pkg{spikeSlabGAM} \citep{scheipl_spikeslabgam_2011}; \pkg{BVSNLP} \citep{nikooienejad_bvsnlp_2020}, \pkg{ptycho} \citep{stell_ptycho_2015}; \pkg{BayesSUR} \citep{zhao_bayessur_2021}; \pkg{BGLR} \citep{perez_genome-wide_2014}; \pkg{MBSGS} \citep{liquet_mbsgs_2017}; and \pkg{mombf} \citep{rossell_mombf_2023}. A check of \pkg{cranlogs} \citep{csardi_cranlogs_2019} revealed that \pkg{projpred} is the most popular of these in terms of number of downloads (last check: April 26, 2023).
\section{The R2D2 prior}\label{appendix:R2D2}
\begin{table}[!h]
    \centering
    \addtolength{\tabcolsep}{-0.4em}
    \begin{tabular}{llccccc}
        \toprule
        Case study & Dataset & $n$ & $p$ & $\xi$ & $\mu_{R^2}$ & $\varphi_{R^2}$ \\ \midrule
        Section~\ref{sec:high-corr} & Highly-correlated & $500$ & $100$ & $1$ & $0.3$ & $5$ \\
        & predictors & & & & &  \\
        Section~\ref{sec:weakly-relevant} & Weakly-relevant & $500$ & $50$ & $10$ & $0.3$ & $5$ \\
        & predictors & & & & & \\
        Section~\ref{sec:nutrimouse} & Prostate cancer & $102$ & $5966$ & $1$ & $0.3$ & $5$ \\
        Section~\ref{sec:over-confidence} & Linear regression & $100$ & $60$ & $0.2$ & $0.3$ & $20$ \\
        & simulation & & & & & \\
        \bottomrule \\
    \end{tabular}
    \caption{R2D2 prior hyperparameter values used across case studies.}
    \label{tab:r2d2-hyperparams}
\end{table}
Fitting high-dimensional models on relatively sparse data is a difficult task in general. Joint shrinkage priors implore the practitioner to encode a prior jointly over the predictor-related parameter space and the predictive space. The R2D2 prior \citep{zhang_bayesian_2022,r2d2,aguilar_intuitive_2023,yanchenko_r2d2_2023} allows the statistician to encode some prior belief on the model's $R^2$ and the number of predictors necessary to achieve it. In doing so, one mitigates the risk of over-fitting the data present when using more simple independent Gaussians over the regression coefficients. Formally, the R2D2 model as we use it is defined as:
\begin{IEEEeqnarray}{rl}
    y_i \;&\sim \normal(\beta_0 + \sum_{k=1}^{p}x_{k,i}\beta_k,\sigma^2) \nonumber \\
  \beta_0 \;&\sim \studentt_3(0, 2.5) \nonumber \\
  \sigma \;&\sim \studentt^{+}_3(0, 2.5) \nonumber \\
  \beta_k \;&\sim \normal(0, \sigma^2 \tau^2 \phi_k) \nonumber \\
  R^2 \;&\sim \betadist(\mu_{R^2}, \varphi_{R^2}) \nonumber \\
  \phi \;&\sim \dirichlet(\xi,\dotsc,\xi) \nonumber \\
  \tau^2 \;&= \frac{R^2}{1-R^2}. \nonumber
\end{IEEEeqnarray}
By our notation, $\studentt_\nu(\mu, \sigma)$ denotes Student's $t$-distribution with $\nu$ degrees of freedom, location $\mu$, and scale $\sigma$. The super-script $+$ indicates a half-Student-$t$ distribution. Note that we use the mean and a pseudo-precision to parameterise the beta distribution for $R^2$ as opposed to the conventional shape parameters $a > 0$ and $b > 0$. The relationship between the two parameterisations is as follows:
\begin{IEEEeqnarray}{rl}
\mu_{R^2} \;&= \frac{a}{a+b}, \nonumber\\
\varphi_{R^2} \;&= a + b. \nonumber
\end{IEEEeqnarray}
The hyperparameters used throughout the three case studies are tabulated in Table~\ref{tab:r2d2-hyperparams}. The Dirichlet concentration parameter $\xi$ (which we choose to be constant across dimensions) controls sparsity in the regression coefficients: larger $\xi$ imply more uniform regression coefficients (in magnitude), typically with more predictors explaining the variance of the response. In Section \ref{sec:over-confidence}, when assessing the calibration, the R2D2 prior was applied to all predictors except the one for which the calibration was being assessed. This predictor under inspection was given a standard normal prior.